\documentclass{article}
\usepackage[utf8]{inputenc}

\usepackage{geometry}
\usepackage{wrapfig}
\usepackage{multicol,caption}
\usepackage{titlesec}
\usepackage[T1]{fontenc}
\usepackage{times}
\usepackage{setspace}

\usepackage{amsmath}
\usepackage{latexsym}
\usepackage{amsfonts}
\usepackage{mathrsfs}
\usepackage{upgreek}
\usepackage{graphics}
\usepackage{graphicx}
\usepackage{epstopdf}
\usepackage{epsfig}
\usepackage{amsbsy}
\usepackage{amsfonts}
\usepackage{amsmath}
\usepackage{amssymb}
\usepackage{bm}
\usepackage{color}

\title{A model for proppant dynamics in a perforated wellbore}

\author{E.V. Dontsov$^1$}
\date{
    $^1$egor@resfrac.com, ResFrac Corporation, Palo Alto, USA
}
    
\begin{document}
\maketitle


\begin{abstract}
    \noindent This paper presents a model to simulate behavior of particle-laden slurry in a horizontal perforated wellbore with the goal of quantifying fluid and particle distribution between the perforations. There are two primary phenomena that influence the result. The first one is the non-uniform particle distribution within the wellbore's cross-section and how it changes along the flow. The second phenomenon is related to the ability of particles to turn from the wellbore to a perforation. Consequently, the paper considers both of these phenomena independently at first, and then they are combined to address the whole problem of flow in a perforated wellbore. A mathematical model for calculating the particle and velocity profiles within the wellbore is developed. The model is calibrated against available laboratory data for various flow velocities, particle diameters, pipe diameters, and particle volume fractions. It predicts a steady-state solution for the particle and velocity profiles, as well as it captures the transition in time from a given state to the steady-state solution. The key dimensionless parameter that quantifies the latter solution is identified and is called dimensionless gravity. When it is small, the particles are fully suspended and the solution is uniform. At the same time, when the aforementioned parameter is large, then the solution is strongly non-uniform and resembles a flowing bed state. A mathematical model for the problem of particle turning is developed and is calibrated against available experimental and computational data. The key parameter affecting the result is called turning efficiency. When the efficiency is close to one, then most of the particles that follow the fluid streamlines going into the perforation are able enter the hole. At the same time, zero efficiency corresponds to the case of no particles entering the perforation. Solutions for the both sub-problems are combined to develop a model for the perforated wellbore. Results are compared (not calibrated) to a series of laboratory and field scale experiments for perforated wellbores. Comparison with the available computational results is presented as well. In addition, the comparison is presented in view of the parametric space defined by the dimensionless gravity and turning efficiency. Such a description allows to explain seemingly contradictory results observed in different tests and also allows to highlight parameters for which perforation orientation plays a significant role.
\end{abstract}


\section{Introduction}

The problem of slurry flow in a perforated pipe is primarily relevant for oil and gas industry, where it occurs during hydraulic fracturing treatment of rock formations. In particular, a slurry consisting of a carrying fluid and proppant particles is pumped down a wellbore, while the perforations represent the entry points into rock formation or  hydraulic fracture initiation points. Typically, the perforations are arranged in clusters, or several closely spaced holes. There can be more than ten clusters and typically there are up to six perforations within each cluster. The goal is to quantify the fluid and particle distribution between the perforations depending on particle and fluid properties, as well as perforation geometry and orientation. 

One of the earliest experimental studies~\cite{Grues1982} considered slurry flow in a perforated vertical well, targeting conventional hydraulic fracturing treatments. Authors observed that the amount of particles entering the perforation depends on particle size and viscosity, while the dependence on the perforation diameter was relatively small. In general, smaller particles and higher viscosities tend to increase the amount of proppant entering the perforation. Another interesting paper is related to laboratory investigation of slurry flow in T-junctions~\cite{Nasr1989}. Authors did not specifically target petroleum applications and considered equal pipe diameters in all directions, but nevertheless results in this paper are very useful for calibrating the model. Sensitivities to particle size, orientation, velocity, and concentration are presented. Recently, Computational Fluid Dynamic (CFD) simulations have been employed to investigate the problem of particle turning into a single perforation~\cite{Wu2016,Wu2018,Wu2019,WuPhD}. Results demonstrate that under certain conditions there is no sensitivity to particle size and carrying fluid viscosity, which seemingly contradicts the observations of~\cite{Grues1982}. But, as will be shown in this paper, these two groups of authors use very different parameters, which lead to dominance of different mechanisms and results in fundamentally different sensitivities to problem parameters. Finally, the effect of deviated or inclined wellbores is investigated in~\cite{Zhang2019}. It is shown that the result is independent of the wellbore orientation for high injection rates, while the inclination becomes important for low injection rates.

Many authors investigated the problem of slurry flow in a perforated wellbore. Most of the studies perform laboratory or large scale experiments, but some employ Computational Fluid Dynamic (CFD) simulations. A series of small scale laboratory experiments was done by~\cite{Ngameni2017,NgameniMS,Ahmad2019a,Ahmad2019b,AhmadPhD}. Three cluster geometry with various perforation configuration was considered. The effect of particle size, flow velocity, particle volume fraction, as well as the addition of High Viscosity Friction Reducer (HVFR) was investigated. What is interesting about this laboratory data set is that there are multiple cases, in which the first cluster receives more proppant, while the downstream clusters receive less particles. At the same time, once the flow velocity is increased, the first cluster receives less proppant, while the downstream clusters get more particles. This change of the trend is explained later in this paper by comparing the measurements to the developed model. Further extension of the laboratory investigation for the multi-cluster geometry is performed in~\cite{Ahmad2021}. The pipe with four clusters is used and the effect of various HVFRs was investigated. One important novelty of this experimental study lies in the fact that the amount of particles flowing from each individual perforation was measured. This makes the data much reacher and allows to better observe the effect of perforation orientation within a cluster. 

The study~\cite{Liu2021} used CFD modeling approach to replicate the measurements of~\cite{Ahmad2021}. It was observed that the addition of HVFR made the particle distribution nearly uniform and less dependent on perforation orientation. Thus, the experiments~\cite{Ahmad2021} demonstrated that HVFR significantly increased the ability of fluid to suspend particles. The authors in~\cite{Liu2021} also noted difficulties matching the observations with CFD modeling and the need to increase the apparent viscosity in the modeling. In addition, it is shown in~\cite{Liu2021} that the effect of stress shadow from the previous stage is significant and can strongly affect the resultant slurry distribution between the clusters, which in turn alters the corresponding proppant distribution as well. Another interesting observation is made in~\cite{Liu2021} that consecutive inline perforations improve the overall efficiency. The reason lies in the non-uniform particle distribution within the wellbore's cross-section immediately after a perforation. Some particles miss the perforation, which leads to an increased particle concentration imemediately downstream from the perforation. Thus, if another perforation is placed nearby with the same orientation, the locally increased particle concentration leads to bigger amount of proppant entering the next hole.

With regard to large scale modeling, one of the relatively early studies~\cite{Crespo2013} performed a series of yard tests to investigate the effect of various parameters on proppant distribution between perforations. Later, the study~\cite{Bokane2014} used CFD simulations to model the large scale tests from~\cite{Crespo2013}. The study~\cite{Almul2020} utilized a CFD model to analyze the proppant distribution within a hydraulic fracturing stage for realistic field parameters and accounting for perforation erosion. One interesting observation made in the paper is that particle concentration profile in the wellbore is non-uniform in general and changes from nearly uniform for the first cluster to strongly non-uniform for the last cluster. Another CFD modeling study~\cite{Wang2022} focused on the developing of machine learning algorithm to have an ability to quickly predict particle distribution for various configurations. This is perhaps the first study that explicitly outlined that the problem of slurry flow in a perforated wellbore has two sub-problems, one related to the non-uniform particle distribution in the wellbore, and one related to proppant turning. Also, the study~\cite{Wang2022} pointed out that fluid streamlines located closely to perforation enter the perforation. At the same time, only a fraction of proppant streamlines located within this zone enters the perforation. This is a very important observation and it will be used for constructing the model in this paper. It is also important to mention field scale measurements discussed in~\cite{Kolle2022,Snider2022}. Authors considered field scale stage length, wellbore diameter, and particle diameters. Three perforation configurations are considered: 8 clusters with 6 holes and 60$^\circ$ phasing, 13 clusters with 3 holes and 120$^\circ$ phasing, and 15 clusters with one perforation per cluster located at the bottom of the pipe. Results demonstrate different degree of uniformity of particle distribution between the clusters, with the last case having a single perforation per cluster located at the bottom of the well showing the least uniform placement. CFD modeling of these field scale results as well as some sensitivities are done in~\cite{Benish2022,SniderPPT}. 

The modeling effort of the problem is relatively unfilled. As was mentioned before, there are several sucessfull attempts to develop proxy models based on fitting~\cite{Wang2022,WuPhD}. One available model that stands up a little is presented in~\cite{Sinkov2021}. It utilizes a suspension flow model called Saskatchewan Research Council (SRC) model~\cite{Gillies2000,Gillies2004}. This is a two-layer model that assumes that there are two layers of particle concentration in the wellbore~\cite{Miedema2017}. One layer resembles a flowing bed and another one above it has lower particle concentration. This model is combined with a machine learning algorithm for the particle turning problem in~\cite{Sinkov2021} to address the whole problem of slurry dynamics in a perforated wellbore.

As becomes evident from all of the aforementioned studies, many researchers attempt to solve the whole problem of slurry dynamics in a perforated wellbore in one shot. Only a few realize that there are actually two sub-problems involved. While there are studies that focused predominantly on the proppant turning phenomenon, there are not many publications related to examining the particle distribution in the pipe. One of the notable exceptions, which is going to be used vigorously in this study, is the work by~\cite{GilliesPhD}. This study summarizes laboratory experiments that measure particle and velocity profiles in pipes for different velocities, pipe diameters, particle diameters, and particle concentrations. Arguably, the sub-problem of particle concentration profile in a turbulent flow in a pipe is more important for the resultant particle distribution between the clusters, than the problem of proppant turning. Therefore, the importance of this study is arguably higher than that of many others. 

The paper is organized as follows. Section~\ref{secprobform} first formulates the problem of slurry flow in a perforated wellbore and also outlines typical parameters that occur in the field. Section~\ref{secslurryflow} develops a mathematical model for the first sub-problem, namely, for the problem of turbulent slurry flow in a wellbore with the goal of finding the particle and velocity profiles. Section~\ref{secturn} develops a mathematical model for the second sub-problem, namely, for the problem of proppant turning into a perforation. Section~\ref{secperfwell} presents mathematical formulation for the whole problem of slurry dynamics in a perforated wellbore. Section~\ref{secmodelcal} contains calibration of both sub-models as well as the comparison with the available experimental and computational data for the multi-cluster geometry. Section~\ref{secdisc} discusses the results, as well as outlines potential experiments that can further fill the gaps in the understanding of this complex problem. And finally, summary of the results and observations is given in Section~\ref{secsumm}.

\section{Problem formulation}\label{secprobform}

\begin{figure}[h]
\centering \includegraphics[width=0.9\linewidth]{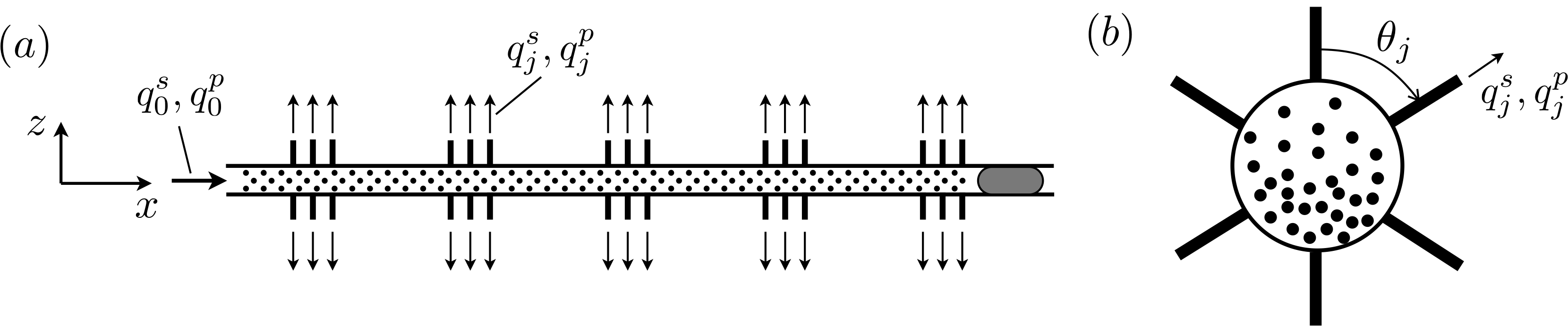}
\caption{Schematics for the slurry flow problem in the wellbore $(a)$ and wellbore cross-section $(b)$.}
\label{fig1}
\end{figure}

Schematics for the problem is shown in Fig.~\ref{fig1}. Suspension consisting of fluid mixed with particles flows in a horizontal wellbore and exits the pipe through a series of perforations. The latter perforations are organized in clusters or several closely spaced holes. The end of the pipe is sealed by a plug. The slurry enters the rock formation and forms hydraulic fractures after flowing through the perforations. The slurry rate $q^s_0$ and the particle rate $q^p_0$ at the entrance of the perforated region are known. The wellbore diameter is $d_w$. Each perforation is assumed to be circular with the diameter $d_p$. Perforation erosion is not included into the model at this stage. Each perforation is also quantified by the azimuth angle $\theta_j$ ($1\leqslant j\leqslant N_p$, where $N_p$ is the number of perforations). The latter angle is calculated clock-wise from the vertical axis. Location of every perforation is also assumed to be known. Fluid is characterized by its density $\rho_f$ and viscosity $\mu$, while particles are described by density $\rho_p$ and radius $a$. The goal is to determine the distribution of slurry and proppant flow rates $q^s_j$ and $q^p_j$ between the clusters.

While the problem seems to be trivial in the sense that the solution $q^s_j=q^s_0/N_p$ and $q^p_j=q^p_0/N_p$ seems to be a good approximation, this is not actually the case. In contrast, there are three coupled problems, each of which requires a detailed investigation. The first problem is related to the actual flow of slurry in a wellbore. The flow is fully turbulent and slows down towards the toe of the stage (the right side on Fig.~\ref{fig1}$(a)$). This causes the particles to settle in the wellbore leading to the non-uniform particle concentration in the cross-section, as illustrated in Fig.~\ref{fig1}$(b)$. This clearly affects the amount of proppant entering the perforations since the entry holes that are located in the lower part of the pipe take more particles due to settling. The second problem is also related to the difference in mass densities between particles and fluid. The slurry needs to take a sharp turn in order to enter a perforation. In this scenario, heavier proppant particles tend to miss the perforation and to continue flowing in the wellbore. This, in turn, contributes to the non-uniform particle concentration distribution between the perforations. Finally, there is perforation erosion. The more amount of proppant flows through the perforation, the more it erodes. This is also affected by the horizontal velocity component, which introduces a heel bias (i.e. to the left in Fig.~\ref{fig1}$(a)$) of the erosion pattern. The eroded perforations allow for bigger slurry flow rate and hence can take more proppant. This demonstrates that the seemingly trivial problem in reality is much more difficult and consists of three coupled problems that all affect the final result. It should also be noted that the effect of hydraulic fractures is neglected for simplicity, even though it can be significant. For instance, of there is a screenout in a fracture, then it cannot freely take more slurry, which effectively removes the perforation or even the whole perforation cluster from the system. Finally, due to the overall complexity of the problem, perforation erosion is not included in this study. Therefore, only two phenomena, namely proppant settling in the wellbore and the inertial effect  at the turning point, are considered in this study.

It is instructive to outline typical problem parameters that occur in the field. These will be used to estimate parameters and to simplify the model by focusing only on dominant effects. The total slurry flow rate is taken as $q^s_0 = 90$~bpm~$=0.24$~m$^3$/s. Note that the slurry flow rate in the wellbore slows down significantly towards the toe of the stage and also varies depending on the number of clusters used. The number of perforation clusters and holes per cluster varies, but typically there are between 4 and 15 clusters and up to 6 holes per cluster. The wellbore diameter is taken as $d_w = 4.5$~in~$=0.114$~m, while the perforation diameter as $d_p = 0.4$~in~$=0.01$~m. Fluid properties can vary noticeably, but for the purpose of this study the fluid viscosity is taken as $\mu = 1$~cp~$=1\times10^{-3}$~Pa$\cdot$s, while the fluid mass density $\rho_f = 1000$~kg/m$^3$. Various proppant sizes are used. The often used 40/70 sand has a distribution of particle sizes, but for the purpose of this study the median particle radius of $a=0.18$~mm is used, while the 100 mesh sand is assumed to be twice as small. The particle intrinsic mass density is $\rho_p = 2650$~kg/m$^3$. Typical average particle concentration is 2~ppg (lb/gal), which translates into the average flowing volume fraction of $q^p_0/q^s_0=0.08$. With these parameters in mind, Reynolds number in the pipe can reach $10^6$ at the entrance of the stage, while it reduces to almost $10^4$ right before the last perforation. The average slurry velocities vary from 10s of meters per second at the heel to a fraction of a meter per second at the toe. This confirms that the flow is fully turbulent throughout the whole zone of interest. Note that the friction reducers can delay the transition to turbulent flow and thus there is a higher chance of having a laminar or close to laminar flow near the last few perforations. The average velocity in the perforation is close to 50~m/s, while the Reynods number is approximately $10^5$, which also implies that the flow through perforations is fully turbulent. The turbulence makes the whole problem more challenging, but, at the same time, more interesting to solve.

\section{Slurry flow in a wellbore}\label{secslurryflow}

\subsection{Micro motion of particles in a turbulent flow}
  
Before proceeding with the primary governing equations for the proppant transport in a wellbore, it is instructive to consider the micro motion of particles in a turbulent flow. The turbulent flow is dynamic and has vortices, which causes particles to move at the micro scale, and, in particular, introduces a slip velocity between the two phases. This slip velocity cannot be neglected because it affects the overall drag at the macro scale due to non-linear nature of the turbulent friction.

While the fluid flow in the pipe is fully turbulent, the local flow around the particle can either be laminar or turbulent. Therefore, the particle drag should consider both laminar and turbulent components. As a result, the governing equation for a particle is (see e.g.~\cite{Crowe1998})
\begin{equation}\label{particledyn}
\rho_p V_p\dfrac{d\boldsymbol{v_p}}{dt} = \dfrac{1}{2}\rho_f V_p\dfrac{d(\boldsymbol{v_f}-\boldsymbol{v_p})}{dt} +\rho_f V_p\dfrac{d\boldsymbol{v_f}}{dt}+ 6\pi \mu a(\boldsymbol{v_f}-\boldsymbol{v_p})+C_d\dfrac{\pi a^2\rho_f}{2}|\boldsymbol{v_f}-\boldsymbol{v_p}| (\boldsymbol{v_f}-\boldsymbol{v_p}).
\end{equation}
The first term on the right hand side represents the effect of added mass, the second term is the buoyancy force related to fluid acceleration, and the last two terms capture the viscous and turbulent drag force, respectively. In the above equation, $a$ is particle radius, $\rho_p$ is particle mass density, $\rho_f$ is fluid density, $\boldsymbol{v_p}$ is particle velocity vector, $\boldsymbol{v_f}$ is fluid velocity vector, and $V_p=4\pi a^3/3$ is the particle volume. In view of~(\ref{particledyn}), the total drag coefficient for a particle is approximated by
\begin{equation}\label{dragcoef}
C_{tot} = C_d + \dfrac{24}{\text{Re}_p},\qquad \text{Re}_p = \dfrac{2a\rho_f|\boldsymbol{v_f}-\boldsymbol{v_p}|}{\mu},
\end{equation}
where $\text{Re}_p$ is particle Reynolds number, while $C_d=0.47$ is the drag coefficient for a sphere for the fully turbulent case. By introducing the slip velocity, equation~(\ref{particledyn}) can be simplified to
\begin{equation}\label{particledynslip}
\Bigl(\rho_p\!+\!\frac{1}{2}\rho_f\Bigr)\dfrac{d\Delta \boldsymbol{v}}{dt} = -(\rho_p\!-\!\rho_f) \dfrac{d\boldsymbol{v_f}}{dt}-\dfrac{9\mu}{2a^2}\Delta \boldsymbol{v}-C_d\dfrac{3\rho_f}{8a}|\Delta\boldsymbol{v}| \Delta\boldsymbol{v},\qquad \Delta\boldsymbol{v} = \boldsymbol{v_p}-\boldsymbol{v_f}.
\end{equation}
This form of the equation can now be used to analyze motion of a particle. It shows that the source of particle acceleration is fluid acceleration and the density contrast, while the force that prevents the relative motion is the drag force.

Consider particle dynamics under circular fluid motion, which represents a vortex. Let the fluid velocity vector be $\boldsymbol v_f =  p_Vv_w (\cos(\omega t),\sin(\omega t))$, where it is assumed that the characteristic vortex velocity is proportional to the average flow velocity $v_w$ and $p_V$ is a numeric parameter. The characteristic frequency can be estimated by relating the wall shear stress to the pressure drop as
\begin{equation}\label{tauwall}
\tau_w = \mu\omega = -\dfrac{d_w}{4}\dfrac{\partial p}{\partial x} =  \dfrac{f_D \rho_f v_w^2}{8}.
\end{equation}
The above equation provides an estimate for the shear rate or vortex frequency at the wall. It is assumed that the average shear rate inside the flow is of the same order. The right hand side of the equation represents turbulent pipe friction with $f_D$ being the friction factor. This friction coefficient is approximately constant for turbulent flow under consideration, and, according to Moody's diagram~\cite{Mood1944}, $f_D=0.02\div0.08$ and depends on roughness. As was mentioned before, the addition of HVFRs can delay the transition to a fully turbulent flow and to effectively reduce the friction factor.

Since $|\boldsymbol v_f |$ is time independent, equation~(\ref{particledynslip}) essentially becomes linear for the circular motion and can be solved to find
\begin{equation}\label{slipsoldv}
|\Delta v| = \dfrac{ \Delta\bar \rho\, \Omega}{\sqrt{1+\Omega^2}}\, p_Vv_w,\qquad |v_p| = |v_f+\Delta v| =  \dfrac{\sqrt{1+\Omega^2(1\!-\!\Delta\bar \rho)^2}}{\sqrt{1+\Omega^2}}\, p_Vv_w.
\end{equation}
where the dimensionless density contrast, dimensionless frequency, and particle Reynolds number are given by 
\begin{equation}\label{slipparams}
\Delta \bar\rho = \dfrac{\rho_p\!-\!\rho_f}{\rho_p\!+\!\frac{1}{2}\rho_f},\qquad\Omega =  \dfrac{2a^2\omega \Bigl(\rho_p\!+\!\frac{1}{2}\rho_f\Bigr)}{9\mu \Bigl(1+\dfrac{C_d}{24} \text{Re}_p\Bigr)},\qquad  \text{Re}_p = \dfrac{2a\rho_f |\Delta v|}{\mu}.
\end{equation}

Based on the considered field parameters, the typical values of $\Omega$ are $10\div 10^3$. This implies that $\Omega\gg 1$ and therefore the solution~(\ref{slipsoldv}) can be simplified to  
\begin{equation}\label{slipsolsimple}
|\Delta v| = \Delta\bar \rho\,p_V v_w,\qquad |v_p| = |v_f+\Delta u| =  (1\!-\!\Delta\bar \rho)\, p_Vv_w.
\end{equation}
It is important to note that the considered circular motion is also supplemented by a constant horizontal flow. Therefore, the result for the particle velocity actually applies only to the vertical direction.

The obtained result~(\ref{slipsolsimple}) has two main implications. First of all, the vertical particle velocity is reduced relative to the fluid velocity. And the degree of reduction is proportional to $1\!-\!\Delta\bar\rho$. For typical applications including sand in water, this factor is approximately equal to 0.5. But for sand in the air, on the other hand, this factor can be very small. The second implication of the result is that there is always a relative motion between the particles and fluid at the micro scale. Consequently, when there is a relative macro scale motion, the friction is governed by an apparent viscosity, which in turn depends on the micro motion. The latter apparent viscosity can be calculated as
\begin{equation}\label{apparvisc}
\mu_a = \mu \Bigl(1+\dfrac{C_d}{24} \text{Re}_p\Bigr),\qquad \text{Re}_p = \dfrac{2a\rho_f \Delta \bar \rho p_V v_w}{\mu}.
\end{equation}
Depending on the value of $\text{Re}_p$, the apparent viscosity can be either equal to viscosity or be significantly larger. This apparent viscosity is able to strongly reduce particle settling in the wellbore and provides an additional stopping mechanism for particles entering perforations.

It is also worth describing how the effect of non-Newtonian fluid rheology is included in the model. For the sake of simplicity, the viscosity is calculated using the characteristic value of the shear rate $\dot \gamma = \Delta \rho p_V v_w/(2a)$. Also, the cutoff viscosity equal to that of water is used to limit the apparent viscosity for high shear rates.

\subsection{Mathematical model for proppant transport in a wellbore}

The purpose of this section is to formulate the governing equations for proppant transport in the wellbore. In particular, the goal is to determine spatial variation of particle volume fraction and velocity, as well as their dynamics in the flow. It is first assumed that there is no macro scale slip velocity in the horizontal direction, or, in other words, the horizontal velocities for particles and fluid coincide. At the same time, there can be a macro scale slip velocity in the vertical direction due to settling. 

Given that the flow is fully turbulent, it becomes very difficult to develop a rigorous model. Therefore, the approach consists of a series of assumptions that allow to obtain a relatively simple result, which is then fitted to the available experimental data. First, it is assumed that the solution (i.e. particle concentration profile) depends solely on the vertical coordinate and that dynamics of particles in the vertical direction is governed by the linearized balance of linear momentum (see e.g.~\cite{Dont2014b}):
\begin{equation}\label{linmomvert}
\phi\rho_p\dfrac{\partial v_z}{\partial t} = -\dfrac{\partial p_p}{\partial z}-\phi(\rho_p\!-\!\rho_f)g - \dfrac{ 9\mu_a \phi v_z}{2a^2}.
\end{equation}
Here $\phi$ is particle volume fraction, $v_z$ is vertical component of particle velocity, $p_p$ is particle pressure, $z$ is the vertical coordinate pointing upwards, $\rho_p$ and $\rho_f$ are particle and fluid densities, $g$ is gravitational constant, $a$ is particle radius, and $\mu_a$ is apparent fluid viscosity that accounts for the micro motion of particles in turbulent flow~(\ref{apparvisc}). Equation~(\ref{linmomvert}) captures the dominant effects associated with particle buoyancy as well as viscous drag force. Stoke's law is used to calculate the drag force, which applies for laminar flow. This is justified since the vertical slip velocities are relatively small. By neglecting the acceleration and pressure gradient terms in~(\ref{linmomvert}), the slip velocity for 40/70 mesh proppant can be estimated as $v_s = 2a^2(\rho_p\!-\!\rho_f)g/(9\mu)=0.03$~m/s. This corresponds to Reynolds number of only 3.7 and therefore laminar approximation is still adequate. The micro motion of particles is included via apparent viscosity. Also, hindered settling is neglected here for simplicity. 

Particle volume balance can be written as
\begin{equation}\label{massbalance}
\dfrac{\partial S\phi }{\partial t} + \dfrac{\partial S\phi v_x}{\partial x} + \dfrac{\partial S\phi v_z}{\partial z} = 0,\qquad S = \sqrt{1\!-\!\dfrac{4z^2}{d_w^2}},
\end{equation}
where $S$ is the pipe's shape factor that accounts for the circular cross-section, $v_x$ and $v_z$ are the horizontal and vertical velocity components. Note that the origin of the $z$ coordinate is in the center of the pipe.

The next item in the model is constitutive model for slurry that provides the expression for the particle pressure and shear resistance (or effective viscosity). A suspension rheology was proposed in~\cite{Boyer2011} based on laboratory experiments. These experiments assumed laminar flow, which is clearly not the case for the problem under consideration. Therefore, an extension for turbulent flow is proposed here. In particular, the adopted expressions for the participle pressure and shear stress are
\begin{eqnarray}\label{rheology}
p_p &=& \eta_n(\phi)\mu\dot\gamma+f_T\phi
\rho_p (1\!-\!\Delta \bar \rho)^2 v_x^2,\quad \eta_n = \dfrac{\phi^2}{(\phi_m\!-\!\phi)^2},\\
\tau &=& \eta_s(\phi)\mu\dot\gamma,\quad \eta_s(\phi) = 1+2.5\phi_m{\cal A}(\phi)^{-1}+\Bigl(\mu_1 + \dfrac{\mu_2-\mu_1}{1\!+\!I_0{\cal A}(\phi)^{-2}}\Bigr){\cal A}(\phi)^{-2},\qquad {\cal A}(\phi) =  \dfrac{\phi_m}{\phi}-1,\notag
\end{eqnarray}
where $\phi_m=0.585$, $\mu_1=0.32$, $\mu_2 = 0.7$, $I_0=5\times10^{-3}$, while $\dot \gamma$ is shear rate. The difference from the original rheology~\cite{Boyer2011} lies in the term proportional to $f_T$, which is nonlinear in velocity and captures the effect of turbulence. Conceptually, this term represents particle pressure stemming from the same logic as in the kinetic theory of gases. Turbulent flow in the horizontal direction creates chaotic motion in the vertical direction too. This motion qualitatively resembles molecule motion in gas. Therefore, there is particle pressure that is proportional to average square vertical velocity, concentration, and density. The dimensionless coefficient $f_T$ is treated as a fitting parameter. Also note that~(\ref{slipsolsimple}) is used to account for the difference between the fluid and particle micro scale velocities that are relevant for particle pressure. It will be shown later that the turbulence term is crucial for capturing the experimental data.

The remaining unknown parameter in the slurry rheology~(\ref{rheology}) is the shear rate $\dot \gamma$. The estimate of $v_w/d_w$ does not apply because the flow is turbulent, which in turn creates vortices with their own local shear rate. By following the same logic as in deriving~(\ref{tauwall}), the shear rate can be estimated as
\begin{equation}\label{tauwall2}
\dot\gamma =  \dfrac{f_D \rho_f v_x^2}{8 \mu},
\end{equation}
where $f_D$ is now treated as a fitting parameter to match experimental data. Note that it does not matter if $\mu$ or $\mu_a$ is used in~(\ref{rheology}) and~(\ref{tauwall2}) since the product between viscosity and shear rate is actually used. 

Finally, it is also desired to model the non-uniform velocity profile across the pipe's cross-section. Velocity can vary because particle concentration can be non-uniform and higher concentrations lead to higher viscosity, as is apparent from~(\ref{rheology}). To this end, it is assumed that dissipation is spatially uniform in the flow, i.e.
\begin{equation}\label{diss}
D = \eta_s(\phi) \mu \dot\gamma^2 = \text{const.}
\end{equation}
This assumption leads to the following velocity variation
\begin{equation}\label{vdist}
v_x(z) = \dfrac{v_w}{\eta_s(\phi)^{1/4}I},\qquad I = \dfrac{4}{\pi d_w}\int_{-d_w/2}^{d_w/2} \dfrac{S(z) \,dz}{\eta_s(\phi(z))^{1/4}}.
\end{equation}
The assumption of constant dissipation is motivated primarily by the ability of the model to fit laboratory data~\cite{GilliesPhD}. It is possible to use other models in the form of $\eta_s\dot \gamma^p=\text{const.}$ It is found that $p$ in the range from approximately $1/4$ to $3/4$ also works. But the use of dissipation has at least some physical interpretation and therefore it is selected in this study.

\subsection{Dimensionless formulation}

The whole model~(\ref{linmomvert})--(\ref{vdist}) can be written in the dimensionless form as
\begin{eqnarray}\label{system1}
&& \bar \phi G {\cal T} \dfrac{\partial \bar v_z}{\partial \bar t} = -\dfrac{1}{I^2}\dfrac{\partial \Pi_p}{\partial \bar z}  -\bar \phi G(1+\bar v_z),\notag\\
&&\dfrac{\partial S\bar\phi}{\partial \bar t}  + \dfrac{\partial S\bar\phi \bar v_z}{\partial \bar z} + \dfrac{\partial S\bar\phi \bar v_x}{\partial \bar x}  = 0,\qquad S = \sqrt{1\!-\!4\bar z^2},\notag\\
&&\Pi_p =  \dfrac{\eta_n(\bar \phi)}{\eta_s(\bar \phi)^{1/2}}+\dfrac{\bar \phi
}{\eta_s(\bar\phi)^{1/2}}T, \\
&&\bar v_x = \dfrac{1}{\eta_s(\bar \phi)^{1/4}I},\qquad I = \dfrac{4}{\pi}\int_{-1/2}^{1/2} \dfrac{S(\bar z) \,d\bar z}{\eta_s(\bar\phi(\bar z))^{1/4}}.\notag
\end{eqnarray}
Here the dimensionless quantities are defined as
\begin{eqnarray}\label{barquantities}
&&\bar\phi = \dfrac{\phi}{\phi_m},\quad \bar v_x = \dfrac{v_x}{v_w},\quad \bar v_j = \dfrac{v_j}{v_w},\quad \bar v_z = \dfrac{v_z}{v_0},\quad \bar t = \dfrac{t}{t_0},\quad \bar x = \dfrac{x}{L},\quad \bar z=\dfrac{z}{d_w},\notag\\
&&v_0 = \dfrac{2(\rho_p\!-\!\rho_f)g a^2}{9\mu_a},\quad t_0 = \dfrac{9\mu_a d_w}{2(\rho_p\!-\!\rho_f)g a^2},\quad L =v_w t_0.
\end{eqnarray}
Note that both the normalized volume fraction and horizontal velocity depend on $(\bar x,\bar z, \bar t)$. 

There are three dimensionless groups that enter the system~(\ref{system1}):
\begin{equation}\label{dimpars}
{\cal T} =\dfrac{4(\rho_p\!-\!\rho_f)\rho_pg a^4}{81\mu_a^2 d_w},\qquad G = \dfrac{8\phi_m(\rho_p\!-\!\rho_f)g d_w}{f_D\rho_f v_w^2},\qquad T = \dfrac{8(1\!-\!\Delta\bar\rho)^2f_T\phi_m\rho_p }{f_D  \rho_f }.
\end{equation}
The first dimensionless parameter ${\cal T}$ represents the effect of particle inertia during settling, the second parameter $G$ quantifies the effect of gravity, while $T$ represents the relative strength of the turbulence related term in the expression for particle pressure. For typical parameters, ${\cal T} \approx 10^{-3}$, $G = 0.2\div 10^3$, and $T \approx 3$. Note that some of these parameters depend on $f_D$ and $f_T$, and the calibrated values are used for the estimations. The aforementioned parameter $G$ can also be viewed as an inverse Shields number that is used to calculate the initiation of sediment in a flow. As can be seen from the estimations, the settling-related particle inertia can be safely neglected since ${\cal T} \ll 1$. In this case, the system of equations~(\ref{system1}) further reduces to
\begin{eqnarray}\label{system2}
&&\dfrac{\partial S\bar\phi}{\partial \bar t} - \dfrac{\partial }{\partial \bar z} \biggl[ S \Bigl(\bar\phi+\dfrac{1}{G I^2}\dfrac{\partial }{\partial \bar z}\Bigl\{ \dfrac{\eta_n(\bar \phi)}{\eta_s(\bar \phi)^{1/2}}+\dfrac{\bar \phi
}{\eta_s(\bar\phi)^{1/2}}T \Bigr\}\Bigr)\biggr]  + \dfrac{\partial S\bar\phi \bar v_x}{\partial \bar x}= 0,\notag\\
&&\bar v_x(\bar z) = \dfrac{1}{\eta_s(\bar \phi)^{1/4}I},\qquad I = \dfrac{4}{\pi}\int_{-1/2}^{1/2} \dfrac{S(\bar z) \,d\bar z}{\eta_s(\bar\phi(\bar z))^{1/4}},\qquad S = \sqrt{1\!-\!4\bar z^2}.
\end{eqnarray}
As can be seen, the problem is reduced to solving a nonlinear diffusion equation for normalized particle volume fraction, which is then used to calculate spatial velocity distribution.

For large times and in the absence of variation with respect to $\bar x$, the particle concentration reaches a steady-state solution, that can be calculated from
\begin{equation}\label{steadystate}
\dfrac{d }{d \bar z}\Bigl\{ \dfrac{\eta_n(\bar \phi_s)}{\eta_s(\bar \phi_s)^{1/2}}+\dfrac{\bar \phi_s
}{\eta_s(\bar\phi_s)^{1/2}}T \Bigr\} = -G I^2\bar\phi_s.
\end{equation}
In order to better understand qualitative behavior of the solution, the viscosity factor $\eta_s$ is further approximated as $\eta_s(\bar \phi_s)\approx (1\!-\!\bar\phi)^{-2}$, which introduces maximum error of up to 40\% for concentrations $\bar\phi < 0.9$. Since $\eta_s^{1/2}$ occurs in the equation~(\ref{steadystate}), then the error is practically under 20\%. In this case, the differential equation can be solved as
\begin{equation}\label{steadystatesol}
-\log (1\!-\!\bar\phi_s)+ \dfrac{1}{1-\bar\phi_s} + T\bigl(\log(\bar\phi_s)-2\bar \phi_s\bigr) = -G I^2\bar z + C,
\end{equation}
where $C$ is an integration constant. The latter integration constant can be found from the specified value of the average volume fraction as
\begin{equation}\label{avphi}
\langle \bar\phi_s\rangle = \dfrac{4}{\pi}\int_{-1/2}^{1/2} S \bar\phi_s\, d\bar z,
\end{equation}
where $\langle \bar\phi_s\rangle$ is the known average volume fraction in the wellbore. For small particle concentrations, equation~(\ref{steadystatesol}) can be reduced to
\begin{equation}\label{steadystatesol1}
\bar \phi_s = Ce^{-G\bar z/T}.
\end{equation}
This solution shows that there is always an exponential ``tail'', even if there is a strong particle segregation. For large particle concentrations, solution~(\ref{steadystatesol}) becomes
\begin{equation}\label{steadystatesol2}
\bar\phi_s = 1-\bigl(-GI^2\bar z + C\bigr)^{-1}.
\end{equation}
This limit outlines form of the solution in the bed-like particle flow. The above solution is valid as soon as $\bar\phi_s>0$. Once the solution reaches a point $\bar\phi_s=0$, then $\bar\phi_s\equiv0$ above this point.

\subsection{Reduced model}

While the system of equations~(\ref{system2}) gives the complete or full model, it requires solving a two-dimensional equation, which needs significantly more computational effort than classical wellbore simulation approaches. Therefore, the purpose of this section is to outline a reduced model, that is based on the steady-state solution~(\ref{steadystate}), but yet still captures the dominant effect of time dependence.

To proceed with the reduced model, let's assume that $\bar \phi(\bar t,\bar z)=\bar\phi_s(G_a(\bar t),\bar z)$, i.e. the solution can always be represented as a steady-state solution~(\ref{steadystate}) with an apparent $G_a$. The primary governing equation for the reduced model is the vertically averaged particle balance equation~(\ref{massbalance})
\begin{equation}\label{massbalanceav}
\dfrac{\partial \langle \bar\phi\rangle}{\partial \bar t} + \dfrac{\partial \bar q^p\langle \bar\phi\rangle}{\partial \bar x} = 0,
\end{equation}
where $\langle \bar\phi\rangle$ is the average particle volume fraction. Then, upon substituting the steady-state solution with an apparent $G_a$ into~(\ref{system2}), one has
\begin{eqnarray}\label{system2a}
S\dfrac{\partial \bar\phi_s}{\partial G} \dfrac{\partial G_a}{\partial \bar t} - \Bigl(1-\dfrac{G_a}{G}\Bigr)\dfrac{\partial S \bar\phi_s}{\partial \bar z}   + S\dfrac{\partial \bar\phi_s \bar v_x}{\partial G}\dfrac{\partial G_a}{\partial \bar x}= 0.
\end{eqnarray}
By neglecting the effect of $\bar v_x$ and applying the first moment averaging to the above equation, the result reads 
\begin{eqnarray}\label{system2b}
\dfrac{\partial G_a}{\partial \bar t} + \dfrac{\partial G_a}{\partial \bar x} + \Bigl(\dfrac{G_a}{G}-1\Bigr) \langle \bar \phi\rangle \biggl[\dfrac{4}{\pi}\int_{-1/2}^{1/2}\bar z S\dfrac{\partial \bar\phi_s}{\partial G}\,d\bar z \biggr]^{-1}= 0.
\end{eqnarray}
Instead of calculating the expression in square brackets, the following equation is proposed
\begin{equation}\label{Geq}
\dfrac{\partial G_a}{\partial \bar t} + \dfrac{\partial G_a}{\partial \bar x} + \dfrac{G_a\!-\!G}{\bar t_G} = 0.
\end{equation}
This equation essentially states that in Lagrangian coordinates (i.e. when the observation point follows the flow) the apparent dimensionless gravity $G_a$ exponentially approaches the true value of $G$ within the relaxation time $\bar t_G$ (this is normalized time with the time scale $t_0$~(\ref{barquantities})). Technically, $\bar t_G=Gf(G_a,\langle\bar\phi\rangle,T)$, but for simplicity the functional dependence is reduced to $\bar t_G=f(G,\langle\bar\phi\rangle,T)$.

Finally, the steady-state solution for vertical variation of particle volume fraction is calculated from
\begin{eqnarray}\label{steadystatefin}
&&\dfrac{d }{d \bar z}\Bigl\{ \dfrac{\eta_n(\bar \phi_s)}{\eta_s(\bar \phi_s)^{1/2}}+\dfrac{\bar \phi_s
}{\eta_s(\bar\phi_s)^{1/2}}T \Bigr\} = -G_a I^2\bar\phi_s,\qquad \dfrac{4}{\pi}\int_{-1/2}^{1/2} S\bar\phi_s d\bar z = \langle \bar\phi\rangle\notag,\\
&&\bar q^p = \dfrac{4}{\pi\langle \bar\phi\rangle}\int_{-1/2}^{1/2} \dfrac{S\bar \phi_s \,d\bar z}{\eta_s(\bar\phi_s)^{1/4}},\qquad I = \dfrac{4}{\pi}\int_{-1/2}^{1/2} \dfrac{S(\bar z) \,d\bar z}{\eta_s(\bar\phi_s)^{1/4}},\qquad S = \sqrt{1\!-\!4\bar z^2}.
\end{eqnarray}
where the functions $\eta_n$ and $\eta_s$ are given in~(\ref{rheology}), while $G$ and $T$ are specified in~(\ref{dimpars}). Note that the dimensionless proppant flow rate $\bar q^p$ is introduced to capture the difference between the fluid and proppant average velocities.

\begin{figure}[h]
\centering \includegraphics[width=0.6\linewidth]{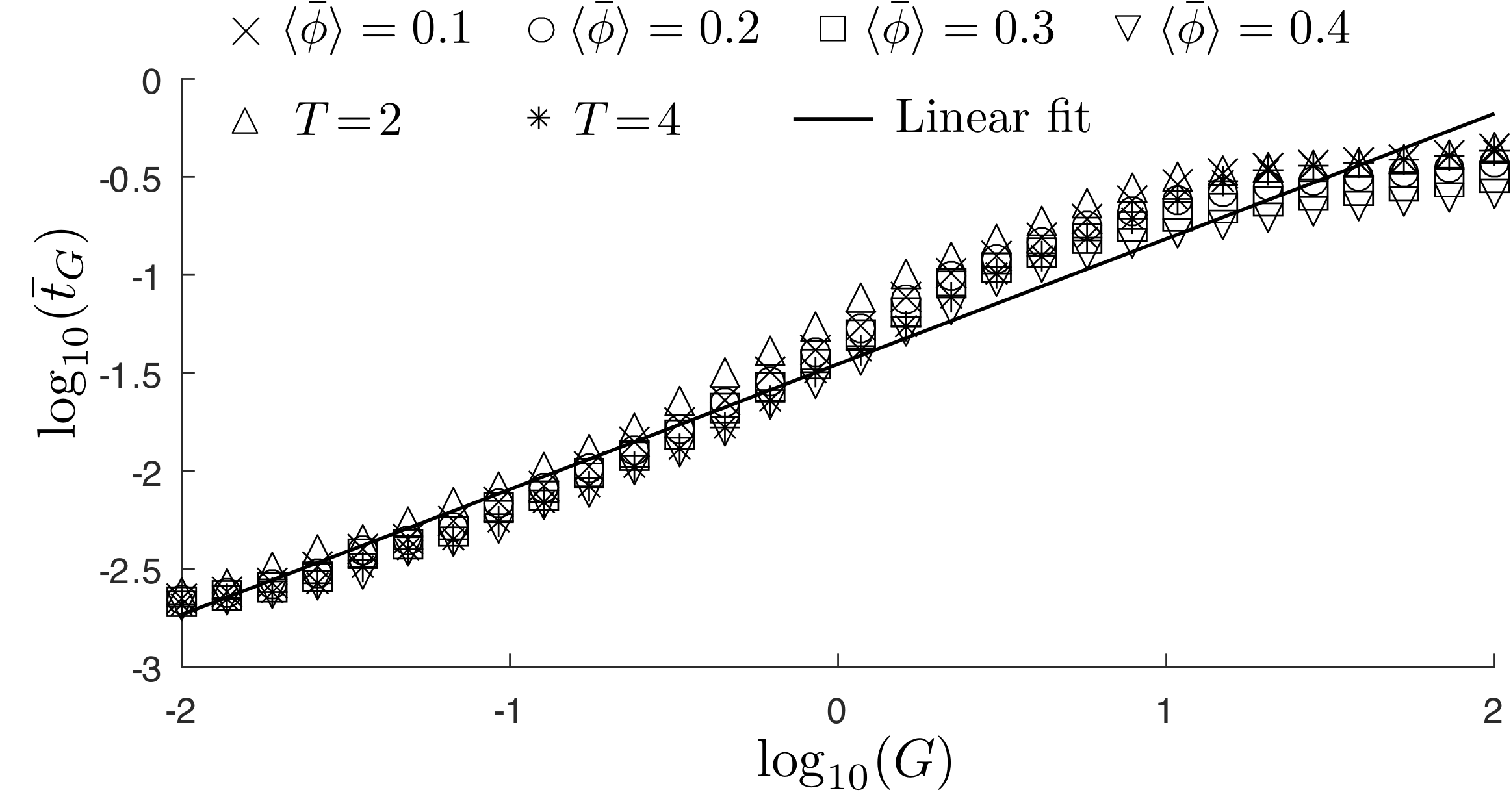}
\caption{Variation of the dimensionless relaxation time versus $G$. Markers show the calculated data, while the straight line indicates the fitting~(\ref{tGfit}). The first four data points correspond to $T=2.81$ and various $\langle \bar\phi\rangle$, while the last two data points correspond to $\langle \bar\phi\rangle=0.1$ and various values of $T$.}
\label{fig2}
\end{figure}

In order to calibrate the relaxation time $t_G$, the two-dimensional system~(\ref{system2}) is solved for the case of $\partial (\cdot)/\partial \bar x\equiv 0$, i.e. by focusing only on temporal variation. The following difference measure is calculated
\begin{equation}\label{Ddef}
D(\bar t) = \sqrt{\dfrac{\int_{-1/2}^{1/2}(\bar\phi(\bar t,\bar z)-\bar \phi_s(\bar z))^2\,d\bar z}{\int_{-1/2}^{1/2}\bar\phi_s(\bar z)^2\,d\bar z}},
\end{equation}
where $\bar\phi(\bar t,\bar z)$ is the solution of~(\ref{system2}), while $\bar \phi_s(\bar z)$ is the steady-state solution stemming from~(\ref{steadystatefin}) (this solution is calculated with the ``true'' value of $G$). The time dependence is then fitted by the exponential function $D(\bar t) = D_0\exp(-\bar t/\bar t_G)$. Fig.~\ref{fig2} plots the variation of the calculated value of $\bar t_G$ versus $G$ for different values of the average volume fraction $\langle\bar\phi\rangle$ and $T$. As can be seen from the results, the obtained data points follow a straight line on a log-log plot reasonably well. The resultant variation of the relaxation time is
\begin{equation}\label{tGfit}
\bar t_G = 0.035\, G^{0.64}.
\end{equation}
Results also demonstrate that the obtained fitting is universal and does not depend on the average volume fraction and the parameter $T$.

\begin{figure}[h]
\centering \includegraphics[width=0.99\linewidth]{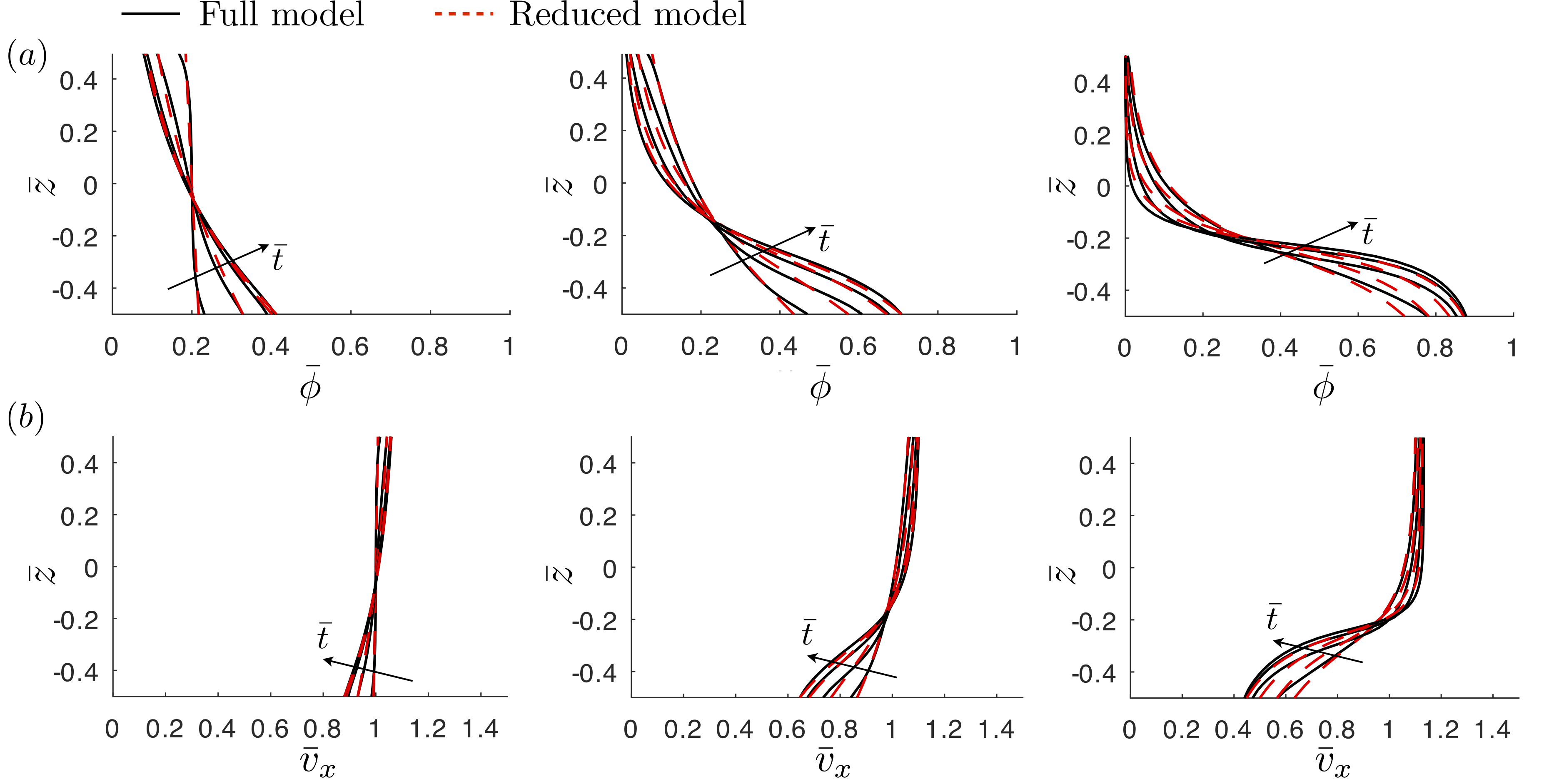}
\caption{Comparison between the numerical results calculated using the full model~(\ref{system2}) and the reduced model~(\ref{Geq}), (\ref{steadystatefin}), and~(\ref{tGfit}) for the case of no spatial variation with respect to $\bar x$. The upper panel $(a)$ shows the volume fraction profiles, while the lower panel $(b)$ indicates the corresponding velocity profiles for different points in time. Moving from left to right, the values of $G$ increase, corresponding to situations with stronger gravitational effects. The left plots correspond to $G\!=\!5$, while the initial condition is calculated using the steady-state solution with $G_i\!=\!0$. The middle plots correspond to $G\!=\!15$ and $G_i\!=\!5$, while the right plots show the results for $G\!=\!45$ and $G_i\!=\!15$. Different lines correspond to various values of the dimensionless time $\bar t = \{0.01,0.1,0.3,1\}$. The same average volume fraction $\langle\bar \phi\rangle = 0.2$ is used for all cases, while $T=2.81$ for all cases as well.}
\label{fig3}
\end{figure}

As an illustration, Fig.~\ref{fig3} compares the results of simulations with the ``full'' model~(\ref{system2}) and with the ``reduced'' model~(\ref{Geq}), (\ref{steadystatefin}), and~(\ref{tGfit}). The plots from left to right correspond to different values of $G=\{5,15,45\}$. The initial condition is taken as the steady-state solution with $G_i=\{0,5,15\}$. The average volume fraction is kept the same for all cases $\langle\bar\phi\rangle=0.2$. The full model is shown by the solid black lines, while the dashed red lines correspond to the results of the reduced model. The panel $(a)$ shows the normalized particle volume fractions, while panel $(b)$ depicts the velocity profiles. Results demonstrate that the difference between the two solutions is relatively small. Thus, the reduced model can be readily used in lieu of the full model due to the simplicity of implementation and better computational performance. Results also illustrate how the solution varies with $G$. The concentration profile is noticeably asymmetric for $G\!=\!5$, and for $G\!=\!45$ it practically reaches the ``flowing bed'' state, i.e. a state in which most of the proppant is located in the lower part of the wellbore and its concentration is close to the maximum allowed value.

\begin{figure}[h]
\centering \includegraphics[width=0.9\linewidth]{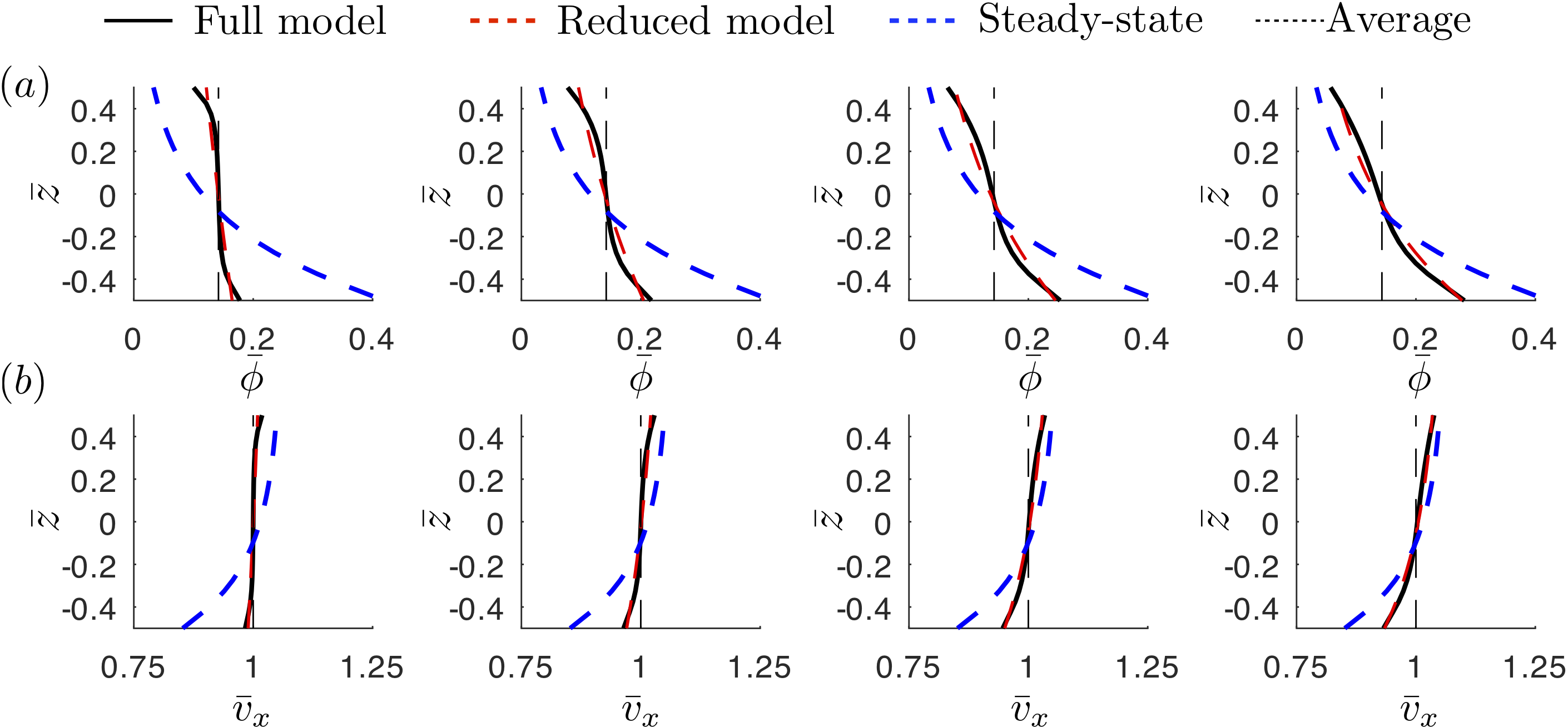}
\caption{Comparison between the full and the reduced model for the example of suspension flow in a pipe without perforations. Panel $(a)$ shows the concentration profiles, while the panel $(b)$ shows the velocity profiles. Pictures from left to right correspond to different evaluation points, the most left being the closest to the inlet and the most right being the furthest. The black solid lines correspond to the full model, the dashed red lines show results of the reduced model, the dashed blue lines indicate the steady-state solution, while the thin dashed black lines show the averaged values.}
\label{fig4}
\end{figure}

To further compare the ``full'' and the ``reduced'' model, consider flow in a pipe without perforations. The initial concentration is uniform, but the proppant settles as the flow progresses along the pipe. The injection rate is $15$~bpm = $0.04$~m/s$^3$, the pump time is 1 minute, 100 mesh proppant is used, the pipe diameter is 4~in=10.2~cm, while the evaluation points are located at $\{5,15,25,35\}$~ft = $\{1.5,4.6,7.6,10.7\}$~m from the injection point. Fig.~\ref{fig4} plots the normalized particle concentration profile and the velocity profile at the evaluation points. The solid black lines correspond to the full solution, the dashed red lines indicate the reduced model, while the dashed blue lines plot the steady-state solution, i.e. the solution that assumes that particle settling happens instantaneously. The thin dashed black lines correspond to the average quantities and are provided for the reference. Results clearly show a gradual transition from a uniform particle volume fraction and velocity near the inlet to the nearly steady-state solution far away from the inlet. As was observed for the previous example, the full and the reduced models agree reasonably well. Also note that this example examines spatial variation of the solution along the well, which is in contrast to the results shown in Fig.~\ref{fig3} that focus exclusively on temporal variation.

\section{Proppant turning from wellbore to a perforation}\label{secturn}

\begin{figure}[h]
\centering \includegraphics[width=0.85\linewidth]{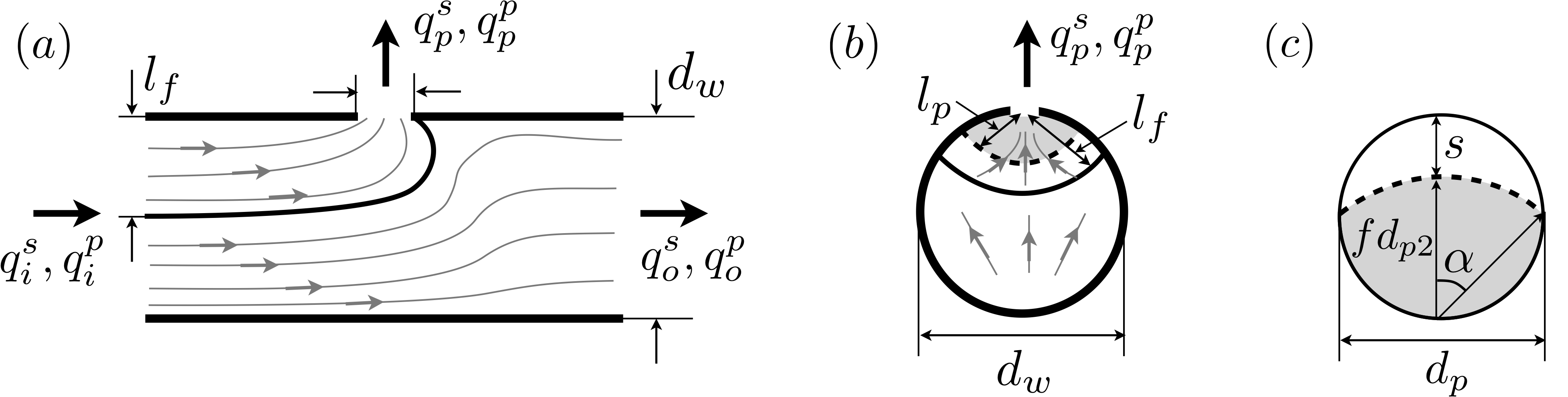}
\caption{Schematics of the problem of slurry flow near a perforation.}
\label{fig5}
\end{figure}

The second part of the problem is related to sharp particle turning from the wellbore to perforation. Fig.~\ref{fig5} shows schematics of the problem with the focus on the flow near a perforation. The panel $(a)$ shows side view of the wellbore, the panel $(b)$ shows the wellbore's cross-section, while the panel $(c)$ depicts the perforation. The total slurry flow rate at the inlet (left side) is denoted by $q^s_i$, the outlet (right side) flow rate is $q^s_o$, while $q^s_p$ is the flow rate through the perforation. The proppant flow rate at the inlet is denoted by $q^p_i$, at the outlet by $q^p_o$, and through the perforation by $q^p_p$. Let $d_w$ be the diameter of the wellbore, while $d_{p}$ be the perforation diameter. Fluid streamlines are schematically shown in Fig.~\ref{fig5}$(a)$ and $(b)$ by gray lines with arrows. It is assumed that portion of the fluid that is located within the distance $l_f$ from the perforation leaves the wellbore and contributes to the perforation flow rate $q^s_p$. The rest of the fluid does not enter the perforation. Only a fraction of the proppant that is located within the distance $l_f$ enters the perforation. Therefore, it is assumed that proppant located within the distance $l_p$ enters the perforation, see the dashed black lines. The reason for that lies in the slip velocity between the fluid and proppant. The displacement of the latter slip is denoted by $s$. Note that the fact that proppant located only near the perforation leaves the wellbore was observed in~\cite{Kolle2022} via CFD simulations and the area outlined by $l_p$ was called the ``ingestion area''. Similarly, authors in~\cite{Wang2022} plotted the fluid and proppant streamlines, which confirmed that the ``ingestion'' area is nearly circular and that there is a difference between those areas for fluid and proppant, as outlined by $l_f$ and $l_p$. Fig.~\ref{fig5}$(c)$ shows cross-section of the perforation. The shaded area corresponds to the fluid streamlines that are initially located within $l_p$ from the perforation, see also the shaded area in Fig.~\ref{fig5}$(b)$. It is assumed that the slurry flow rates are known and the goal is to determine the proppant flow rates.

To find the equation for $l_f$, the total slurry flow within $l_f$ should be made equal to the slurry flow through the perforation. By accounting for the normalized wellbore velocity introduced in section~\ref{secslurryflow}, this condition reads
\begin{equation}\label{Afcalc}
\dfrac{4}{\pi d_w^2}\int_{A_f} \bar v_x\, dA=\dfrac{q^s_p}{q^s_i},
\end{equation}
where $A_f$ is the value of the area outlined by $l_f$, see Fig.~\ref{fig5}$(b)$. The above equation~(\ref{Afcalc}) is implicit in terms of $l_f$, and therefore the solution is calculated numerically. For typical parameters, the ratio between $l_f$ and wellbore diameter $d_w$ is $\bar l_f = l_f/d_w=0.1\div1$. Therefore, the first perforation takes slurry only within the zone that is located relatively close to perforation, while the last perforation takes slurry from the whole wellbore.

Once $l_f$ is known, the next question is what happens to proppant within this zone, and, in particular, how to calculate $l_p$. In order to find the answer, consider the problem of fluid slowing down before the perforation. The motion only along the wellbore is considered. For simplicity let's assume that $v_f=v_w(1-t/t_T)$, where the stopping time is $t_T = p_T \bar l_p d_w/v_w$ ($\bar l_p=l_p/d_w$) and $p_T$ is a calibration parameter. Then, equation~(\ref{particledynslip}) becomes
\begin{equation}\label{turngov}
\Bigl(\rho_p\!+\!\frac{1}{2}\rho_f\Bigr)\dfrac{d\Delta {v}}{dt} = (\rho_p\!-\!\rho_f) \dfrac{v_w}{t_T}-\dfrac{9\mu_a}{2a^2}\Delta{v}-C_d\dfrac{3\rho_f}{8a}\Delta{v}^2,
\end{equation}
where $\mu_a$ is the apparent viscosity from~(\ref{apparvisc}). Assuming zero initial slip velocity, equation~(\ref{turngov}) can be solved to obtain
\begin{eqnarray}\label{turngovsol}
&&\Delta{v}(t) = \dfrac{1}{2R_a}\dfrac{R_*^2\tanh(\tau)}{\sqrt{1\!+\!R_*^2}+\tanh(\tau)} v_w, \qquad  \tau=\Delta \bar \rho\dfrac{2R_a\sqrt{1\!+\!R_*^2}}{R_*^2}\dfrac{t}{ t_T},\qquad R_*= \dfrac{2\beta R_a}{\sqrt{p_T \bar l_p}},\notag\\
&&\beta = \sqrt{\dfrac{8a  \bigl(\rho_p\!-\!\rho_f\bigr)}{3 C_d\rho_f d_w }},\qquad R_a = \dfrac{R}{1\!+\!R\Delta\bar \rho p_V},\qquad R = \dfrac{C_d }{24} \dfrac{2 a\rho_f v_w }{\mu},\qquad \Delta\bar\rho = \dfrac{\rho_p-\rho_f}{\rho_p+\tfrac{1}{2}\rho_f}.
\end{eqnarray}
It is further assumed that the particle velocity (which is equal to the slip velocity since the horizontal fluid velocity is zero at the perforation entrance) cannot exceed slurry velocity after perforation. The latter average slurry velocity after the perforation is
\begin{equation}\label{vo}
v_o = \dfrac{q^s_i-q^s_p}{q^s_i} v_w.
\end{equation}
To provide a smooth cut-off, a harmonic average is used, i.e.
\begin{eqnarray}\label{turngovsolcutoff}
\Delta{v_c} = \dfrac{\Delta v \,v_0}{\Delta v +v_0} =  \dfrac{p_1p_2\tanh(\tau)}{p_1\sqrt{1\!+\!R_*^2}+(p_1\!+\!p_2)\tanh(\tau)} v_w,\qquad p_1 = \dfrac{q^s_i-q^s_p}{q^s_i},\qquad p_2 = \dfrac{R_*^2}{2R_a}.
\end{eqnarray}
The above expression for the velocity can be integrated in time to calculate the total slip displacement as
\begin{equation}\label{slipsol}
s(t_T) = \dfrac{p_T \bar l_p d_w}{\tau(p_3^2-1)}\dfrac{ p_1p_2}{p_1+p_2} \Bigl(p_3\log\bigl[p_3\cosh(\tau)+\sinh(\tau)\bigr]-\tau -p_3\log(p_3)\Bigr),\qquad p_3 = \dfrac{p_1\sqrt{1+R_*^2}}{p_1+p_2}.
\end{equation}
Finally, it is the dimensionless slip that is relevant for the analysis. The latter dimensionless slip is defined as $\bar s = s/d_p$, i.e. the slip normalized by the perforation diameter.

With the reference to Fig.~\ref{fig5}$(c)$, the ratio of the slurry flow through the area outlined by $l_p$ to the flow outlined by $l_f$ is related to the slip as
\begin{equation}\label{lp}
\dfrac{\int_{A_p}\bar v_x\,dA}{\int_{A_f}\bar v_x\,dA} = {\cal A}\bigl(1\!-\! \bar s\bigr),\qquad {\cal A}(f) = 1+\dfrac{4\alpha}{\pi} f^2 -\dfrac{2\alpha+\sin(2\alpha)}{\pi},\qquad \alpha = \cos^{-1}(f).
\end{equation}
Here ${\cal A}$ is the geometric factor representing the ratio between the shaded area in Fig.~\ref{fig5}$(c)$ to the total perforation area. Equation~(\ref{lp}) is solved numerically for $\bar l_p$.

The obtained value of $\bar l_p$ can be used to calculate the proppant flow through the perforation as
\begin{equation}\label{qpp}
q^p_p = \dfrac{\int_{A_p}\phi \bar v_x\,dA}{\int_{A_f}\bar v_x\,dA}q^s_p.
\end{equation}
Note that the value of $q_p^p$ can can vary depending on perforation orientation since the volume fraction $\phi$ and velocity $\bar v_x$ depend on the vertical coordinate. Finally, to calculate the slurry and proppant flows at the outlet, the balance equations can be used, i.e.
\begin{equation}\label{qo}
 q^s_i = q^s_p+q^s_o,\qquad  q^p_i = q^p_p+q^p_o.
\end{equation}
The solution for the problem can be outlined as follows. Given the perforation orientation, solutions for $\phi$ and $\bar v_x$, inlet slurry and proppant flows $q^s_i$, $q^p_i$, as well as perforation slurry flow $q^s_p$, and other problem parameters, equation~(\ref{Afcalc}) is first solved to find $\bar l_f$. Then~(\ref{lp}) is solved for $\bar l_p$. After that~(\ref{qpp}) is used to obtain proppant flow through the perforation $q^p_p$, and finally~(\ref{qo}) allows to calculate the slurry and proppant flows after the perforation. 

To better understand solution for the dimensionless slip, let's consider its dependence on $\tau$ and its limits. For the case of $\tau\ll 1$, equation~(\ref{slipsol}) reduces to
\begin{equation}\label{slipsol0}
\bar s_0 =  p_1 \Bigl(1  +\bar p_1\log(\bar p_1) - \bar p_1\log(1\!+\!\bar p_1)\Bigr)\dfrac{ p_T \bar l_p d_w}{d_p},\qquad \bar p_1 = \dfrac{p_1}{\Delta \bar \rho},\qquad \tau_0=\dfrac{\Delta \bar \rho \sqrt{p_T \bar l_p} }{ \beta}.
\end{equation}
The result shows that the slip does not depend on $\tau$, but instead depends on the parameter $p_1$, $\Delta \bar \rho$, and the dimensionless group $p_T \bar l_p d_w/d_p$. Physically, this limit corresponds to the situation of pure inertial turning, i.e. when the effect of drag force is negligible. Therefore, this limit represents the maximum possible slip. The result depends on the fitting parameter $p_T$, which provides useful sensitivity for calibrating the model. There is also no sensitivity to particle size, viscosity, and velocity since the drag force is negligible in this case. At the same time, there is a strong dependence on the perforation and wellbore diameters. Bigger wells and smaller perforations lead to bigger slip.

The limit of $\tau\gg 1$ corresponds to domination of the drag force and~(\ref{slipsol}) becomes
\begin{equation}\label{slipsolinf}
\bar s_\infty =  \dfrac{\Delta \bar \rho}{2\tau_\infty}\dfrac{ p_T \bar l_p d_w}{d_p} = \dfrac{ d_w}{d_p} \dfrac{ \beta^2 R}{ (1\!+\!R\Delta\bar \rho p_V)},\qquad  \tau_\infty=\dfrac{\Delta \bar \rho\, p_T \bar l_p}{ 2\beta^2} \dfrac{1\!+\!R\Delta\bar \rho p_V}{R}.
\end{equation}
This result depends on $\tau$, but most importantly, when $\tau$ is substituted, the final expression does not depend on $p_T$ or $\bar l_p$. At the same time, in this drag dominated case, the dimensionless slip depends on particle size, as well as on viscosity. If the drag is turbulent, i.e $R\gg 1$, then there is no viscosity dependence, while the slip is proportional to the particle size. At the same time, if the drag is laminar, i.e. $R\ll 1$, then the slip is inversely proportional to viscosity and scales as the particle size squared. Thus, there is much stronger particle size dependence in this case.

\begin{figure}[h]
\centering \includegraphics[width=0.85\linewidth]{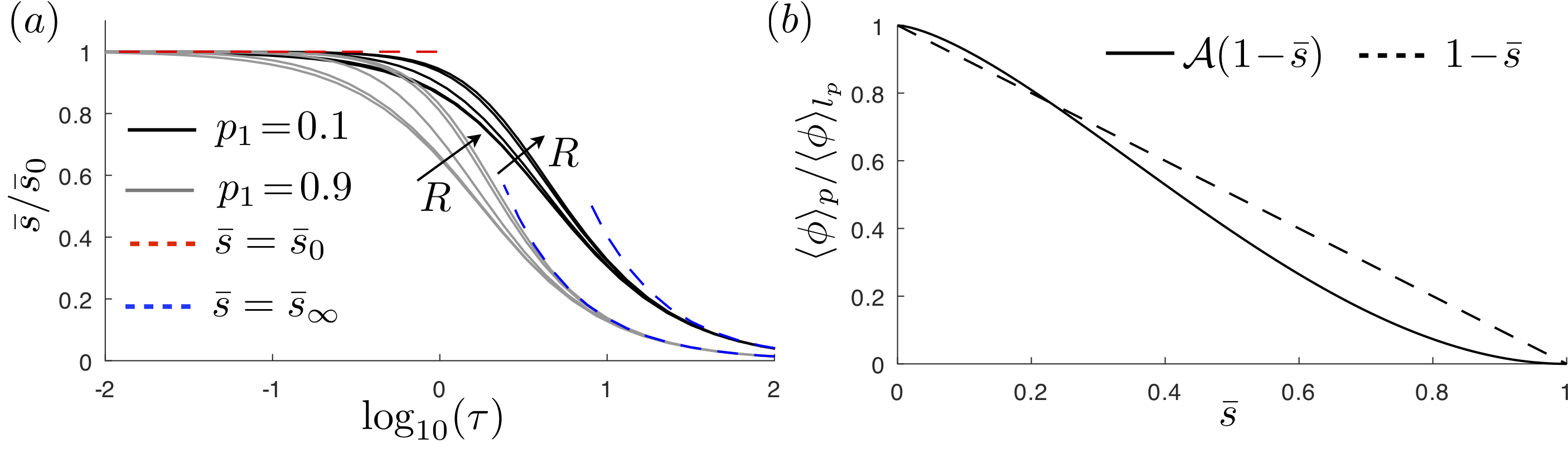}
\caption{$(a)$: Normalized particle slip magnitude versus dimensionless particle turning time $\tau$ for two values of perforation flow fraction $p_1=\{0.1,0.9\}$ and five values of scaled particle Reynolds number $R=\{0.01, 0.1, 1, 10, 100\}$ (see~(\ref{turngovsol}) and~(\ref{turngovsolcutoff}) for the definitions of these parameters). The dashed red and blue lines show the solutions~(\ref{slipsol0}) and~(\ref{slipsolinf}), respectively. $(b)$: Variation of proppant entry efficiency versus dimensionless slip (solid black line). The dashed black line shows $1\!-\!\bar s$ for the reference.}
\label{fig6}
\end{figure}

To illustrate the behavior of slip, Fig.~\ref{fig6}$(a)$ plots the variation of the normalized slip $\bar s/\bar s_0$ versus $\tau$ for different values of the parameters $p_1$ and $R$. The density contrast is taken as $\Delta \bar \rho\!=\!0.52$. The asymptotic solutions~(\ref{slipsol0}) and~(\ref{slipsolinf}) are shown by the red and blue dashed lines, respectively. As can be seen, the solution for slip transitions from its maximum value for small $\tau$ to the large $\tau$ asymptote. The dependence on $R$ is relatively mild. The dependence on $p_1$ is stronger, but is also relatively mild. Note that the solution is normalized by $\bar s_0$, which has a strong dependence on $p_1$. Also, the dependence on $R$ is actually hidden in $\tau$, i.e. the value of $\tau$ strongly depends on $R$ for $R\ll 1$.

To better understand how the dimensionless slip affects the amount of proppant that enters the perforation, let's focus on the situation, in which the particle volume fraction $\phi$ is approximately constant within the zone outlined by $l_p$. In this case, equations~(\ref{lp}) and~(\ref{qpp}) can be combined to obtain
\begin{equation}\label{qpp2}
\eta=\dfrac{\langle\phi\rangle_p}{\langle\phi\rangle_{l_p}}=\dfrac{q^p_p}{\langle\phi\rangle_{l_p}q^s_p} = {\cal A}\bigl(1\!-\! \bar s\bigr).
\end{equation}
Here $\langle\phi\rangle_p$ is the average volume fraction of the slurry that enters the perforation, while $\langle\phi\rangle_{l_p}$ is the particle volume fraction within $l_p$. In other words, the ratio $\eta={\langle\phi\rangle_p}/{\langle\phi\rangle_{l_p}}$ represents the fraction of proppant located within $l_p$ that enters the perforation or entry efficiency. Fig.~\ref{fig6}$(b)$ plots the variation of the latter ratio versus the normalized slip $\bar s$. The dashed line shows the solution $1\!-\!\bar s$ for the reference. Clearly, the dimensionless slip determines the amount of proppant that enters the perforation. If the slip is small, then all particles enter the perforation. At the same time, if the dimensionless slip approaches one, then practically no particles enter the perforation.

\begin{figure}[h]
\centering \includegraphics[width=0.5\linewidth]{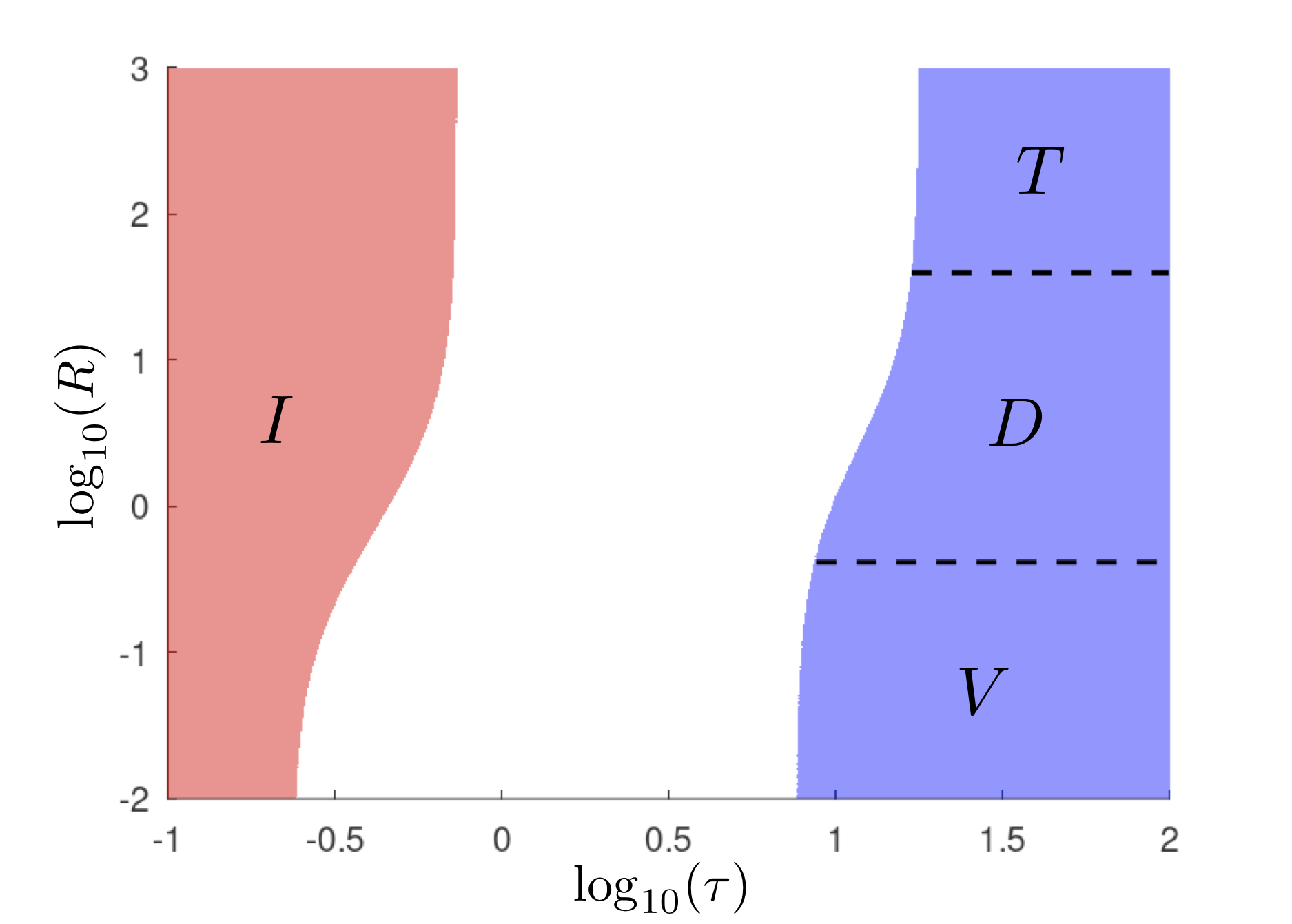}
\caption{Parametric space for particle slip, $\tau$ is the dimensionless turning time, $R$ is scaled particle Reynolds number (see~(\ref{turngovsol} for definitions). The red region $I$ outlines zone of applicability of the inertia dominated solution~(\ref{slipsol0}). The blue region $D$ outlines zone of applicability of the drag dominated solution~(\ref{slipsolinf}). The drag dominated solution has two sub-limits: $T$ - turbulent drag and $V$ - viscous drag.}
\label{fig7}
\end{figure}

It is also useful to investigate the variation of the dimensionless slip in the $(\tau,R)$ parametric space. Fig.~\ref{fig7}  shows such a parametric space for $p_1=0.9$ and $\Delta\bar\rho = 0.52$. The red zone (labeled as $I$) indicates the zone of applicability of the inertia dominated solution~(\ref{slipsol0}). If the parameters fall into this zone, then the slip does not depend on particle size and viscosity. The blue zone (labeled as $D$) indicates the zone of applicability of the drag dominated solution~(\ref{slipsolinf}). The slip is generally smaller within this zone and also depends on particle size and viscosity in some cases. Within the drag dominated zone, there are two sub-zones that correspond to the viscous drag limit, i.e. $R<R_V$ and $R_V = 0.4$, and turbulent drag limit, i.e. $R>R_T$ and $R_T=40$. The particle size dependence is linear within the $T$ region and there is no viscosity dependence. At the same time, the particle size dependence is quadratic within the $V$ region and there is strong viscosity dependence. Note that the diagram is constructed using the tolerance of 0.1 to outline the aforementioned regions. It should also be mentioned that the sensitivity can only be observed when the dimensionless slip is neither close to one nor small. If the dimensionless slip is close to one, then no particles practically enter the perforation and the sensitivity to various parameters is hard to observe. Similar logic applies when the dimensionless slip is close to zero. Almost all particles enter the perforation in this case and therefore the overall change in the behavior with respect to parameters is minimal.

The parametric space shown in Fig.~\ref{fig7} has several applications. First of all, it allows to put the calibration points on the plot and to understand which limits have been validated and which ones still require validation. At this point, the model has only one fitting parameter $p_T$ and the solution is not sensitive to it in the drag dominated $D$ region. Therefore, the validation needs to be done outside of $D$ region. The second use of the diagram is that it allows to understand how fluid viscosity and particle size affect the solution. It might be tempting to increase viscosity to attempt having more uniform proppant distribution between clusters, but if the parameters are such that the corresponding location in the parametric space is far away from the $V$ region, then it won't affect the result. The same is applicable to particle size. Smaller particles tend to turn to the perforation easier, but this applies only outside of the inertia dominated zone $I$. If the parameters fall into the $I$ zone, then reducing the particle size will not help. To summarize, these examples indicate that the constructed parametric space can be used to better understand sensitivities and helps designing the desired proppant distribution between clusters.

\section{Slurry flow in a perforated wellbore}\label{secperfwell}

Previous two sections~\ref{secslurryflow} and~\ref{secturn} addressed the problem of slurry flow in a wellbore and the problem of particles turning to perforation, respectively. Solutions to these two separate problems are the building blocks for the actual problem of slurry flow in a perforated wellbore. Both particle distribution within the wellbore cross-section and the effect of slip affect the result. With the reference to Fig.~\ref{fig1}, the goal is to calculate slurry and proppant flow rates through perforations $q^s_j$ and $q^p_j$ for the given inlet rates $q^s_0$ and $q^p_0$, as well as location of the perforations $x_j$ and orientation of each perforation $\theta_j$.

A relatively simple model is adopted to calculate slurry distribution between perforations. It is assumed that the slurry is incompressible, and that the effect of pipe friction and pressure in the fractures can be neglected. Therefore, the flow distribution is determined exclusively by the perforation friction drop. The latter pressure drop is the same for all perforation holes, and therefore the governing system of equations can be written as
\begin{equation}\label{perfpdrop}
 \sum_{j=1}^{N_p} q^s_j = q^s_0,\qquad \Delta p_p = \dfrac{\rho_f (q^s_j)^2}{2C_j^2A_j^2},\qquad j = 1.. N_p.
\end{equation}
Here $q^s_j$ is the slurry flow rate into $j$th perforation, $q^s_0$ is the initial flow rate in the wellbore, see Fig.~\ref{fig1}. The perforation pressure drop $\Delta p_p$ is unknown, but it is the same for all perforations. To decouple the flow distribution from proppant transport, it is further assumed that the slurry density can be approximated by that of clear fluid. The value of the density is actually not important for the flow distribution, so the actual assumption is that the slurry density is approximately the same for all perforations. Finally, $A_j=\pi d_{p}^2/4$ is perforation area, while $C_j$ is discharge coefficient. 

The system of equations~(\ref{perfpdrop}) can be solved to obtain the expressions for the perforation pressure drop and flow rates as
\begin{equation}\label{perfpdropsol}
 \Delta p_p = \dfrac{\rho_f(q^s_0)^2}{ 2\Bigl[\sum_{j=1}^{N_p} ,C_jA_j\Bigr]^2},\qquad q^s_j= \dfrac{C_jA_jq^s_0}{\sum_{i=1}^{N_p}  C_iA_i},\qquad j = 1.. N_p.
\end{equation}
This, in turn, allows to find spatial variation of the wellbore slurry flow rate as
\begin{equation}\label{qsol}
q^s_w(x) = q^s_0 - \sum_{j=1}^{N_p}H(x\!-\!x_j)q_j,
\end{equation}
where $q^s_w(x)$ is the wellbore slurry flow rate, $x_j$ is the location of the $j$th perforation, while $H(\cdot)$ is Heaviside function. Finally, the average wellbore velocity is simply $v_w(x)=4q^s_w(x)/(\pi d_w^2)$. 
 
Once the wellbore velocity is known, the next step is to calculate the updated average particle volume fraction in the wellbore. To this end, the dimensionless coefficients $M_j$ are introduced such that $q^p_j = M_j \langle \phi\rangle q^s_j$. Thus, the meaning of these coefficients is the ratio between the actual proppant flow rate and the proppant flow rate in the case of uniform particle distribution in the wellbore and no particle slippage. Each perforation is characterized by such a dimensionless parameter. In this situation, equation~(\ref{massbalanceav}) can be generalized as
\begin{equation}\label{proptrans}
\dfrac{\partial \langle \phi\rangle}{\partial t} + \dfrac{\partial \bar q_p v_w\langle \phi\rangle }{\partial x} +\sum_{j=1}^{N_p}\langle\phi\rangle M_j v_j \delta(x\!-\!x_j) = 0,
\end{equation}
where $v_j = 4q^s_j/\pi d_w^2$ is the change of the average wellbore velocity at the $j$th perforation. Transport equation for the apparent dimensionless gravity~(\ref{Geq}) becomes
\begin{equation}\label{Geqdim}
\dfrac{\partial G_a}{\partial t} + v_w\dfrac{\partial G_a}{\partial x} + \dfrac{G_a\!-\!G}{ t_G} = 0.
\end{equation}
Note that this equation can also be written in the conservative form, but in this case there will be an additional loss term. Finally, the particle distribution within the wellbore is calculated from~(\ref{steadystatefin}) and the value of $G_a$ from~(\ref{Geqdim}). The same equations provide the expression for $\bar q_p$. While $M_j$ are calculated using the procedure outlined in Section~\ref{secturn}.

\begin{figure}[hp]
\centering \includegraphics[width=0.95\linewidth]{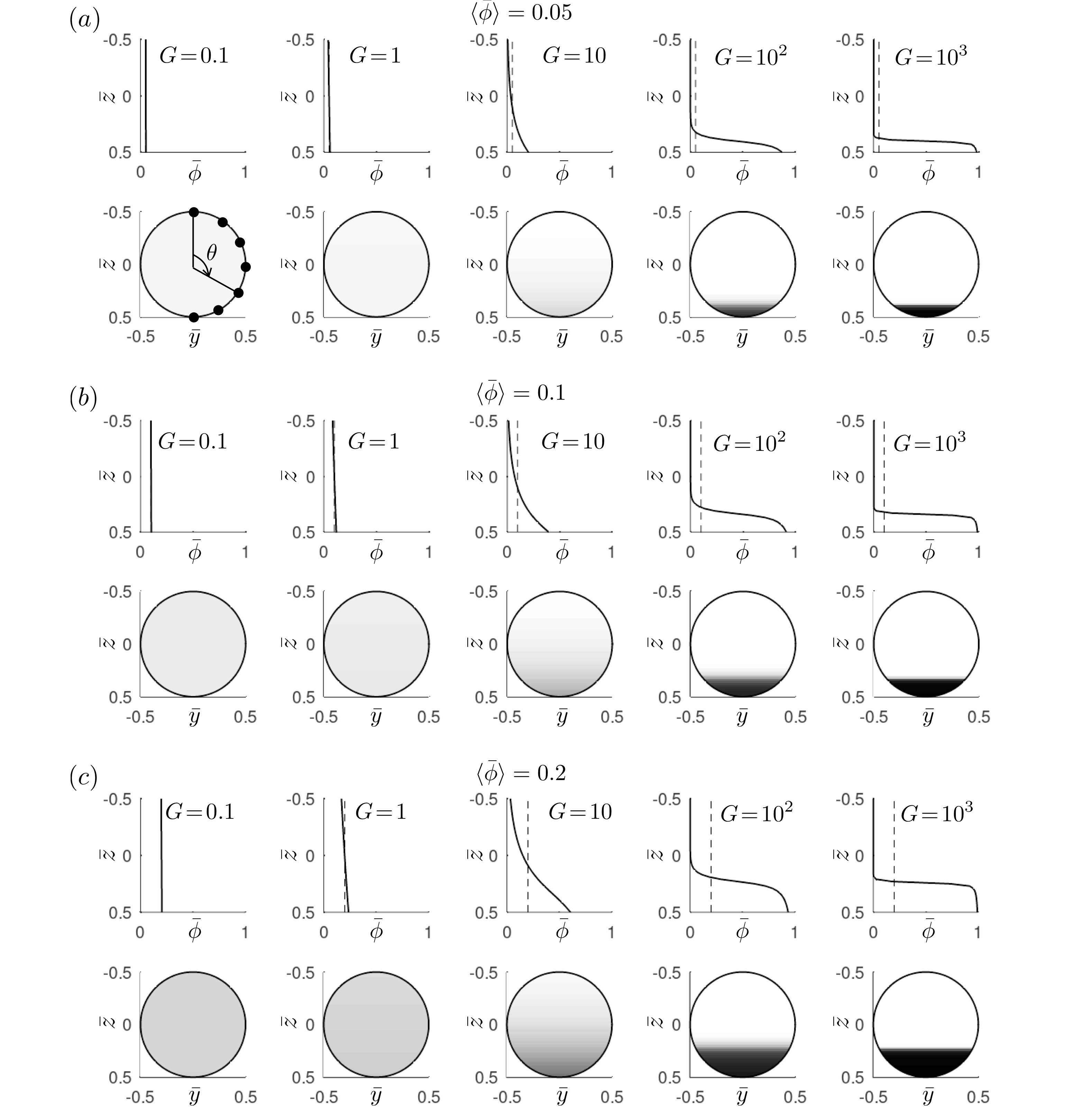}
\caption{Variation of the normalized particle volume fraction versus $G$ for different values of average volume fraction. $(a)$: $\langle\bar\phi\rangle = 0.05$. $(b)$: $\langle\bar\phi\rangle = 0.1$. $(c)$: $\langle\bar\phi\rangle = 0.2$.}
\label{fig8}
\end{figure}

To better understand the solution behavior, Fig.~\ref{fig8} plots the particle distribution solution~(\ref{steadystatefin}) for different values of $G$ and normalized average particle volume fraction in the wellbore $\langle \bar\phi\rangle$. Proppant density is taken as $\rho_p=2650$~kg/m$^3$, which translates into $T=2.81$. The particle volume fraction of $\langle\bar\phi\rangle = 0.05$ (panel $(a)$) corresponds to approximately 0.7~ppg, $\langle\bar\phi\rangle = 0.1$ (panel $(b)$) to 1.4~ppg, while $\langle\bar\phi\rangle = 0.2$ (panel $(c)$) to 3~ppg. Two types of visualizations are shown. The top pictures indicate the linear plots, or variation along the vertical line passing through the center of the wellbore. The bottom pictures show actual wellbore cross-sections with color filling indicating the value of $\bar \phi$. The black colored regions correspond to $\bar\phi=1$, while the white regions to $\bar\phi=0$. Results clearly demonstrate that particle volume fraction is nearly uniform for $G<1$ and the solution reaches a nearly flowing bed state for $G>10^2$. The higher the average particle volume fraction, the bigger the height of the flowing bed. Note that the model is unable to model formation of the immobile bed. However, with the reference to Fig.~\ref{fig3}, the slurry velocity is significantly reduced for large values of $G$, which can practically lead to nearly immobile bed.

\begin{figure}[h]
\centering \includegraphics[width=0.95\linewidth]{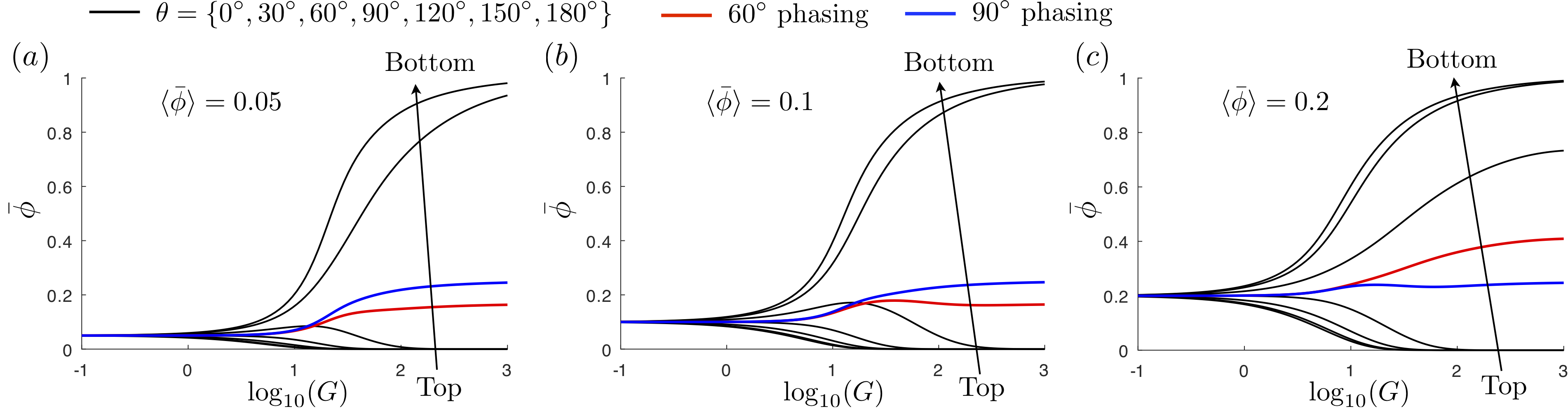}
\caption{Variation of the particle volume fraction entering the perforation versus $G$ for different perforation orientations (solid black lines). The case of 6 perforations with 60$^\circ$ phasing is shown by the red line. The case with 4 perforations with 90$^\circ$ phasing is shown by the blue line. Different panels correspond to various average proppant volume fractions in the wellbore. $(a)$: $\langle\bar\phi\rangle = 0.05$. $(b)$: $\langle\bar\phi\rangle = 0.1$. $(c)$: $\langle\bar\phi\rangle = 0.2$.}
\label{fig9}
\end{figure}

To further investigate behavior of particles in the perforated wellbore, Fig.~\ref{fig9} plots the variation of normalized particle concentration versus $G$ for different perforation orientations $\theta=\{0^\circ,30^\circ,60^\circ,90^\circ,120^\circ,150^\circ,180^\circ\}$ and different average particle volume fractions. Note that the actual volume fraction at these orientations is plotted, i.e. $l_p=0$, see Fig.~\ref{fig5}$(b)$ for the definition of $l_p$. The red lines correspond to the average over 6 perforations with $60^\circ$ phasing, starting with $0^\circ$. The blue lines correspond to the average over 4 perforations with $90^\circ$ phasing, starting with $0^\circ$. To make a better connection with the previous result in Fig.~\ref{fig9}, the panel $(a)$ on the latter figure shows the definition of perforation azimuth $\theta$ and locations of the considered values of $\theta$ by the circular black markers. Results clearly show that phasing starts playing a crucial role for large values of $G$, while when the particle volume fraction is uniform, then there is no variation with respect to phasing. The bottom perforation holes receive significantly more particles for large $G$, while the perforations located above the flowing bed receive virtually no particles in this case. Recall, that these plots correspond to the situation of $l_p=0$, i.e. plot the particle volume fraction right at the perforation. Once non-trivial $l_p$ is introduced, then there is going to be some smearing of the results, especially for the perforations located on the sides. Both 60$^\circ$ and $90^\circ$ phasing strategies lead to less dramatic variation with respect to $G$ due to the averaging effect. Nevertheless, the average tends to be higher than the average value, which is especially noticeable for low particle volume fractions. Also, even though the average value is relatively uniform, the actual particle volume fractions in individual perforations vary dramatically for large values of the dimensionless gravity $G$.

\begin{figure}[h]
\centering \includegraphics[width=0.85\linewidth]{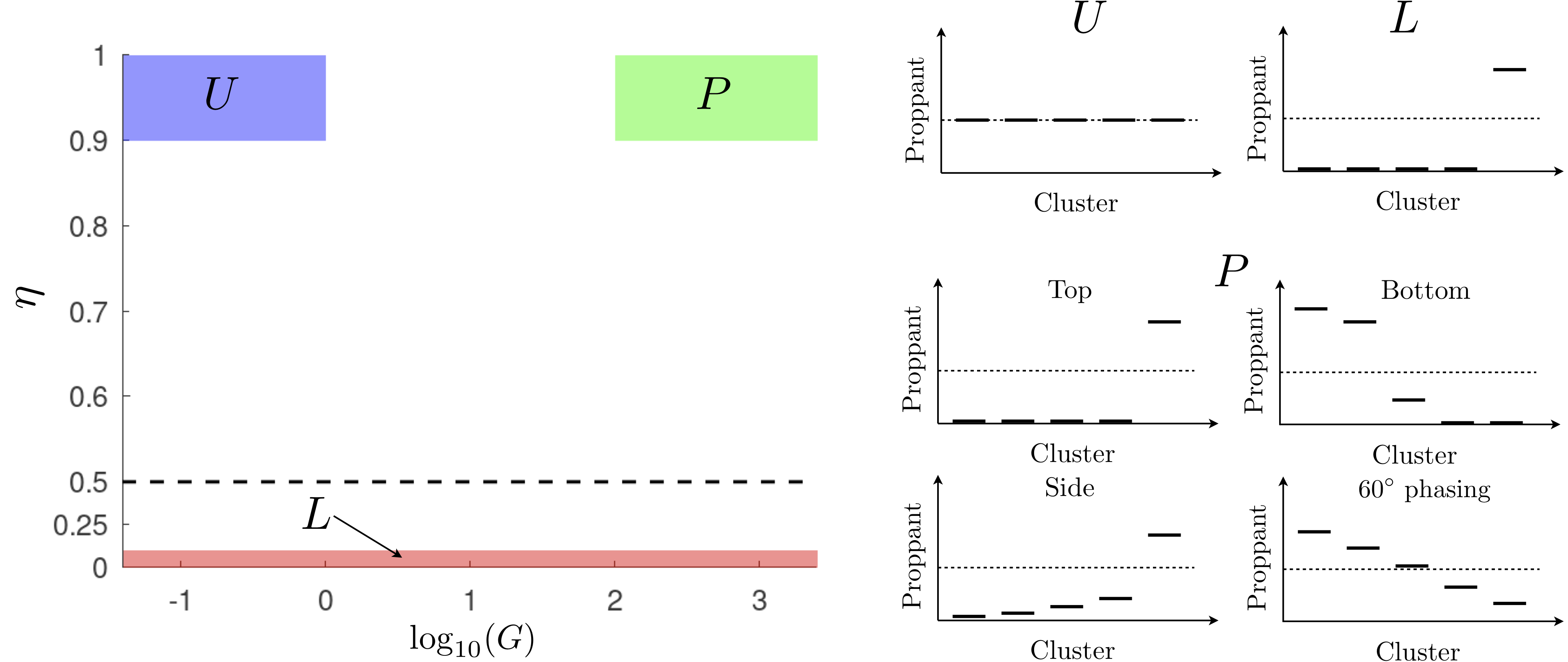}
\caption{Parametric space for slurry in a perforated wellbore. The limiting case of uniform particle distribution is shown by the blue region $U$. The limiting case of ``last cluster takes all'' is shown by the red region $L$. The limiting case of strong phasing sensitivity is shown by the green region $P$. The panels on the right schematically show proppant distribution between clusters for the limiting cases.}
\label{fig10}
\end{figure}

To better understand slurry behavior for the perforated wellbore, i.e to examine particle distribution between the clusters, Fig.~\ref{fig10} plots $(\eta,G)$ parametric space for the problem. Here the horizontal axis $G$ determines particle distribution in the wellbore (see Fig.~\ref{fig8}), while the vertical axis $\eta$ quantifies the particle turning efficiency, as defined in~(\ref{qpp2}). 

For small values of $G$ and $\eta$ close to 1, particle distribution inside the well is uniform and practically all particles are able to turn to the perforation. This leads to uniform particle distribution between the clusters. This region is schematically indicated by the blue zone and is denoted by $U$. There is no dependence on perforation phasing in this case since there is no particle concentration variation within the wellbore's cross-section. 

The second limiting case corresponds to very small values of turning efficiency $\eta$. In this scenario, almost all particles miss the perforation. As a result, all proppant goes into the last perforation cluster because proppant has to go somewhere. Mathematically, $p_1=0$ for the last perforation hole, see~(\ref{turngovsolcutoff}). This makes the slip zero and enforces all proppant in the wellbore to go into the last perforation. Note that the model does not consider screen-out, but it will likely happen in this limit. This situation is referred to as ``last cluster takes all'' or $L$ limit (see the red zone in Fig.~\ref{fig10}). Practically, this is a very unlikely scenario or at least very undesired. There is no phasing dependence in this case as well.

Finally, the $P$ or strong phasing dependence limit corresponds to large $G$ and high turning efficiency. As follows from the name, there is strong dependence on phasing in this case. With the reference to Fig.~\ref{fig9}, it is clear that if the perforations are located at the top, then the situation becomes similar to the $L$ limit, i.e. when the last cluster receives all the proppant. When the perforations are located on the side or $90^\circ$, the situation is somewhat similar, but a bit better. It may also become better when the effect of smearing due to finite $l_p$ is included. So, to this end, for the sake of schematics, it is assumed that some proppant enters the perforation. Then, the particle volume fraction in the wellbore increases towards the toe and slightly more proppant enters the next cluster. But the last cluster still takes most of the proppant. When the peforations are oriented downwards, they take the maximum allowable proppant, then the average volume fraction in the wellbore decreases. If the latter average volume fraction drops to zero, then the downstream clusters receive no proppant. If it does not drop to zero, then some more clusters may receive some proppant, but eventually the last clusters receive no proppant. The situation with $60^\circ$ and $90^\circ$ phasing is similar to the case of the perforations located at the bottom, but it is significantly milder. After the averaging over the perforations, the first cluster receives more proppant than the average volume fraction in the wellbore, as is evident from Fig.~\ref{fig9}. Then the proppant concentration in the wellbore drops. Thus, the next cluster receives less proppant. The volume fraction in the wellbore drops again. The process continues until the end of the stage is reached or the volume fraction in the well becomes zero. Note that the effect of screen-out is not included in this analysis. 

The considered limiting cases and the associated behavior represent a qualitative analysis, which is provided to better understand the limits. In most practical cases, the situation changes dramatically from one cluster to another, and thus these limiting cases are not fully realized. It is also interesting to observe that the reduced turning efficiency tends to make the proppant volume fraction in the first cluster lower than the average in the wellbore. At the same time, non-uniform particle distribution in the wellbore's cross-section paired with $60^\circ$ or $90^\circ$ phased perforations tend to promote the opposite behavior of larger proppant fraction in the first cluster. This also applies to the perforations located at the bottom or lower part of the well. Thus, there is a potential to balance these two phenomena to achieve a relatively uniform particle distribution between clusters. 

Overall, in order to qualitatively understand the behavior for a general case, three things need to be considered. The first thing is the variation of $G$ along the well. Typically, $G$ is small at the heel and becomes very large at the toe for practical cases. Thus, there is almost no effect of phasing near the heel and significant effect of phasing at the toe. As will be shown shortly in next section, the value of turning efficiency is approximately constant for all clusters, but its magnitude is important. The last thing to pay attention to is particle volume fraction in the wellbore. If first perforation clusters receive less proppant than the initial average value in the wellbore, then the average concentration in the wellbore increases and thus next clusters tend to receive more proppant than their upstream neighbors. At the same time, if first few clusters receive more proppant than average, then slurry becomes more dilute downstream, which causes the toe clusters to receive less particles. Clear understanding of these three phenomena allows to understand dynamics of slurry in a perforated wellbore.

With regard to the three aforementioned physical processes, a practical case with downward oriented perforations has the following qualitative behavior. The first few clusters are not going to be affected by the perforation orientation since $G$ is small, but there is going to be some particle slip and therefore the turning efficiency is $\eta\!<\!1$. Thus, the aforementioned first few clusters are going to receive less proppant than average and the concentration in the wellbore is going to gradually increase. Then, as the slurry slows down, the effect of phasing becomes important. Since the perforations are oriented downwards, then they tend to receive more proppant than average, which leads to the decreasing concentration received per cluster and also the decreasing trend for the particle concentration in the wellbore. Therefore, since the upward trend is being replaced by the downward trend, there is going to be a maximum of proppant received per cluster somewhere in the middle of the stage. The location of this maximum depends on actual parameters and it can be shifted more towards the toe or to the heel based on the design parameters. After the maximum, the particle volume fraction should decay quickly since the dimensionless gravity continues to grow rapidly and thus the effect of phasing becomes only stronger downstream. This example demonstrates how a relatively simple analysis enables one to understand behavior of slurry in a perforated wellbore. The same qualitative analysis can be applied to other situations as well. For instance, for the vertically oriented perforations, the behavior for the first several clusters is similar since the values of $G$ are small. Then, there is going to be a competition between the increasing concentration in the wellbore and the reduced particle concentration in the top of the wellbore due to settling. Depending on the parameters, this can lead to either a local decrease in the amount of proppant per cluster or just weakening the upward trend. At the end of the day, the particle concentration in the wellbore is going to continuously increase and the last cluster is going to receive the disproportionate large amount of proppant. Naturally, the situation with horizontally oriented perforations is somewhere in the middle between these two cases. Later in the paper, Fig.~\ref{fig30} is going to quantitatively examine all three these cases.
 
\section{Model calibration}\label{secmodelcal}

The developed model has four fitting parameters: $f_D$, $f_T$, $p_V$ and $p_T$. The first two affect the particle distribution in pipe flow, while the last two quantify the proppant turning phenomenon. The purpose of this section is to make comparisons between the model and available experimental and computational data in order to find the values of these parameters. In particular, the comparison with the following four studies is included~\cite{Grues1982,Nasr1989,WuPhD,GilliesPhD}. The study~\cite{GilliesPhD} considers flow of slurry in a pipe and focuses on experimentally quantifying particle and velocity variation within the pipe. This is the key study to determine the flow-related parameters $f_D$, $f_T$. The studies~\cite{Grues1982,Nasr1989,WuPhD} consider the T-junction geometry or the situation of a single perforation in a wellbore. The study~\cite{WuPhD} employs CFD approach, while the other two~\cite{Grues1982,Nasr1989} focus on experimental measurements.

\begin{figure}[h]
\centering \includegraphics[width=0.9\linewidth]{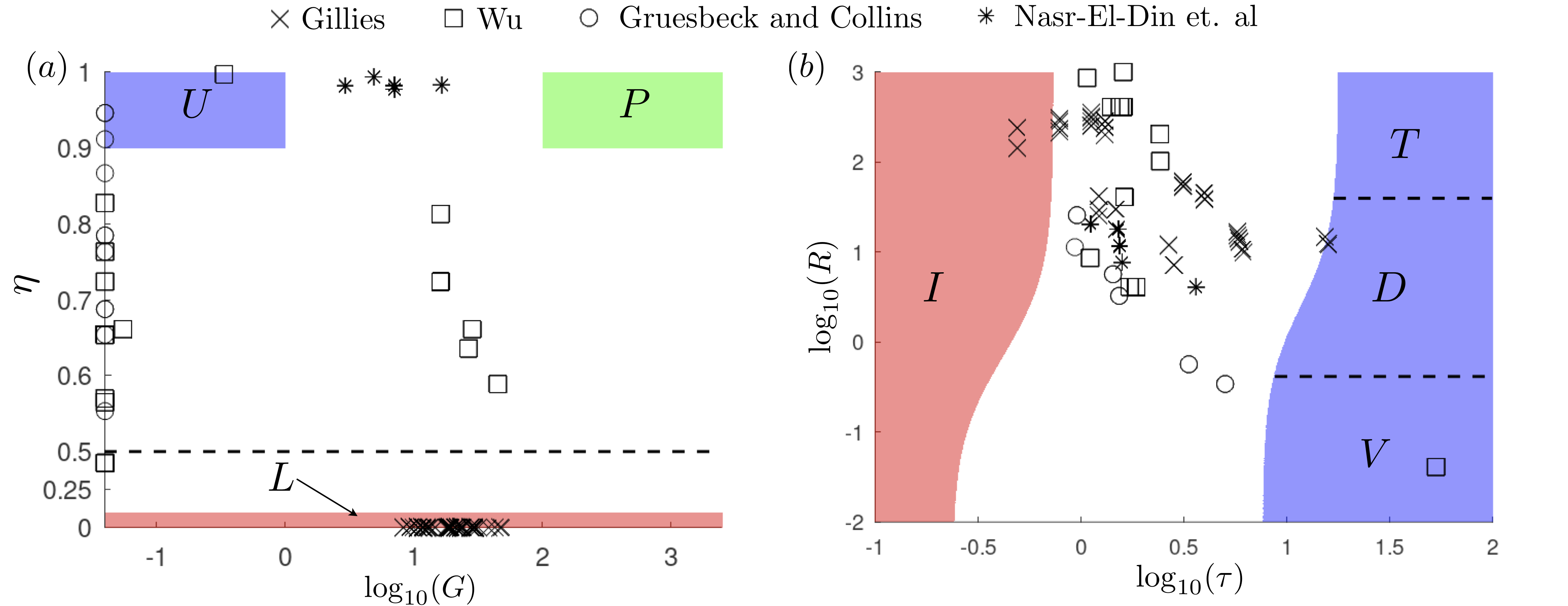}
\caption{Location of the data used for calibration in the parametric spaces. Panel $(a)$ shows $(\eta,G)$ parametric space, while the panel $(b)$ illustrates $(R,\tau)$ parametric space. The crosses correspond to the experimental data~\cite{GilliesPhD}, square markers indicate CFD results from~\cite{WuPhD}, circular markers show the experimental results from~\cite{Grues1982}, while the star markers correspond to the laboratory results~\cite{Nasr1989}.}
\label{fig11}
\end{figure}

To better understand how each of the four studies can be used to calibrate the model, Fig.~\ref{fig11} plots the $(\eta,G)$ and $(R,\tau)$ parametric spaces (see sections~\ref{secturn} and~\ref{secperfwell}) and illustrates the parameters used in~\cite{Grues1982,Nasr1989,WuPhD,GilliesPhD} by markers. Note that the apparent $G$, i.e. $G_a$ is shown.

The study~\cite{GilliesPhD} (crosses) focuses exclusively on pipe flow and therefore the turning efficiency is set to zero. The location on the $(R,\tau)$ diagram is also irrelevant since there is no actual particle turning involved. What is important, on the other hand, is the span of $G$ that the experiments cover. This span covers the most interesting part on the $(\eta,G)$ diagram, that corresponds to the transition from the uniform particle distribution to a nearly bed flow regime. It would also be nice to have the results for lower and higher $G$ for completeness, but for the lower $G$ the result is rather trivial (uniform particle distribution), and for the higher $G$ there is a higher chance of bed formation, which is, first of all, not accounted in this study and, second, is relevant only for the last cluster. 

Parameters corresponding to the work~\cite{WuPhD} are shown by the square markers in Fig.~\ref{fig11}. They span a relatively large range of $\eta$, which is good for calibrating $p_T$. Also, many points correspond to very small values of $G$, which allows to decouple the problem of turning from the problem of non-uniform particle distribution in a pipe. This is also good for calibration. There are results that correspond to the low injection rate and moderate values of $G$. These are less relevant for calibrating the turning, but are important for testing the effect of non-uniform particle distribution in a well. In terms of the $(R,\tau)$ parametric space, many points are in close proximity to the inertial regime $I$, in which there is no sensitivity to particle size and viscosity. This is not so good for calibration. There are some cases that fall into the viscous drag dominated limit $V$, but they correspond to the low injection rate and moderate $G$, which overshadows the effect of turning.

Experimental data by~\cite{Grues1982} is shown by circular markers in Fig.~\ref{fig11}. The data points correspond to a vertical well, hence there is no effect of gravity on particle distribution within the well and very small values of $G$ are assigned.  On the $(R,\tau)$ diagram, the results span diagonally from the nearly $I$ limit to $V$ limit. Thus, they are able to provide sensitivity to viscosity and particle size, which is good for calibration.

Finally, laboratory data~\cite{Nasr1989} is shown by stars. The T-junction of pipes with the same diameter was used. Therefore, the perforation diameter is essentially equal to pipe diameter. This makes the turning efficiency very high, as indicated in Fig.~\ref{fig11}$(a)$. As a result, there is no sensitivity to turning (because practically all particles within $l_f$ zone enter the perforation) and the location in the $(R,\tau)$ parametric space is irrelevant. At the same time, the data points span a relatively wide range of $G$ and therefore this experimental data can be used to indirectly examine the particle distribution in a pipe and to double-check the fitting done based on the data from~\cite{GilliesPhD}.

Without going into details of fitting, the resultant fitting parameters are $f_D=f_T=0.04$, $p_V=0.5$, and $p_T=1.3$. The results presented in the remainder of this section use these parameters.

\subsection{Slurry flow in a pipe}

This section compares results of the model to laboratory experiments that are summarized in~\cite{GilliesPhD}. In particular, particle concentration and velocity profiles are measured for various particle sizes, pipe diameters, average flow velocities, and average volume fractions. Results of 73 measurements are reported in~\cite{GilliesPhD}. The following particle sizes (diameters) were used: 0.1-0.3 mm silica sand, 0.3-1.0 mm silica sand, 2-6 mm gravel, 0-1 mm silica sand, 0-6 mm gravel, and 0-10 mm Westar coal. The sand and gravel densities are the same and equal to 2650~kg/m$^3$, while the coal particles have density of $1474$~kg/m$^3$. As can be seen, various mixtures of particles were used. Since the developed model for the slurry flow in a pipe does not significantly depend on particle size, an average between the two limiting values is taken for estimating the particle radius. The following pipe diameters were used in the experiments: 53.5~mm, 158.5~mm, 263~mm, and 495~mm. There was the so-called ``approach section'' to ensure that the flow is steady-state in the measurement section. Its length was 6~m for 53.2~mm pipe, 15~m for 158.5~mm and 263~mm pipes, and 25~m for 495~mm pipe. These values are used to estimate the dimensionless approach time $\bar t$ for the experimental data. Average flow velocity varied within the range $1.8\div4.9$~m/s, while the volume fraction was $0.06\div0.48$. 

\begin{figure}
\centering \includegraphics[width=0.7\linewidth]{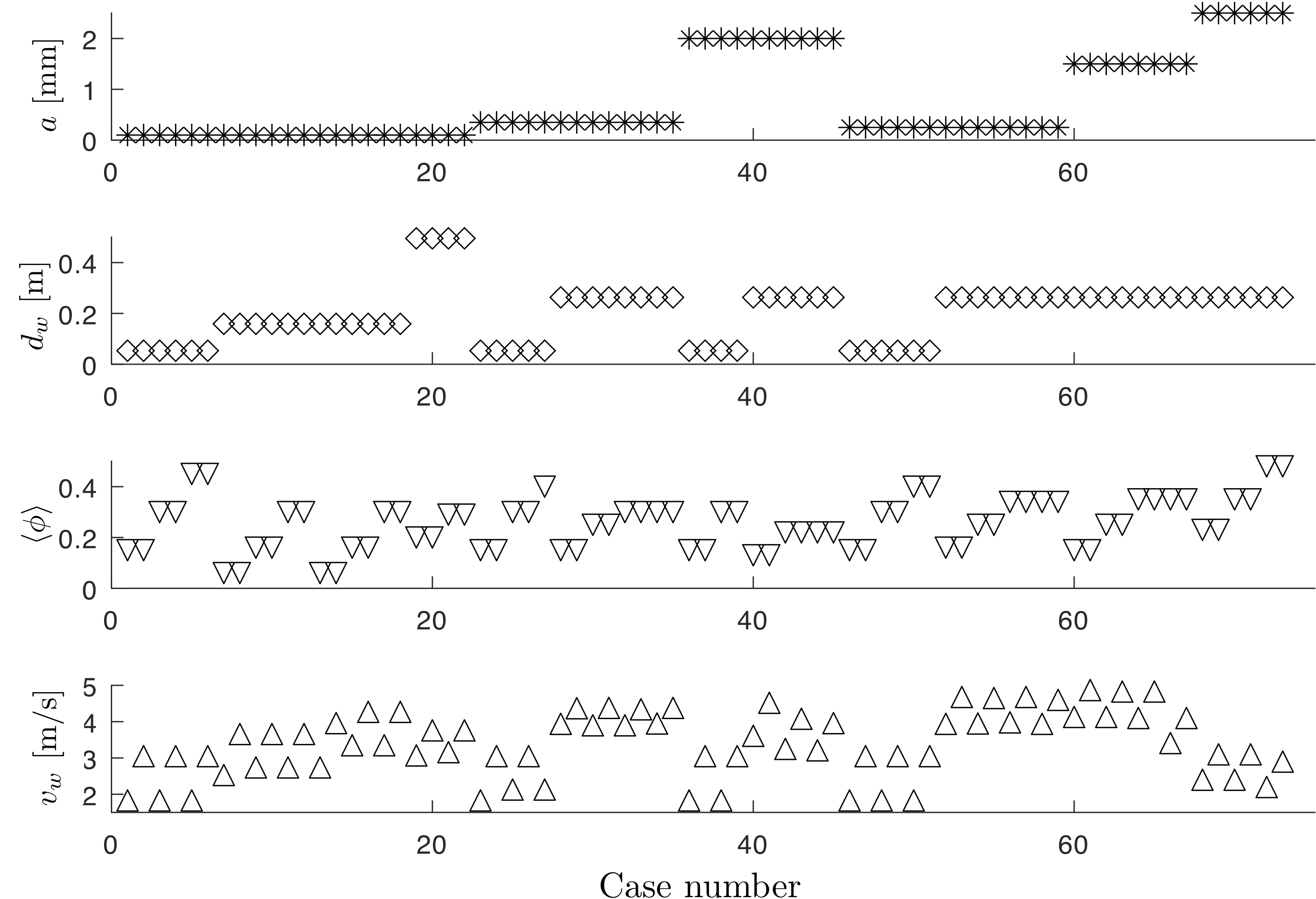}
\caption{Summary of experimental data from~\cite{GilliesPhD}. From top to bottom: $a$ - average particle radius, $d_w$ - pipe diameter, $\langle\phi\rangle$ - average particle volume fraction (not normalized), and $v_w$ - average fluid velocity.}
\label{fig12}
\end{figure}

Fig.~\ref{fig12} summarizes original data for all 73 cases. It plots particle radius $a$, pipe diameter $d_w$, average particle volume fraction $\langle \phi\rangle$, and average flow velocity $v_w$. As can be seen from the data, there are numerous combinations of particle size, pipe diameter, volume fraction, and average velocity. The particle size span is very significant and is much wider than the one typically involved in petroleum applications. Particle volume fractions are also on a high side, while the flow velocities are comparable to typical values in the field.

\begin{figure}
\centering \includegraphics[width=0.7\linewidth]{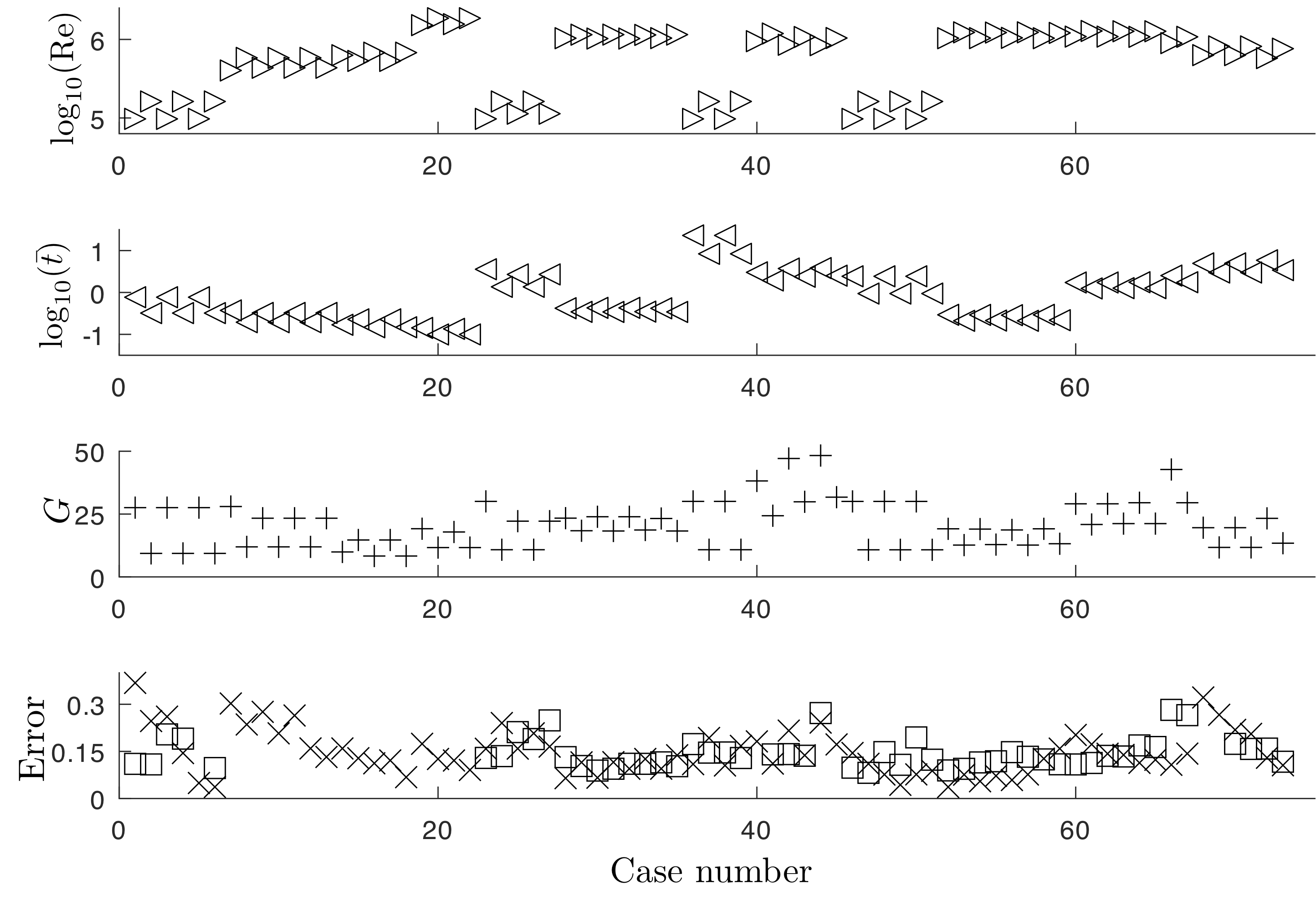}
\caption{Summary of experimental data from~\cite{GilliesPhD}. From top to bottom: $\text{Re}$ - Reylonds number for the pipe flow, $\bar t$ - dimensionless flow time until measurement, $G$ - value of the dimensionless gravity parameter, $\text{Error}$ - relative discrepancy between the experimental data and model for volume fraction (crosses) and velocity (squares).}
\label{fig13}
\end{figure}

It is also important to plot the dimensionless parameters for the data. Fig.~\ref{fig13} shows pipe's Reynolds number $\text{Re} = \rho_f  v_w d_w/\mu$ ($\mu=10^{-3}$~Pa$\cdot$s), dimensionless time $\bar t$, value of the dimensionless gravity $G$, and the error between the model and the laboratory data (crosses correspond to volume fraction and squares for velocity). Pipe Reynolds number varies in a relatively narrow range, but this range fits well within the zone of interest for oil and gas applications. The only potential difference is the lower bound, which in the field practically occurs only for the last toe cluster. With regard to the dimensionless time $\bar t$, it varies quite significantly. In the model it is assumed that the particle concentration is initially uniform and then, the solution is evaluated after the suspension is flowed through the ``approach section''. There are multiple points with $\bar t \ll 1$. This indicates that the approach length selected in the experiments is not sufficient to reach the steady-state flow, at least according to the current model. Nevertheless, since the model can account for the finite approach length, the laboratory results are compared the corresponding non-steady solution. The value of the dimensionless parameter $G$ reaches up to $50$, but most of the points are in the 20th range. As was indicated earlier, this span of $G$ is good for calibrating the model, even though it is more desirable to also have steady-state solution for all cases. Finally, the last plot on Fig.~\ref{fig13} shows the error observed between the model and the experiments. Crosses correspond to particle volume fraction error, while squares represent the velocity error. For the case of particle volume fraction, the error is calculated as
\begin{equation}
\text{Error} = \sqrt{\dfrac{\sum_i S(z_i)^2 (\phi(z_i)-\phi_i)^2}{\sum_i S(z_i)^2 \phi_i^2}},
\end{equation}
where $z_i$ are the measurement points, $S$ is defined in~(\ref{massbalance}), $\phi(z_i)$ is the model value evaluated at $z_i$, while $\phi_i$ is the measured value. Summation in the above equation goes over all evaluation points. The velocity error is calculated in a similar fashion, but the volume fraction is replaced by the velocity value. Results for the error are very reasonable, but not ideal. The vast majority of cases have error less than 0.3. The velocity error is generally less since the velocity has less variation. Note that the values of $f_D$ and $f_T$ are calibrated to minimize the error for all cases simultaneously.

\begin{figure}[h]
\centering \includegraphics[width=0.8\linewidth]{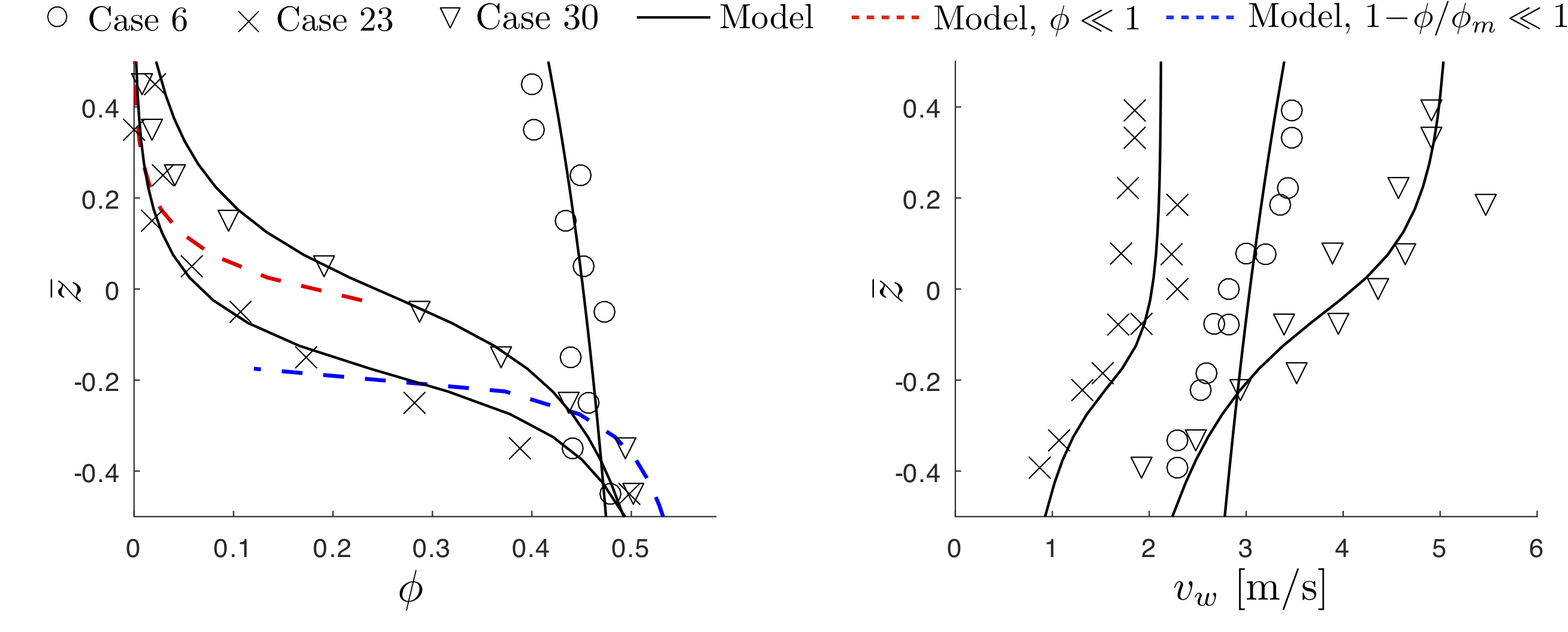}
\caption{Comparison between measurements from~\cite{GilliesPhD} (markers) and the model (black solid lines) for the cases 6, 23, and 30. The left panel shows spatial variation of particle volume fraction, while the right panel shows the distribution of velocity. The dashed lines show the limiting solutions~(\ref{steadystatesol1}) and~(\ref{steadystatesol2}) for the case 23.}
\label{fig14}
\end{figure}

As an illustration, Fig.~\ref{fig14} plots the comparison between the laboratory data and the model for three cases: case 6, case 23, and case 30. The left panel shows the variation of the particle volume fraction, while the right panel shows the variation of the velocity. Markers correspond to measurements, while the black solid lines represent the model. The dashed red and blue lines illustrate the two limiting solutions~(\ref{steadystatesol1}) and~(\ref{steadystatesol2}) (the integration constants in these solutions are selected to closely match the data). To provide a reference, the error for particle concentration for these cases is 0.04, 0.16, and 0.07, while the velocity error is 0.10, 0.13, and 0.09. The value of $G$ is 9, 30, and 24. Results demonstrate a good degree of agreement, even though some differences are clearly visible. Note that it is possible to find the values of $f_D$ and $f_T$ to have a better much for each individual case, but these won't be universal and will require the introduction of more fitting parameters and lead to an overall significantly more complex model. There could also be experimental inaccuracies, so perfect fitting to each case is not necessary. The data clearly follows the low and high concentration solutions, as indicated by the dashed lines for the case 23. This confirms that both the frictional and turbulent terms in the expression for the particle pressure~(\ref{rheology}) are crucial for capturing the experimental data. Recall that the frictional term is responsible for the high concentration solution, while the turbulent term leads to the exponential ``tail'' for low particle concentrations. Finally, the velocity profile is less sensitive to problem parameters, as compared to particle concentration. The velocity variation also matches well for the considered cases, which indicates that the proposed relation between the particle concentration and velocity distribution~(\ref{vdist}) is sufficiently accurate.

\subsection{Particle turning into perforation}

This section evaluates the ability of the model to match the available data for the case of a single perforation. In other words, the problem geometry shown in Fig.~\ref{fig5} is considered. The following three studies~\cite{WuPhD,Grues1982,Nasr1989} are used for comparison. The work~\cite{WuPhD} uses CFD simulations, authors in~\cite{Grues1982} perform laboratory measurements exclusively for vertically oriented well, while~\cite{Nasr1989} has laboratory data for three different orientations (top, bottom, and side), but focuses on the case when perforation diameter is equal to the well diameter. Fluid density $\rho_f=1$~g/cm$^3$ is used in all cases.
  
\begin{figure}[h]
\centering \includegraphics[width=0.99\linewidth]{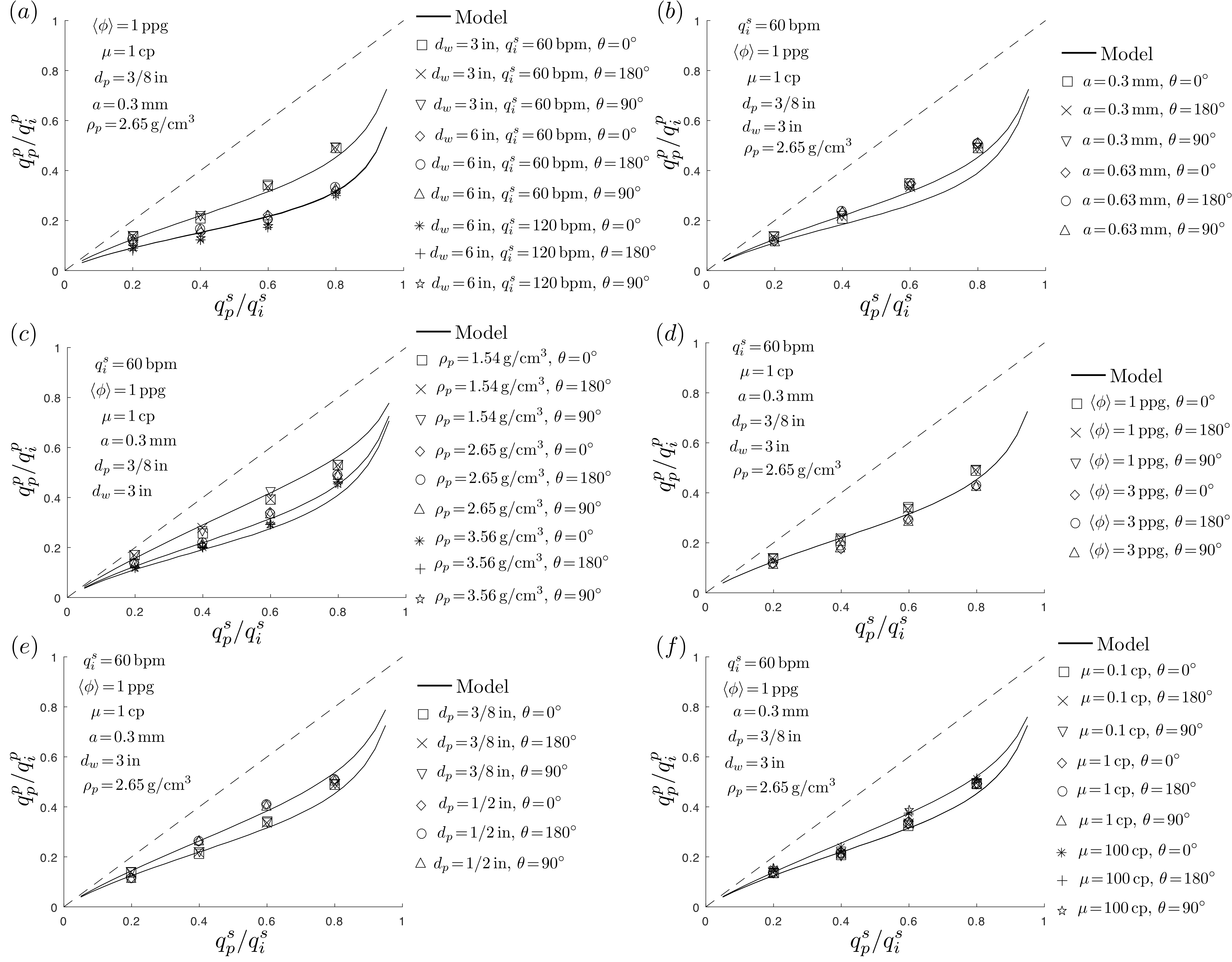}
\caption{Comparison between the model and data from~\cite{WuPhD} for high injection rates.}
\label{fig15}
\end{figure}

Results of comparison with the data from~\cite{WuPhD} is shown in Fig.~\ref{fig15}. In particular, the results for high rate are shown. All the problem parameters are shown on the figure. The horizontal axis is the slurry flow rate fraction, i.e. the ratio between the slurry flow rate going to the perforation to the upstream slurry flow rate. The vertical axis is the proppant flow rate fraction or the ratio between the the propant flow rate going to the perforation to the upstream proppant flow rate. The panel $(a)$ shows the sensitivity to wellbore diameter $d_w$, injection rate $q^s_i$, and perforation orientation $\theta$. Clearly, there is no sensitivity to the perforation orientation. This is consistent with the results shown in Fig.~\ref{fig11}, that demonstrate that such high rate cases correspond to very small values of $G$, which leads to uniform proppant distribution within the well. This applies to all panels in the figure. There is practically no sensitivity to rate, while the sensitivity to wellbore diameter is noticeable. There is no sensitivity to rate because the corresponding location in the $(R,\tau)$ parametric space is far away from the $V$ limit. As can be deduced from~(\ref{turngovsol}), large values of $R$ eliminate dependence on the wellbore velocity (or rate) and viscosity. At the same time, the solution for slip~(\ref{slipsol}) exhibits a linear dependence on $d_w$, which introduces a strong dependence on the wellbore diameter. The panel $(b)$ in Fig.~\ref{fig15} shows the sensitivity to particle size $a$ and perforation orientation $\theta$. There is almost no sensitivity to the particle size since location in the $(R,\tau)$ parametric space is close to the $I$ limit, for which there is no sensitivity to $a$. The panel $(c)$ shows the sensitivity to particle mass density $\rho_p$. There is a noticeable variation of the result with $\rho_p$, which is consistent with the fact that the solution in the $I$ limit~(\ref{slipsol0}) depends on the particle mass density. The panel $(d)$ shows the sensitivity with respect to average particle volume fraction and there is no sensitivity since the particles are distributed uniformly in the wellbore for this case. The panel $(e)$ shows sensitivity to perforation diameter $d_p$. The slip~(\ref{slipsol}) does not depend on the perforation diameter. But the normalized slip $\bar s=s/d_p$ is inversely proportional to $d_p$. Thus, there is some sensitivity to $d_p$. Note that the results seem to be not very sensitive because the considered variation of $d_p$ is relatively small. The panel $(f)$ shows sensitivity to viscosity. As was noted before, since the location in the $(R,\tau)$ parametric space is away from the $V$ limit, there is no sensitivity to viscosity. The final note is that all these results strongly depend on the parameter $p_T$ since the slip is proportional to $p_T$ in the $I$ limit~(\ref{slipsol0}). As a result, the value of $p_T$ is calibrated primarily based on these results.

\begin{figure}[h]
\centering \includegraphics[width=0.99\linewidth]{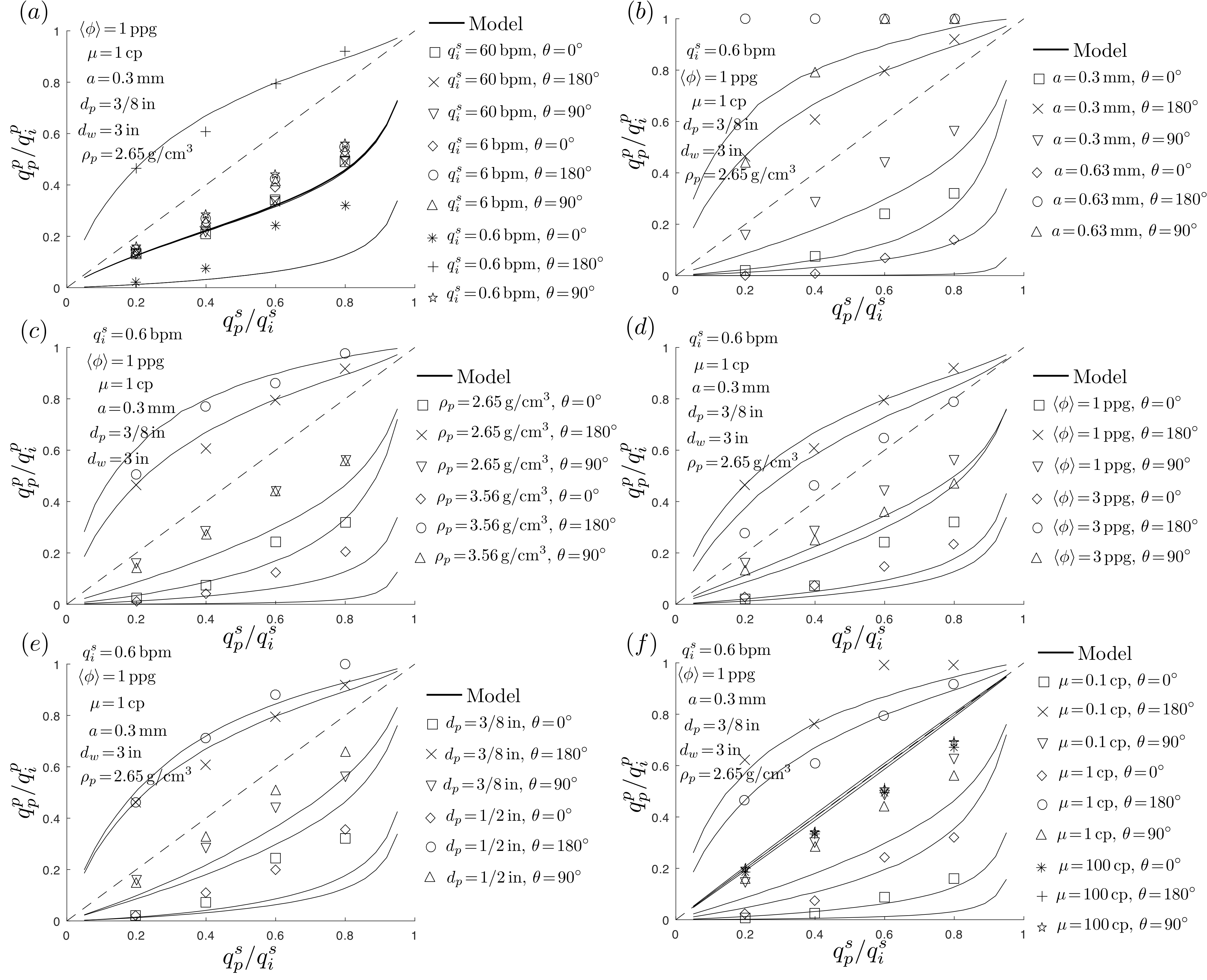}
\caption{Comparison between the model and data from~\cite{WuPhD} for low injection rates.}
\label{fig16}
\end{figure}

Comparison with the data from~\cite{WuPhD} for low injection rates is shown in Fig.~\ref{fig16}. These low rate cases correspond to the data points in Fig.~\ref{fig11} located in the middle of the $(\eta,G)$ diagram. In other words, the values of $G$ are no longer small and therefore spatial particle distribution dominates the result. It is important to note that the particle distribution solution is far away from the steady-state due to the short approach section in the numerical simulations. This approach length is treated as a fitting parameter and is found to be approximately $0.1d_w$. Fig.~\ref{fig16} shows sensitivity to rate, particle size, density, particle volume fraction, perforation diameter, viscosity, and orientation. The model qualitatively agrees with the CFD results, but does not always match quantitatively. As expected, perforations located in the lower part of the well tend to receive more proppant, while the perforations located at the top receive much less proppant. Overall, the purpose of this figure is to highlight that particle volume fraction variation within the well dominates the result for the low rate (or large $G$) scenarios. The particle turning becomes secondary and does not significantly affect the result. Also, the settling dynamics is crucial and thus selecting the approach length to ensure the steady-state solution for particle distribution is very important.

\begin{figure}[h]
\centering \includegraphics[width=0.99\linewidth]{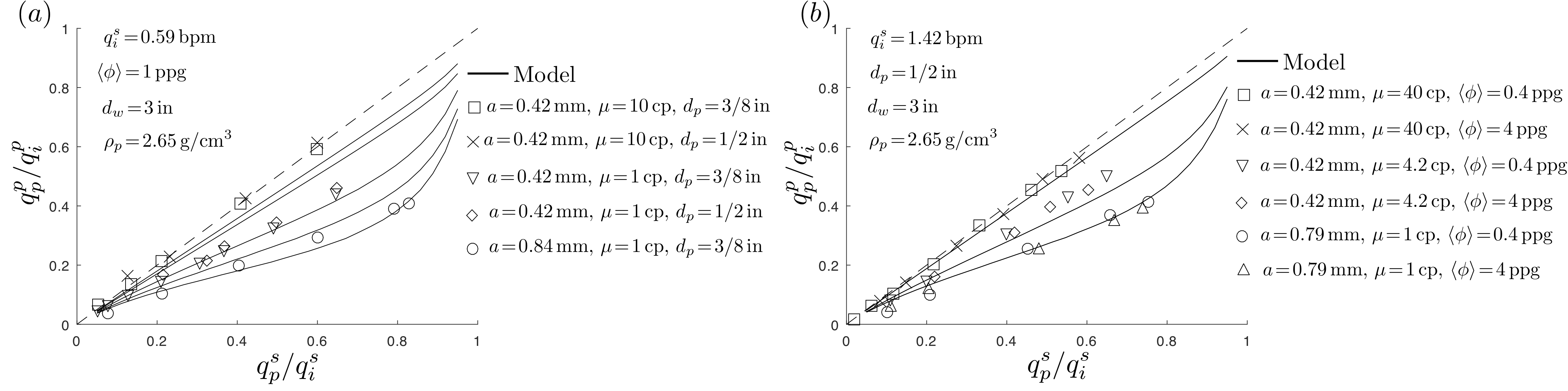}
\caption{Comparison between the model and data from~\cite{Grues1982}.}
\label{fig17}
\end{figure}

Fig.~\ref{fig17} compares model to experimental data from~\cite{Grues1982}. Recall, that~\cite{Grues1982} performed experiments for a vertical well, in which case there is no sensitivity to perforation orientation and particle distribution in the well is uniform. Relatively low injection rates were used. Fig.~\ref{fig17}$(a)$ shows sensitivity to particle size, viscosity, and perforation diameter. With the reference to Fig.~\ref{fig11}$(b)$, the data points span from the nearly $I$ to nearly $V$ limit. Thus, the sensitivity to viscosity and particle size is expected. At the same time, low sensitivity to perforation diameter is not in a full agreement with the model. Fig.~\ref{fig17}$(b)$ shows sensitivity to particle size, viscosity, and particle volume fraction. There is no sensitivity to the particle volume fraction for both the model and the laboratory results. The variation with respect to particle size and viscosity agrees reasonably well between the model and the experimental data. This example clearly demonstrates that there are parameters, for which there is sensitivity to viscosity and particle size. Such a variation is the strongest near the viscous drag $V$ limit in the $(R,\tau)$ parametric space.

\begin{figure}[h]
\centering \includegraphics[width=0.99\linewidth]{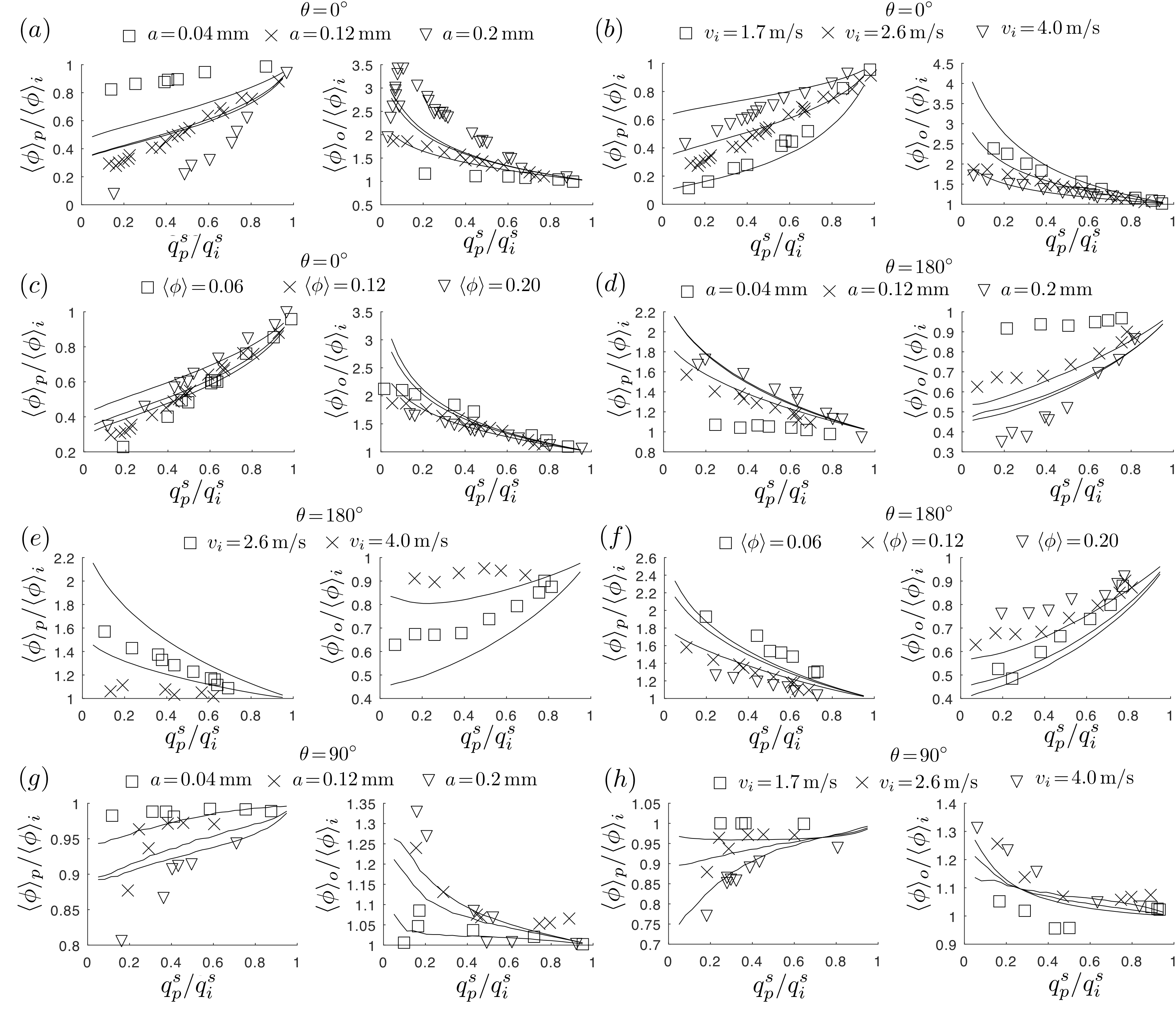}
\caption{Comparison between the model and data from~\cite{Nasr1989} in terms of relative particle volume fractions.}
\label{fig18}
\end{figure}
Fig.~\ref{fig18} compares model to experimental data from~\cite{Nasr1989}. In particular, the figure plots $\langle\phi\rangle_p/\langle\phi\rangle_i$ and $\langle\phi\rangle_o/\langle\phi\rangle_i$ versus $q_p^w/q^s_i$. Here $\langle\phi\rangle_p = q^p_p/q^s_p$ is the average flowing volume fraction in the perforation, $\langle\phi\rangle_i = q^p_i/q^s_i$ is the average flowing volume fraction upstream at the inlet, while $\langle\phi\rangle_o = q^p_o/q^s_o$ is the average flowing volume fraction downstream at the outlet. According to~\cite{Nasr1989}, the length of the approach section is $135d_w$. The default parameters are $d_p=1$~in, $d_w=1$~in, $\langle\phi\rangle=0.12$, $a=0.12$~mm, $v_i=2.6$~m/s, $\mu=1$~cp, and $\rho_p=2.65$~g/cm$^3$. The panel $(a)$ shows sensitivity to particle size for the perforation located at the top. The variation with resect to the particle size is not captured well by the model. One possibility is that some particle size dependence is missing in the model. Alternatively, it can be related to the experimental result as well. The model implicitly assumes that the perforation diameter is much smaller than the wellbore diameter. But in this case, the perforation diameter is equal to wellbore diameter, which can introduce additional effects, which are not accounted for. For instance, this can include settling of particles flowing in the perforation or particle settling in the wellbore during the turning process. The panel $(b)$ shows the sensitivity to the wellbore velocity and the results agree reasonably well. The panel $(c)$ shows the sensitivity to the average particle volume fraction and again the results agree reasonably well. The panel $(d)$ shows the sensitivity to the particle size, but now for the downward oriented perforation. Similarly for the upward located perforation, the variation with respect to particle size is not excellent. The panel $(e)$ shows the sensitivity to the wellbore velocity for the downward oriented perforation, model is able to capture the data reasonably well. The panel $(f)$ shows the sensitivity to the average particle volume fraction and again the results agree reasonably well. Finally, the panels $(g)$ and $(h)$ present sensitivities to particle size and velocity for the case of perforation oriented horizontally. The model agrees with the measurements for these cases. Overall, results demonstrate a good agreement between the model and the laboratory measurements. One exception is the particle size dependence, which is arguably very important for the model. Such dependence is introduced in the model via time needed to reach the steady-state solution. And the solution is not in steady-state for this case. This demonstrates that perhaps further experimental measurements are needed to clarify this issue.

\begin{figure}[h]
\centering \includegraphics[width=0.99\linewidth]{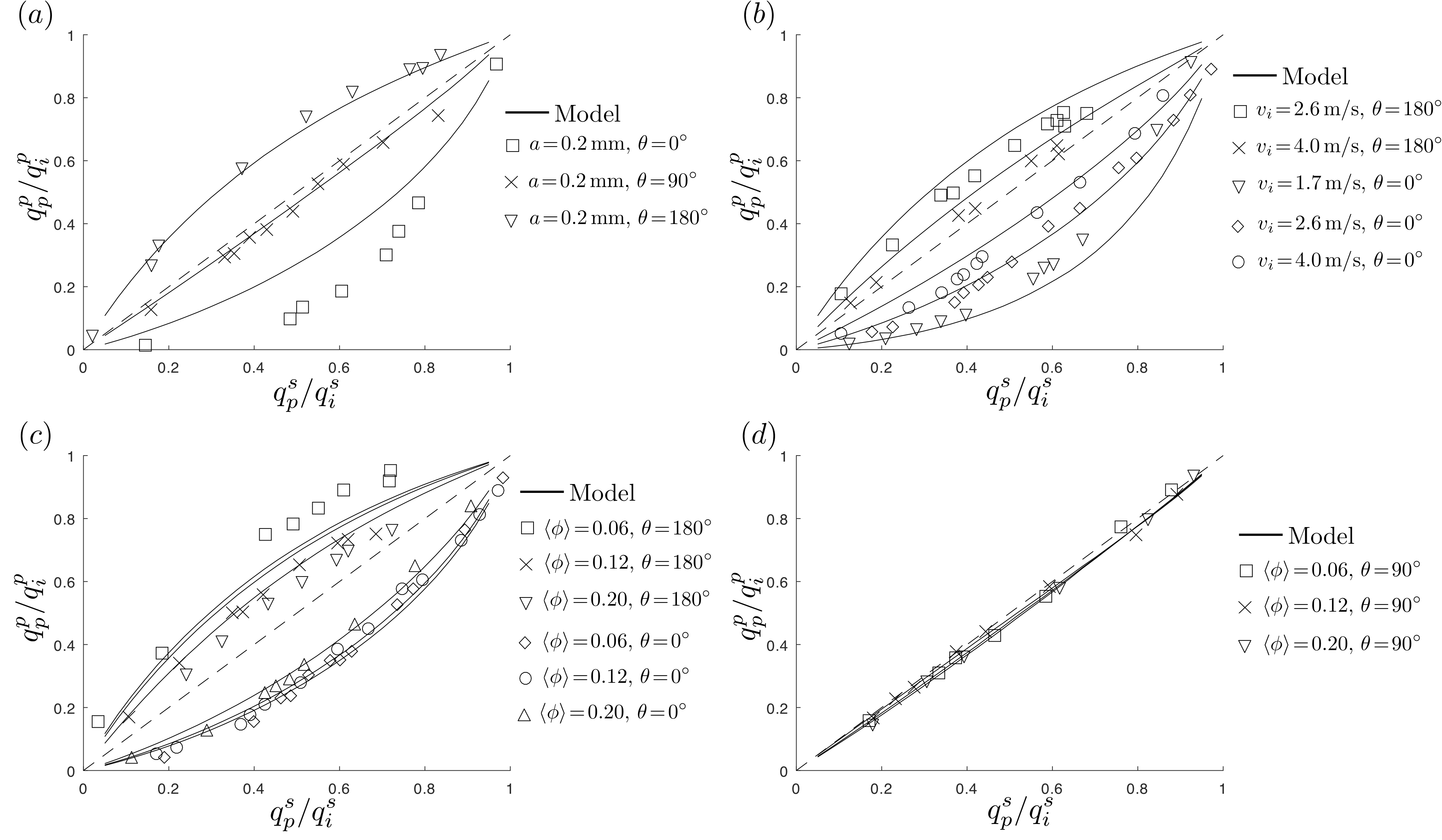}
\caption{Comparison between the model and data from~\cite{Nasr1989} in terms of proppant flow rate fraction.}
\label{fig19}
\end{figure}

Fig.~\ref{fig19} compares model to experimental data from~\cite{Nasr1989} in terms of proppant flow rate fraction. The horizontal axis is the slurry flow rate fraction $q^s_p/q^s_i$, while the vertical axis is the proppant flow rate fraction $q^p_p/q^p_i$. The default parameters are the same as for the previous figure (see the paragraph above). The panel $(a)$ shows the sensitivity to perforation orientation for particle size $a=0.2$~mm. The model and the measurements agree well. The panel $(b)$ shows the sensitivity to perforation orientation and wellbore velocity. Again, the model and the experimental data match reasonably well. Finally, the panels $(c)$ and $(d)$ show the sensitivity to the average particle volume fraction for different perforation orientations. As before, the results show a good degree of consistency. Thus, this figure demonstrates that the sensitivities to orientation, velocity and volume fraction are captured accurately by the model.

\section{Comparison with results for multi-cluster geometry}

This section presents results of comparison between the model and several available experimental and modeling results for the multi-cluster geometry. In particular, comparison is made with the experimental data from laboratory scale experiments for three clusters~\cite{NgameniMS}. The extension of this work for the three cluster geometry, but different pumping conditions and perforation geometry~\cite{AhmadPhD} is also included. After that, a comparison is made with the results shown in study~\cite{Liu2021}, in which laboratory measurements for each individual perforation are reported. Then, field scale experiments from~\cite{Snider2022,Kolle2022} are included. Finally, comparison with the available CFD simulations~\cite{Benish2022,SniderPPT} is performed.

\begin{figure}
\centering \includegraphics[width=0.99\linewidth]{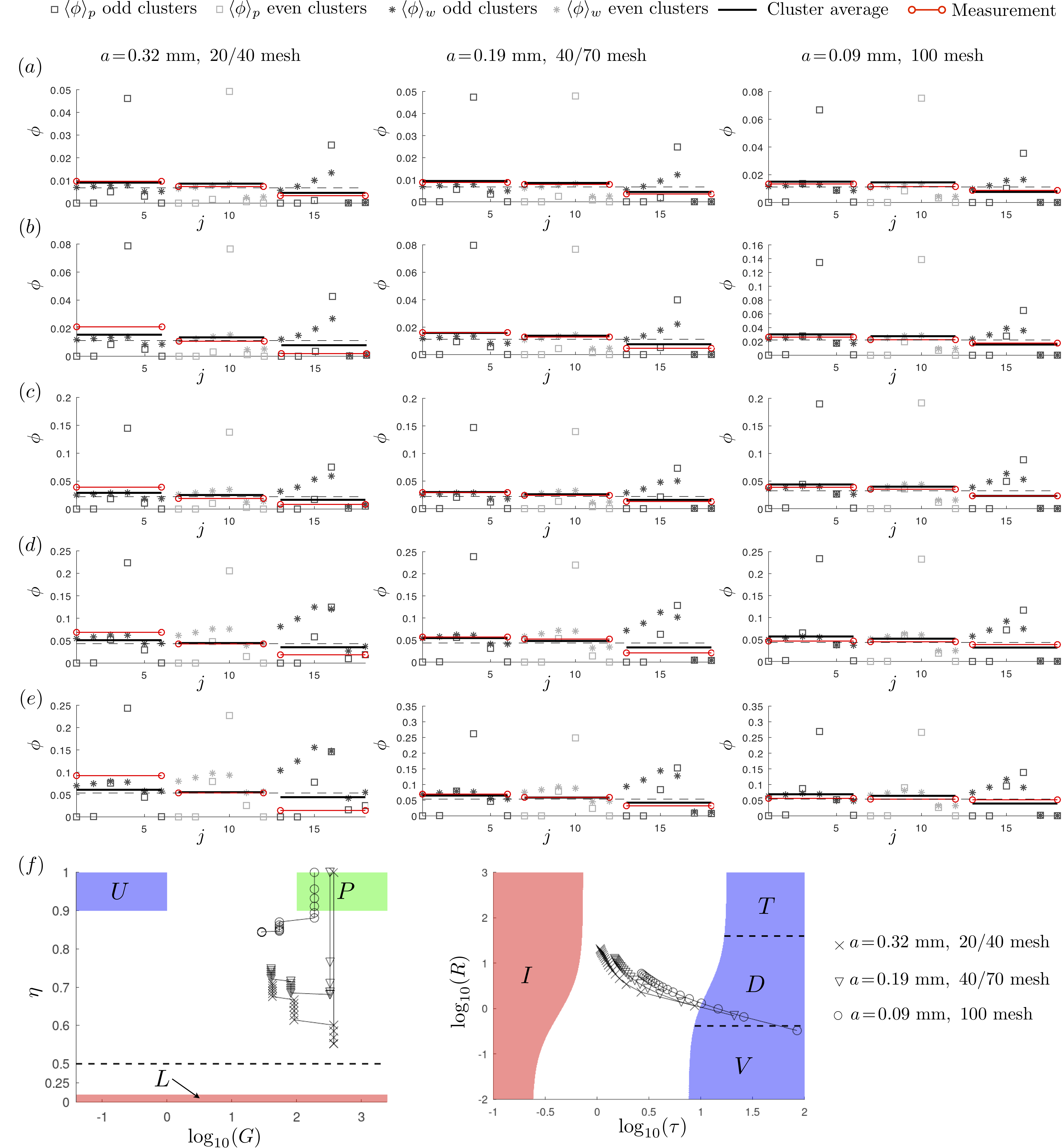}
\caption{Comparison between the model and laboratory data from~\cite{NgameniMS}. The cluster average laboratory data (solid red lines with circular ends) are compared to the cluster average model predictions (solid black lines). Panels $(a)$-$(e)$ show the variation of proppant distribution between perforations for different particle volume fractions and particle size. The dashed lines indicate the initial average volume fraction: $\langle\phi\rangle_0 = \{0.15,0.15,0.25\}$~ppg (panel $(a)$), $\langle\phi\rangle_0 = \{0.25,0.25,0.5\}$~ppg (panel $(b)$), $\langle\phi\rangle_0 = \{0.5,0.5,0.74\}$~ppg (panel $(c)$), $\langle\phi\rangle_0 = \{1,1,1\}$~ppg (panel $(d)$), $\langle\phi\rangle_0 = \{1.25,1.25,1.25\}$~ppg (panel $(e)$). The panel $(f)$ shows the parametric spaces for different particle sizes (location in the parametric spaces does not depend on particle concentration).}
\label{fig20}
\end{figure}

\begin{figure}
\centering \includegraphics[width=0.99\linewidth]{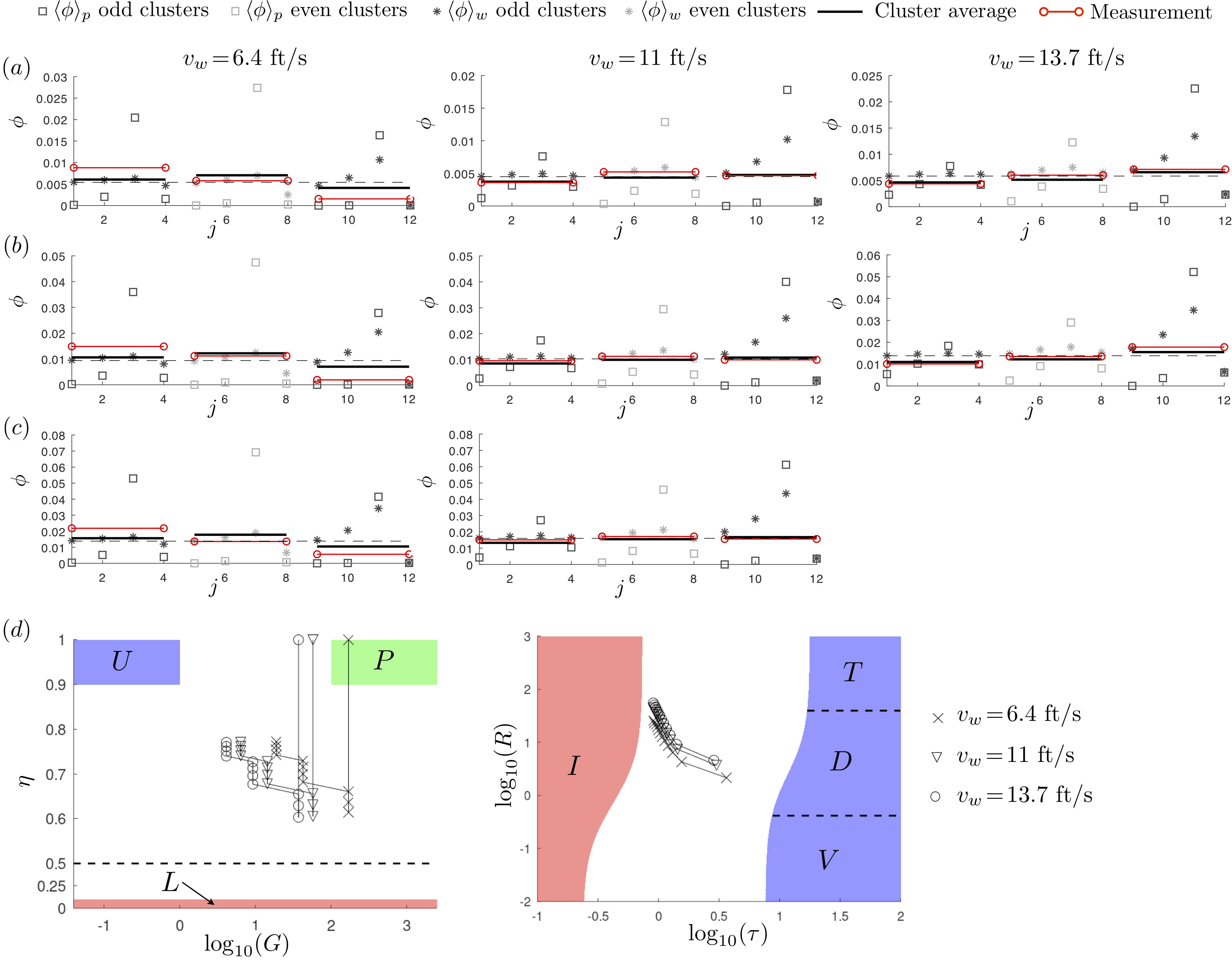}
\caption{Comparison between the model and laboratory data from~\cite{AhmadPhD} for 20/40 mesh sand ($a=0.34$~mm). The cluster average laboratory data (solid red lines with circular ends) are compared to the cluster average model predictions (solid black lines). Panels $(a)$-$(c)$ show the variation of proppant distribution between perforations for different particle volume fractions and wellbore velocity. The dashed lines indicate the initial average volume fraction: $\langle\phi\rangle_0 = \{0.12,0.1,0.13\}$~ppg (panel $(a)$),  $\langle\phi\rangle_0 = \{0.21,0.23,0.31\}$~ppg (panel $(b)$), $\langle\phi\rangle_0 = \{0.31,0.36,N/A\}$~ppg (panel $(c)$). The panel $(d)$ shows the parametric spaces for different velocities (location in the parametric spaces does not depend on particle concentration).}
\label{fig21}
\end{figure}

\begin{figure}
\centering \includegraphics[width=0.99\linewidth]{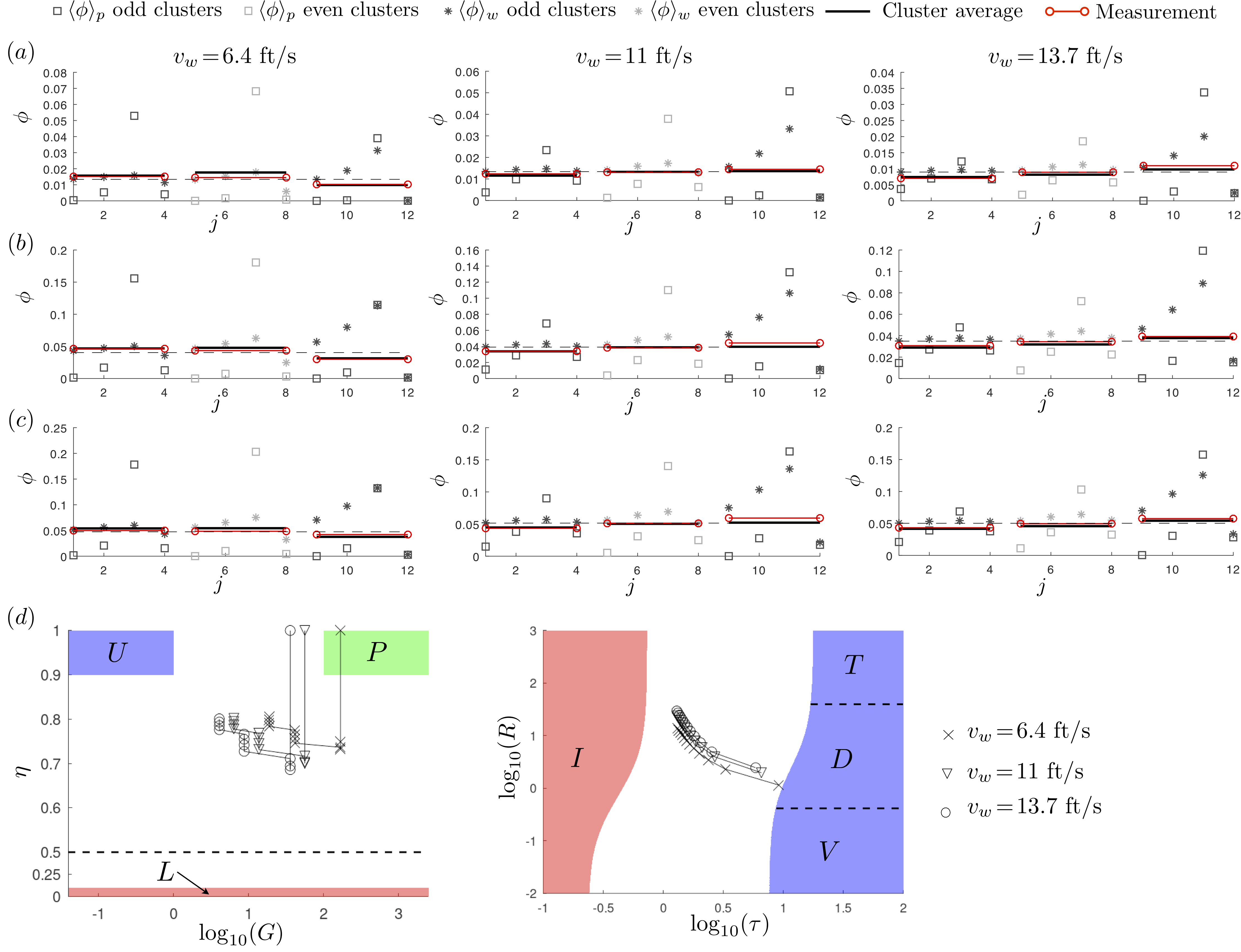}
\caption{Comparison between the model and laboratory data from~\cite{AhmadPhD} for 40/70 mesh sand ($a=0.18$~mm). The cluster average laboratory data (solid red lines with circular ends) are compared to the cluster average model predictions (solid black lines). Panels $(a)$-$(c)$ show the variation of proppant distribution between perforations for different particle volume fractions and wellbore velocity. The dashed lines indicate the initial average volume fraction: $\langle\phi\rangle_0 = \{0.3,0.3,0.2\}$~ppg (panel $(a)$), $\langle\phi\rangle_0 = \{0.93,0.9,0.8\}$~ppg (panel $(b)$), $\langle\phi\rangle_0 = \{1.1,1.2,1.17\}$~ppg (panel $(c)$). The panel $(d)$ shows the parametric spaces for different velocities (location in the parametric spaces does not depend on particle concentration).}
\label{fig22}
\end{figure}

\begin{figure}
\centering \includegraphics[width=0.99\linewidth]{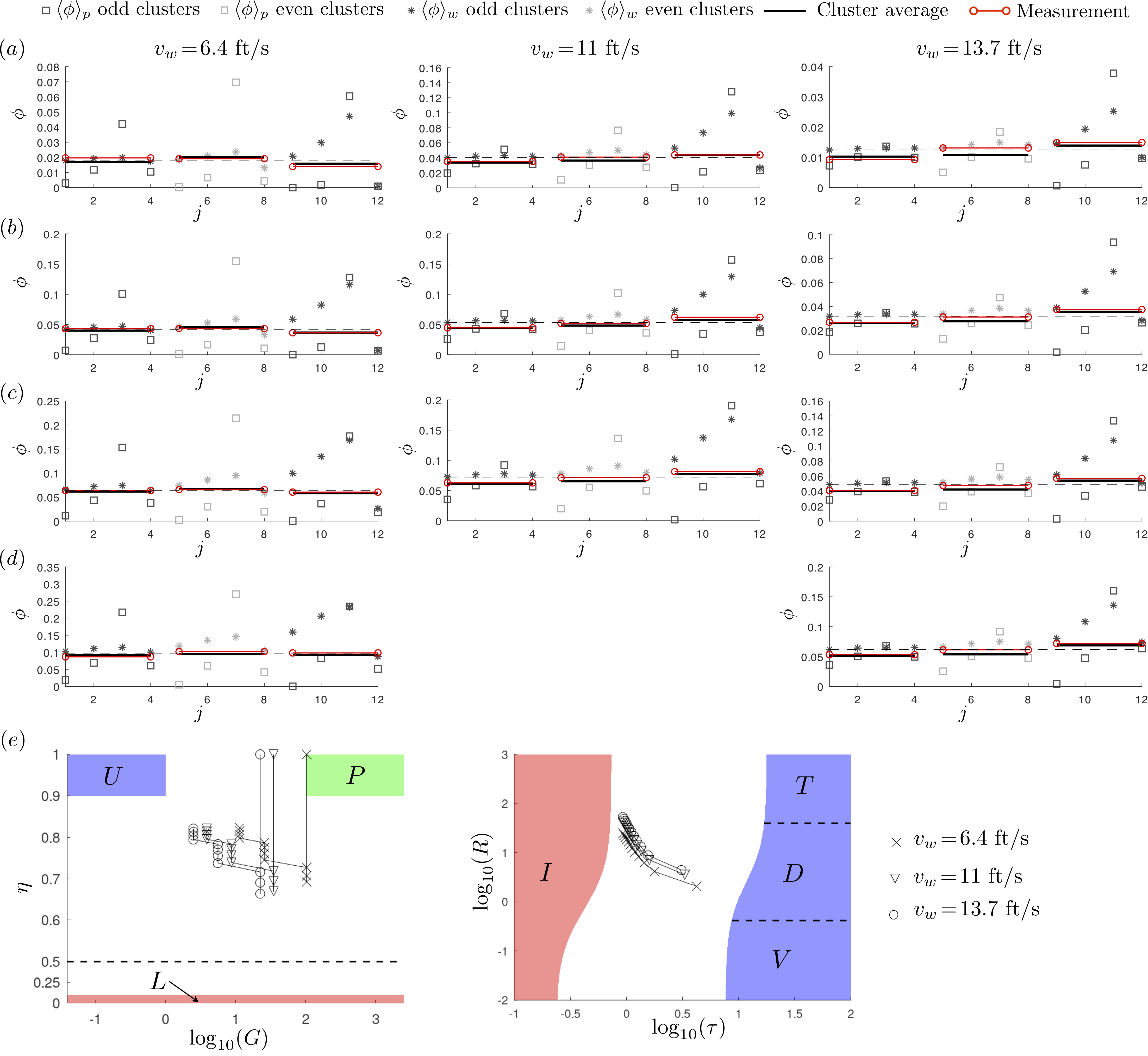}
\caption{Comparison between the model and laboratory data from~\cite{AhmadPhD} for 20/40 mesh ultralight proppant ($a=0.33$~mm, $\rho_p=2000$~kg/m$^3$). The cluster average laboratory data (solid red lines with circular ends) are compared to the cluster average model predictions (solid black lines). Panels $(a)$-$(d)$ show the variation of proppant distribution between perforations for different particle volume fractions and wellbore velocity. The dashed lines indicate the initial average volume fraction: $\langle\phi\rangle_0 = \{0.3,0.7,0.21\}$~ppg (panel $(a)$), $\langle\phi\rangle_0 = \{0.72,0.93,0.55\}$~ppg (panel $(b)$), $\langle\phi\rangle_0 = \{1.14,1.4,0.85\}$~ppg (panel $(c)$),
 $\langle\phi\rangle_0 = \{1.8,N/A,1.1\}$~ppg (panel $(d)$). The panel $(e)$ shows the parametric spaces for different velocities (location in the parametric spaces does not depend on particle concentration).}
\label{fig23}
\end{figure}

\begin{figure}
\centering \includegraphics[width=0.99\linewidth]{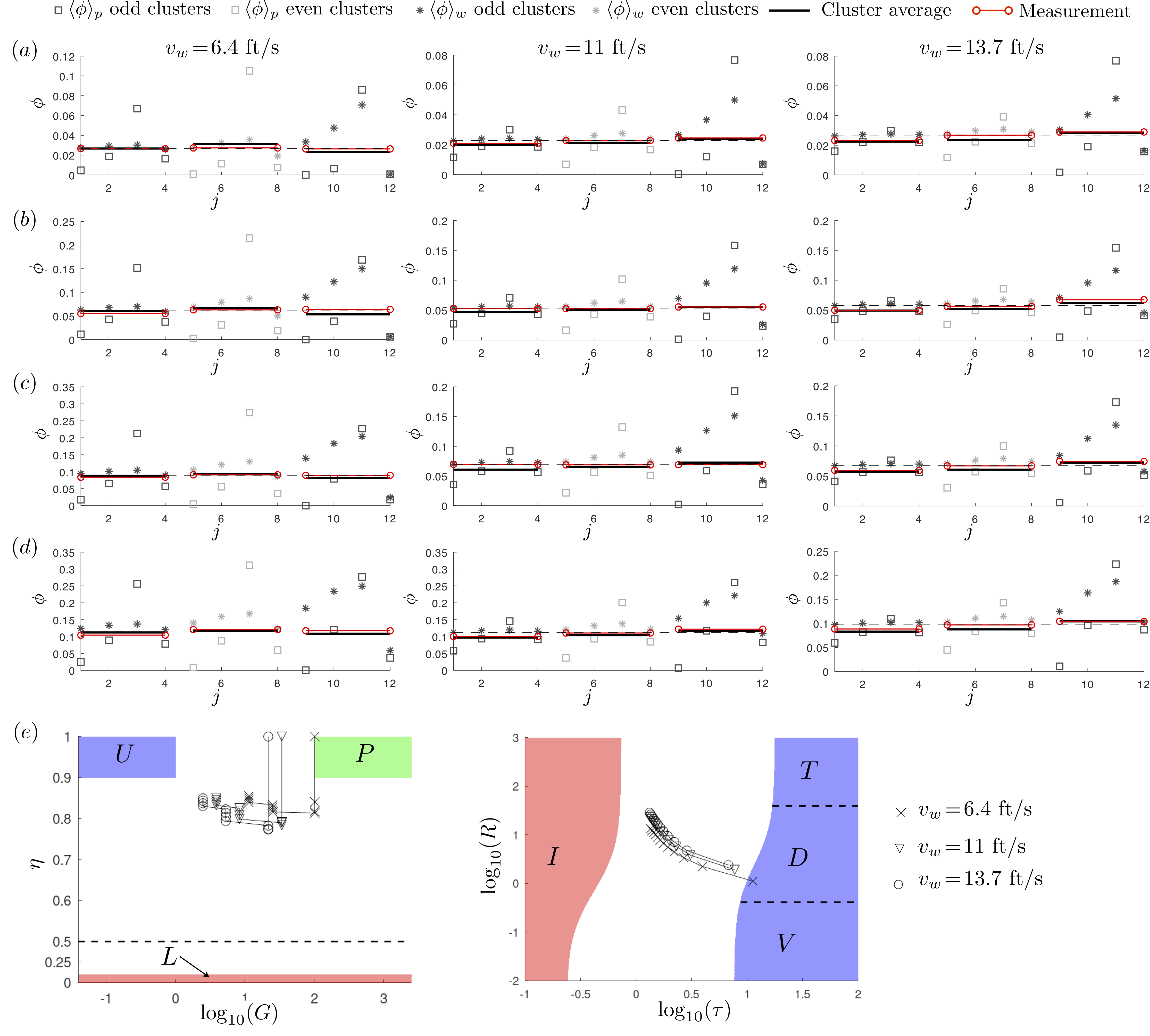}
\caption{Comparison between the model and laboratory data from~\cite{AhmadPhD} for 40/70 mesh ultralight proppant ($a=0.18$~mm, $\rho_p=2000$~kg/m$^3$). The cluster average laboratory data (solid red lines with circular ends) are compared to the cluster average model predictions (solid black lines). Panels $(a)$-$(d)$ show the variation of proppant distribution between perforations for different particle volume fractions and wellbore velocity. The dashed lines indicate the initial average volume fraction: $\langle\phi\rangle_0 = \{0.46,0.39,0.45\}$~ppg (panel $(a)$), $\langle\phi\rangle_0 = \{1.08,0.94,1.02\}$~ppg (panel $(b)$), $\langle\phi\rangle_0 = \{1.64,1.25,1.2\}$~ppg (panel $(c)$), $\langle\phi\rangle_0 = \{2.2,2.1,1.1.79\}$~ppg (panel $(d)$). The panel $(e)$ shows the parametric spaces for different velocities (location in the parametric spaces does not depend on particle concentration).}
\label{fig24}
\end{figure}

\begin{figure}
\centering \includegraphics[width=0.99\linewidth]{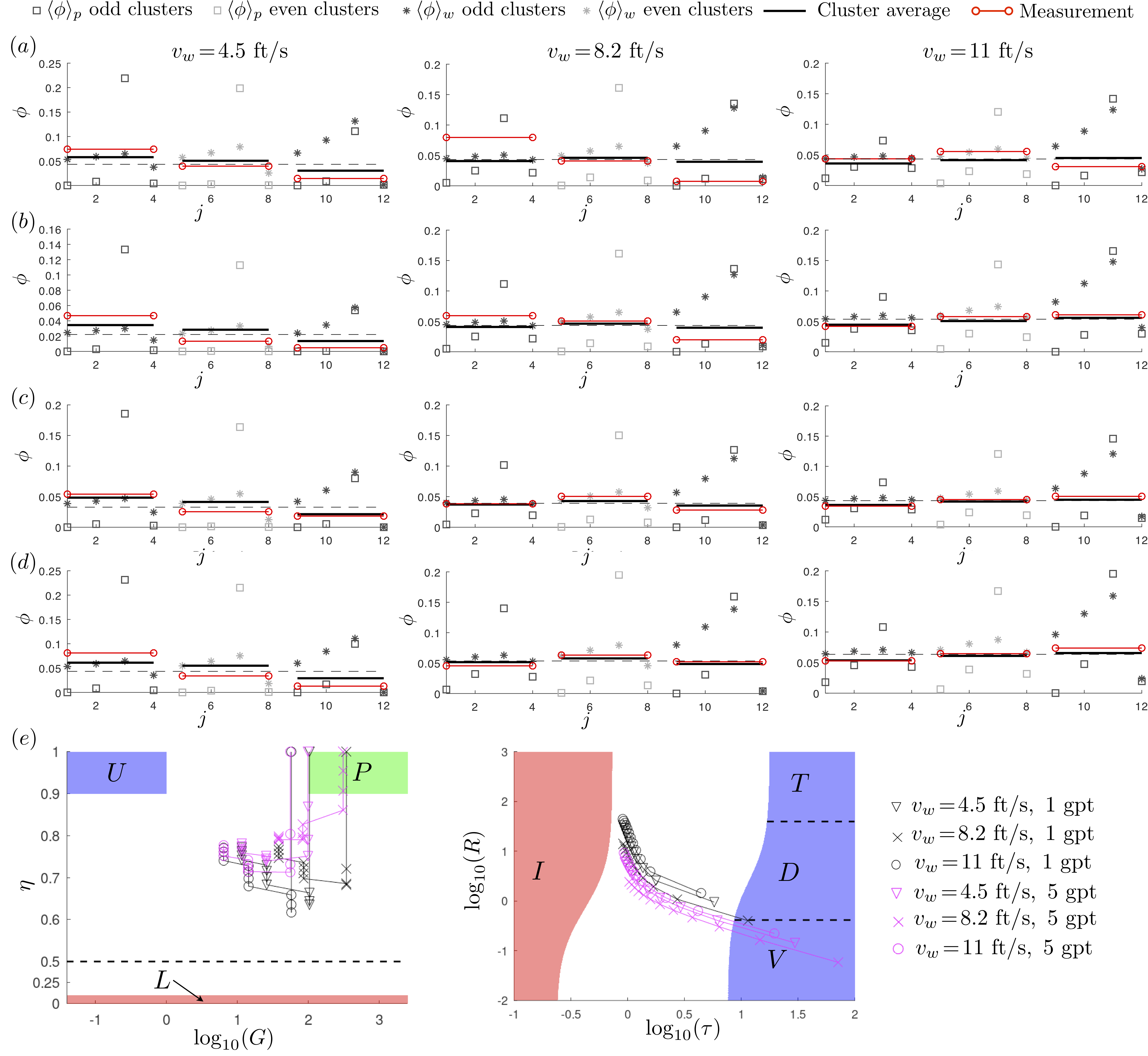}
\caption{Comparison between the model and laboratory data from~\cite{AhmadPhD} for 20/40 mesh sand ($a=0.34$~mm). The cluster average laboratory data (solid red lines with circular ends) are compared to the cluster average model predictions (solid black lines). Panels $(a)$-$(d)$ show the variation of proppant distribution between perforations for different HVFR concentrations and wellbore velocity. The dashed lines indicate the initial average volume fraction, while different panels correspond to various HVFR concentrations: 1~gpt (panel $(a)$), 2~gpt (panel $(b)$), 3~gpt (panel $(c)$), 5~gpt (panel $(d)$). The panel $(e)$ shows the parametric spaces for different velocities and HVFR concentrations.}
\label{fig25}
\end{figure}

\begin{figure}
\centering \includegraphics[width=0.99\linewidth]{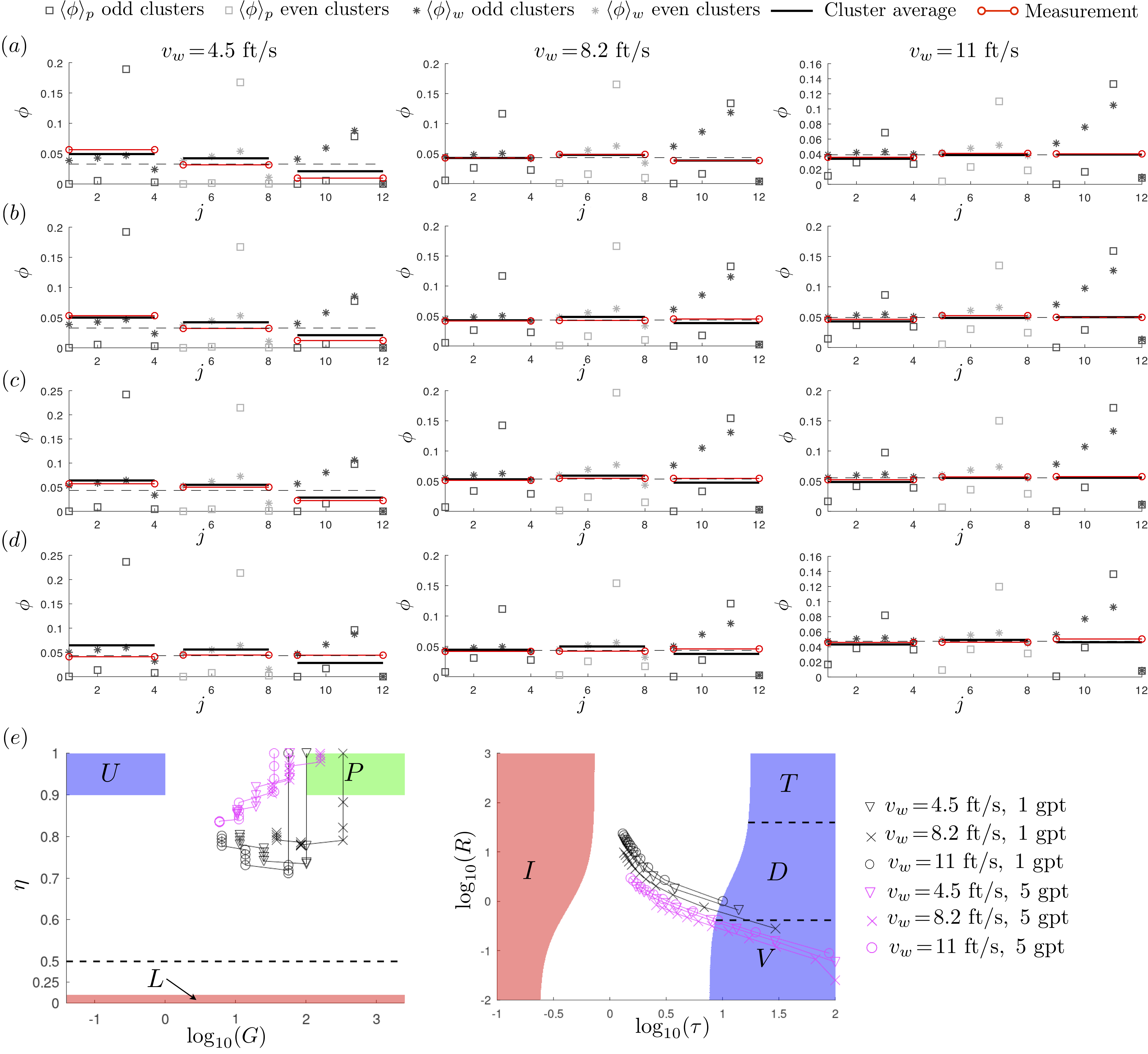}
\caption{Comparison between the model and laboratory data from~\cite{AhmadPhD} for 40/70 mesh sand ($a=0.18$~mm). The cluster average laboratory data (solid red lines with circular ends) are compared to the cluster average model predictions (solid black lines). Panels $(a)$-$(d)$ show the variation of proppant distribution between perforations for different HVFR concentrations and wellbore velocity. The dashed lines indicate the initial average volume fraction, while different panels correspond to various HVFR concentrations: 1~gpt (panel $(a)$), 2~gpt (panel $(b)$), 3~gpt (panel $(c)$), 5~gpt (panel $(d)$). The panel $(e)$ shows the parametric spaces for different velocities and HVFR concentrations.}
\label{fig26}
\end{figure}

\begin{figure}
\centering \includegraphics[width=0.99\linewidth]{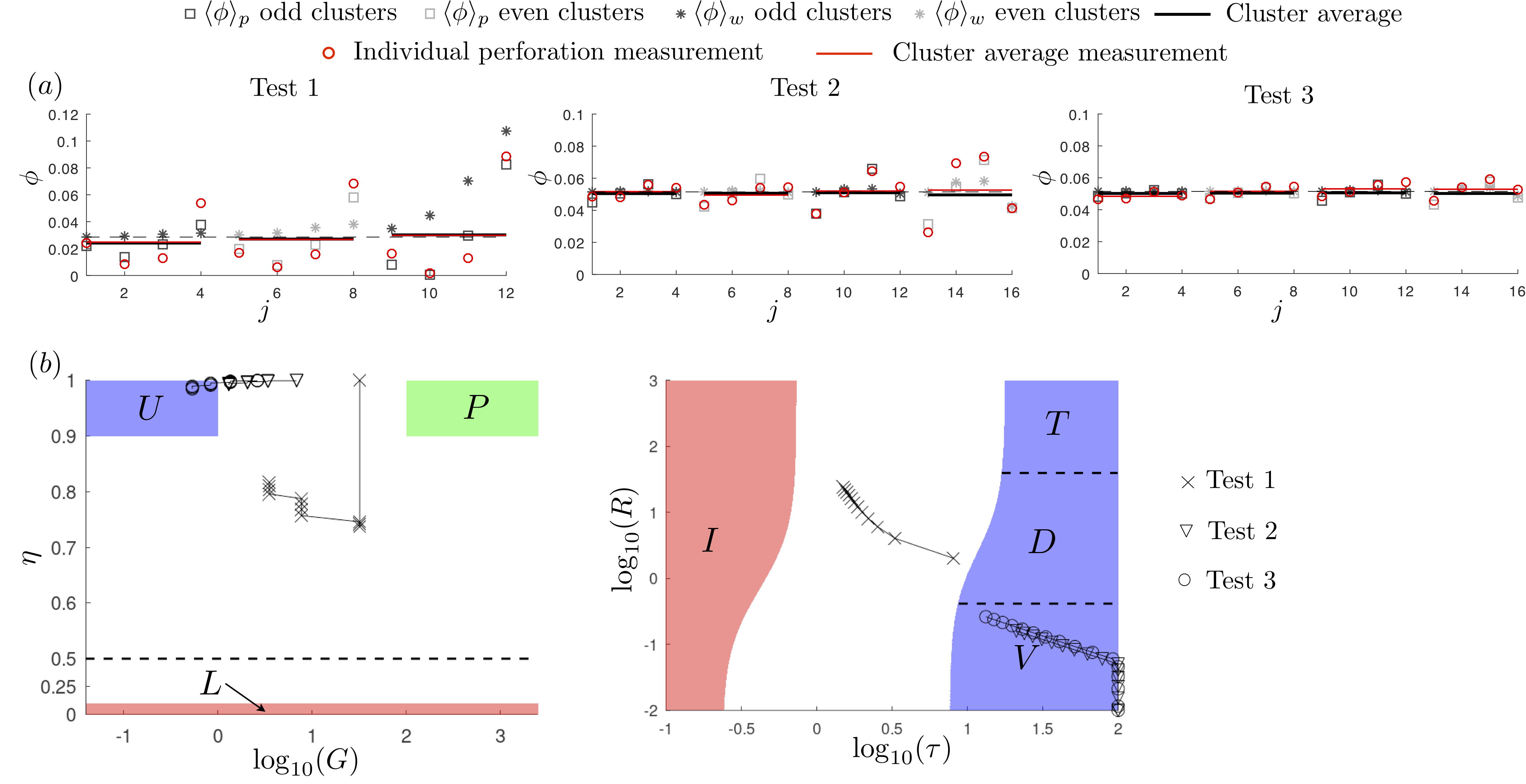}
\caption{Comparison between the model and experimental data from~\cite{Ahmad2021,Liu2021}. The cluster average laboratory data (solid red lines) are compared to the cluster average model predictions (solid black lines). Individual perforation measurements (red circular markers) are compared to the model result for each individual perforation (square markers). Panel $(a)$ shows the variation of proppant distribution between perforations for three different test cases. The dashed lines indicate the initial average volume fraction. The panel $(b)$ shows the parametric spaces for these different test cases.}
\label{fig27}
\end{figure}

\begin{figure}
\centering \includegraphics[width=0.99\linewidth]{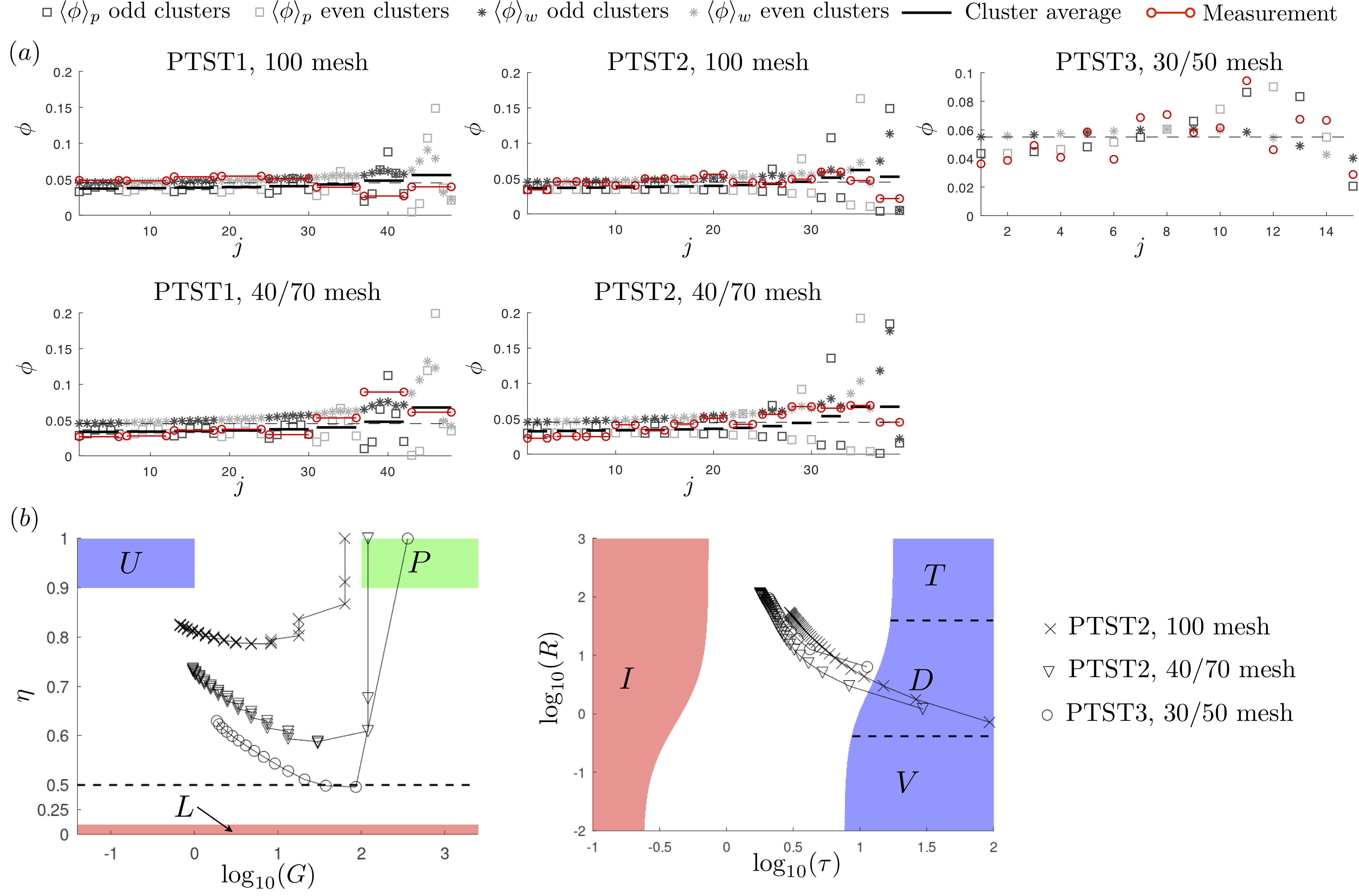}
\caption{Comparison between the model and experimental data from~\cite{Snider2022,Kolle2022}. The cluster average laboratory data (solid red lines with circular ends) are compared to the cluster average model predictions (solid black lines). Panel $(a)$ shows the variation of proppant distribution between perforations for different test cases. The dashed lines indicate the initial average volume fraction. The panel $(b)$ shows the parametric spaces for different test cases.}
\label{fig28}
\end{figure}

\begin{figure}
\centering \includegraphics[width=0.99\linewidth]{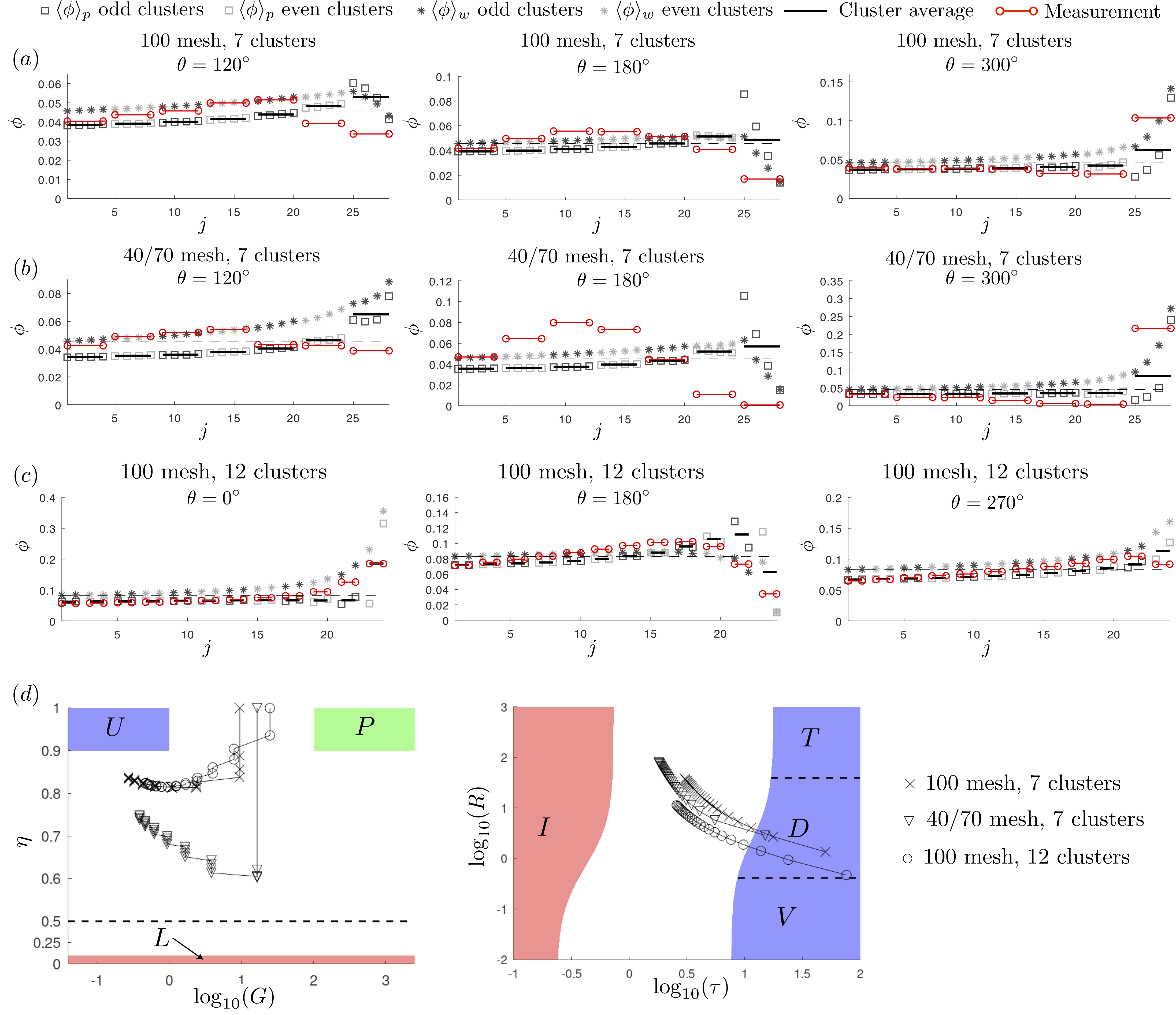}
\caption{Comparison between the model and CFD modeling data from~\cite{Benish2022,SniderPPT}. The cluster average laboratory data (solid red lines with circular ends) are compared to the cluster average model predictions (solid black lines). Panels $(a)$-$(c)$ show the variation of proppant distribution between perforations for different test cases. The dashed lines indicate the initial average volume fraction. The panel $(d)$ shows the parametric spaces for different test cases.}
\label{fig29}
\end{figure}

Results of comparison between the model and available data are shown in Figs.~\ref{fig20}-\ref{fig29}. All the figures contain the plots showing proppant distribution between clusters for different parameters, as well as the parametric spaces in the lower part of each figure indicating the location of the considered cases in the $(\eta,G)$ and $(R,\tau)$ diagrams. The proppant distribution figures plot particle volume fraction $\phi$ versus perforation number $j$. The square markers indicate the actual proppant volume fraction in the slurry flowing through $j$th perforation. Note that different shades of the grey color are used to distinguish between odd and even numbered holes for visualization purposes. The star markers correspond to the average particle volume fraction in the wellbore in the flow prior to $j$th perforation. The black lines depict the average proppant fraction per perforation cluster. The red lines with circular markers correspond to data (either experimental or CFD simulations). Finally, the black dashed lines show the initial average volume fraction in the wellbore. The $(\eta,G)$ and $(R,\tau)$ parametric spaces are shown in the lower part of each figure. Markers indicate values of the parameters for each perforation hole. In general, the trajectory for each case follows from left to right. I.e. the heel clusters tend to have smaller values of $G$ and $\tau$, while the toe clusters have larger values of $G$ and $\tau$. Also, the distance between the perforation holes within the cluster is ignored and therefore all perforations within the same cluster have the same value of $G$. The default fluid is water with properties $\mu=0.001$~Pa$\cdot$s and $\rho_f=1000$~kg/m$^3$, while the default sand density is $\rho_p=2650$~kg/m$^3$. These parameters are used, unless specified otherwise.

Fig.~\ref{fig20} summarizes the comparison between the model and laboratory data from~\cite{NgameniMS}. The latter study considered the flow of slurry in a pipe with the diameter $d_w=2.5$~in with the velocity $5.5$~ft/s. There are three perforation clusters located $5$~ft away from each other. Each perforation cluster has 6 perforations with 60$^\circ$ phasing, i.e. $\theta_j=60j$, $j=\{0,1,2,3,4,5\}$. Perforation diameter is $d_p=0.25$~in. Three different particle sizes are used: 20/40 mesh proppant ($a=0.32$~mm), 40/70 mesh proppant ($a=0.19$~mm), and 100 mesh proppant ($a=0.09$~mm). Water is used as a carrier fluid and various particle concentrations are used for each proppant type. The model assumes uniform slurry distribution between perforations, even though the experiments showed some variation. Results demonstrate a general downward trend, where the first heel cluster receives more proppant than average, while the last cluster receives less proppant than average. The dependence on the particle size and particle volume fraction is relatively minor. However, the laboratory data for 20/40 mesh particles consistently shows stronger heel bias. The model overall captures the observed data. What is also very interesting is that the distribution of particles per perforation varies significantly within the cluster. The holes located in the upper part of the well receive virtually no proppant, while the vast majority of particles come through one perforation located at the bottom. With the reference to the $(\eta,G)$ diagram on the panel $(f)$, this can be explained by the large values of $G$ (even for the first cluster), which cause very asymmetric particle distribution in the wellbore, as is evident from Fig.~\ref{fig8}. This is also consistent with Fig.~\ref{fig9}, which shows that there is a very strong sensitivity of the particle volume fraction on perforation orientation. Various particle sizes also cause different turning efficiency $\eta$. But the latter does not play a significant role since it is overshadowed by the effect of the non-uniform particle distribution in the wellbore. To better understand this, consider the following. Changing the particle size is able to change the turning efficiency by 10s of percent. As shown in Fig.~\ref{fig9}, the variation with respect to perforation orientation is on the order of 10s of percent for $G=1$. Once $G=10$, then the ratio between the amount of proppant received by the bottom and the top perforations exceeds 10. Thus, even for moderate values of $G\geq O(1)$, the effect of particle settling in the wellbore dominates the result. With respect to the $(R,\tau)$ parametric space, the trajectories follow from the nearly inertia dominated parameters for the first perforations to nearly drag dominated parameters for the last perforations. As expected, smaller particles tend to have higher turning efficiency and are closer to the drag dominated limit.
  
Figs.~\ref{fig21}-\ref{fig26} show the results of comparison with the laboratory measurements from~\cite{AhmadPhD}. The wellbore diameter is 1.5~in. Three cluster geometry is used with 6~ft spacing and 4 holes per cluster, phased with 90$^\circ$, i.e. $\theta_j=90j$ $j=\{0,1,2,3\}$. Perforation diameter is $d_p=0.25$~in. Fig.~\ref{fig21} compares the results for 20/40 sand ($a=0.34$~mm) for three different velocities $v_w=\{6.4,11,13.7\}$~ft/s. Fig.~\ref{fig22} compares the results for 40/70 sand ($a=0.18$~mm) for the same three different velocities $v_w=\{6.4,11,13.7\}$~ft/s. Fig.~\ref{fig23} compares the results for 20/40 ultralight proppant ($a=0.33$~mm, $\rho_p=2000$~kg/m$^3$), also for three different velocities $v_w=\{6.4,11,13.7\}$~ft/s. Fig.~\ref{fig24} compares the results for 40/70 ultralight proppant ($a=0.18$~mm, $\rho_p=2000$~kg/m$^3$) for the same three different velocities $v_w=\{6.4,11,13.7\}$~ft/s. Fig.~\ref{fig25} compares the results for 20/40 sand ($a=0.34$~mm) for different high viscosity friction reducer (HVFR) concentrations for three different velocities $v_w=\{4.5,8.2,11\}$~ft/s. Finally, fig.~\ref{fig26} compares the results for 40/70 sand ($a=0.18$~mm) for different HVFR concentrations for three different velocities $v_w=\{4.5,8.2,11\}$~ft/s. Rheology of HVFR fluids are: $k_1=0.0348$~Pa$\cdot$s$^{n_1}$, $n_1=0.57$; $k_2=0.066$~Pa$\cdot$s$^{n_2}$, $n_2=0.54$; $k_3=0.1535$~Pa$\cdot$s$^{n_3}$, $n_3=0.51$; $k_5=0.3446$~Pa$\cdot$s$^{n_5}$, $n_5=0.59$. Here the subscript indicates concentration in the units of gallons per thousand gallons of water (gpt), i.e. 1~gpt, 2~gpt, 3~gpt, and 5~gpt HVFRs were used. 

Results in Fig.~\ref{fig21} demonstrate that there is a strong dependency on wellbore velocity, while the variation with respect to the initial volume fraction is small. The first cluster tends to receive more proppant and the last cluster receives less proppant for the low velocity cases. At the same time, the trend is exactly the opposite for the cases with high velocity. Namely, the first cluster receives less proppant, while the last cluster receives more proppant than average. With the reference to the parametric spaces shown in Fig.~\ref{fig21}$(d)$, this is explained by the shift in the values of $G$ from higher to lower values as the velocity increases. The cases with the highest velocity have more uniform particle distribution within the wellbore to begin with. Thus, the observed lower volume fraction is due to the turning efficiency being less than one. Also, notice how relatively uniform particle distribution is for the first cluster for the $v_w=13.7$~ft/s cases. The variability greatly increases for the lower velocities. Overall, the model captures the experimental data reasonably well. The turning efficiency is practically independent of velocity and on the order of 0.7 for 20/40 mesh sand. Regarding the dimensionless turning time $\tau$, the data for most of the perforations corresponds to nearly inertial $I$ limit, for which there is almost no effect of the drag force during the turning process.

Fig.~\ref{fig22} shows the results for 40/70 mesh sand and summarizes the dependency on the particle volume fraction and velocity. Observations are very similar, as in Fig.~\ref{fig21}. There is a transition in the qualitative behavior for the increasing wellbore velocity. The model captures the observed behavior better, even for the lowest velocity considered. The primary difference between using 20/40 mesh and 40/70 mesh proppant lies in better turning efficiency, which makes the particle distribution more uniform for the high velocity cases.

Fig.~\ref{fig23} shows the results for 20/40 mesh ultralight proppant. A very similar trend is observed, showing a transition from the nearly uniform proppant distribution to the toe bias for the increasing wellbore velocity. The overall agreement with the model is excellent. The ultralight proppant has density of 2000~kg/m$^3$. Such a change in the density reduces the values of $G$, which makes the particle distribution inside the well more uniform. Despite that the particles are relatively large, the turning efficiency is almost the same as for the regular 40/70 mesh sand since the density is reduced. Recall, that the turning efficiency primarily affects the level of proppant reduction for the first cluster for the high velocity cases.

Fig.~\ref{fig24} shows the results for 40/70 mesh ultralight proppant. The use of smaller particles further increases the turning efficiency, which practically does not affect the low velocity cases. But, at the same time, the degree of uniformity of particles for the high velocity cases is better. Recall that the parameters correspond to relatively high values of $G$ for the case of low velocity, in which case the level of particle asymmetry in the wellbore dominates the overall response. For the high velocity cases, on the other hand, the values of $G$ are smaller and hence there is some sensitivity of the result to the turning efficiency.

Fig.~\ref{fig25} examines the effect of various concentrations of HFVR on the result for the case of 20/40 mesh sand. Different wellbore velocities are considered and various HVFR concentrations are used. Overall, there is the same trend of first cluster dominance for the low velocities and last cluster dominance for the high velocities. HVFR concentration does not significantly affect the result, at least according to the model. The agreement between the model and experiments appears to be less accurate for the low HVFR concentrations, while it becomes better for high HVFR concentrations. Again, the same trend versus velocity is observed. As indicated on the parametric spaces, the use of high concentration HVFR leads to better turning efficiency for the second part of the stage. Indeed, wellbore velocity is reduced in this region, which effectively increases the viscosity and elevates the turning efficiency by bringing the parameters closer to the drag dominated limit.

Fig.~\ref{fig26} compares the results for various HVFR concentrations for the case of 40/70 mesh sand. Overall, the results are qualitatively similar to the previously considered 20/40 mesh sand. The agreement between the model and the data is better. In addition, the effect of high concentration HVFR is stronger for smaller particles, as is evident from the values of the turning efficiency shown on the panel $(e)$. As a result, the proppant distribution is more uniform for the case of higher velocity.

Fig.~\ref{fig27} shows results of comparison with laboratory data published in~\cite{Liu2021}. The original reference for the laboratory results is~\cite{Ahmad2021}. Three tests were performed. In the first test, the wellbore diameter is $d_w=1.5$~in, there are three perforation clusters spaced 7~ft apart. Each perforation cluster has 4 holes with orientations $\theta=\{90^\circ,0^\circ,270^\circ,180^\circ\}$ and diameter $d_p=0.25$~in. Tap water with 0.65~ppg 40/70 mesh proppant ($a=0.14$~mm) were pumped with the rate 79~gal/min. Test 2 and 3 have different configuration. The wellbore diameter is $d_w=2$~in, there are 4 clusters spaced 6~ft apart. Each cluster has 4 perforation holes with orientations $\theta=\{0^\circ,90^\circ,180^\circ,270^\circ\}$ and $d_p=0.25$~in. Slurry consisting of 50/140 mesh proppant ($a=0.103$~mm) with 1.2~ppg was injected with 85~gal/min and 140~gal/min rates for the tests 2 and 3, respectively. Two different HVFR concentrations of 0.25~gpt and 0.65~gpt were used for the tests 2 and 3, respectively. As indicated in~\cite{Liu2021}, authors had difficulties matching the result due to possibly complex rheology of the HVFR fluids. Consequently, consistency and power-law indices are adjusted to match the result. The comparison for all test cases is very good (recall, that there is no viscosity adjustment for the first test case). The primary advantage of the laboratory results~\cite{Ahmad2021,Liu2021} lies in the measurement of the amount of proppant per perforation. Therefore, the comparison is made not only in terms of the average per cluster, but for each individual perforation as well. Clearly, there is a strong variation within each cluster, which is captured well by the model. Results for the test cases 2 and 3 are rather trivial and predict a nearly uniform proppant distribution between clusters. For the test case 3, the distribution between individual perforations is also nearly uniform. The turning efficiency is close to 1 for both of these tests due to high viscosity used to match the data and the test 3 practically falls into the $U$ regime shown on the $(\eta,G)$ parametric space.

Fig.~\ref{fig28} compares the modeling results to field scale experiments from~\cite{Snider2022,Kolle2022}. There are three experiments called PTST1, PTST2, and PTST3. For the first two cases, two proppant sizes were used, namely 100 mesh and 40/70 mesh sand. For the third case, only 30/50 mesh sand was used. The wellbore diameter is $d_w=5.5$~in for all tests. For PTST1, there are 8 clusters spaced 15~ft apart. Each cluster has 6 perforations, phased by 60$^\circ$. Perforation diameters are assumed to be all equal and are taken as $d_p=0.33$~in for simplicity. The injection rate is 90~bpm. The initial particle volume fraction is 0.045. Particle size is taken as $a=0.09$~mm for 100 mesh sand and $a=0.19$~mm for 40/70 mesh sand. HVFR fluid is used with the parameters $k=0.13$~Pa$\cdot$s$^n$ and $n=0.47$ (fitted based on the provided data). The test case 2 or PTST2 has the same wellbore diameter, but different cluster arrangement. There are 13 clusters spaced 15~ft apart. Each cluster has 3 perforations phased by 120$^\circ$. Perforation diameter is taken as $d_p=0.33$~in and is the same for all holes. The injection rate is the same 90~bpm, the particle volume fraction is the same, and particles sizes are the same also. The HVFR concentration is different, which resulted in $k=0.02$~Pa$\cdot$s$^n$ and $n=0.7$. The last test case 3 or PTST3 has the same wellbore and perforation diameters, but has 15 clusters with only one perforation per cluster, which is oriented downwards. The injection rate of 67~bpm is used together with water as a carrier fluid. Particle seize corresponds to 30/50 mesh sand and is taken as $a=0.21$~mm. The initial particle volume fraction is 0.055. These input parameters correspond to field data and therefore are more representative. First of all, the first clusters have lower values of $G$ and therefore more uniform particle distribution within the well's cross-section. The values for the last clusters are quite large and can reach $G=100$. The turning efficiency strongly depends on particle size, with smaller particles having higher efficiency. This is because smaller particles are closer to the drag dominated turning $D$ compared to large particles that are closer to the inertia dominated turning $I$. Comparison between the experimental data and the model is overall great, especially for PTST3. The degree of non-uniformity is relatively small for PTST1 and PTST2. The variation for 40/70 mesh proppant is higher and has stronger toe bias. This is captured by the model. At the same time, the ``dive'' of the proppant concentration for the last cluster for both PTST2 results is not as accurate. There are also ``wiggles'' in the data that are not captured and are likely caused by the variation of perforation diameter, which is not accounted for. What is striking is that there is a very strong variation of the amount of proppant per perforation for the last third of the stage. Thus, the effect of perforation hole size variability can be significant. Also, these perforations erode with different rates, which can further affect the result. With regard to PTST3, the agreement is excellent. First clusters receive less proppant due to relatively small turning efficiency. This causes the wellbore concentration to gradually increase. Then, once the flow slows down and asymmetry develops within the wellbore cross-section, more particles enter the perforation holes since they are located predominantly at the bottom of the well. This in turn reverses the growing trend of the proppant concentration in the wellbore and eventually leads to the reduction of the proppant received by toe clusters.

Fig.~\ref{fig29} compares the developed model to CFD simulations provided in~\cite{Benish2022,SniderPPT}. There are three main test cases. The panel $(a)$ corresponds to the test case with 100 mesh proppant and 7 cluster design. The panel $(b)$ corresponds to the test case with 40/70 mesh proppant and also 7 cluster design. In all of these cases $d_w=4.67$~in and there are 7 clusters spaced 25~ft apart. There are 4 oriented perforations per cluster with 0.33~in diameter. Particle mass density is taken as $\rho_p=2500$~kg/m$^3$, the injection rate is 90~bpm, particle concentration of 1~ppg is assigned, and viscosity of 2~cp is used. Various perforation orientations are used, as is indicated on the figure. The panel $(c)$ correspond to different set of input data, for which $d_w=5.2$~in, there are 12 perforation clusters spaced 15~ft apart, each having 2 oriented perforations with the diameter $d_p=0.43$~in. The injection rate is 100~bpm, the particle size is $a=0.15$~mm, viscosity is $\mu=10$~cp, and the particle concentration is 2~ppg. Various perforation orientations are used, as indicated on the figure. Results demonstrate that the agreement for the first cases (panels $(a)$ and $(b)$) is less accurate, while the last case (panel $(c)$) is much more accurate. This is somewhat contradictory since the parameters used for the cases on the panels $(a)$ and $(c)$ are very close to each other, even though come from two different references. The first test cases would fit much better, if the value of $G$ is increased by approximately the factor of 4. This would introduce much more non-uniform particle distribution within the well and cause the characteristic maximum of the proppant received per cluster observed roughly in the middle of the stage. The overall conclusion stemming from this figure is the following. Non-uniform particle variation within the wellbore significantly affects the result. Both the CFD models and the analytical models have to be calibrated specifically to capture this non-uniform particle variation with respect to problem parameters. Once the flow calibration is in agreement, then it is appropriate to compare the model and CFD results for the multi-cluster geometry.

\section{Discussion}\label{secdisc}

\begin{figure}
\centering \includegraphics[width=0.7\linewidth]{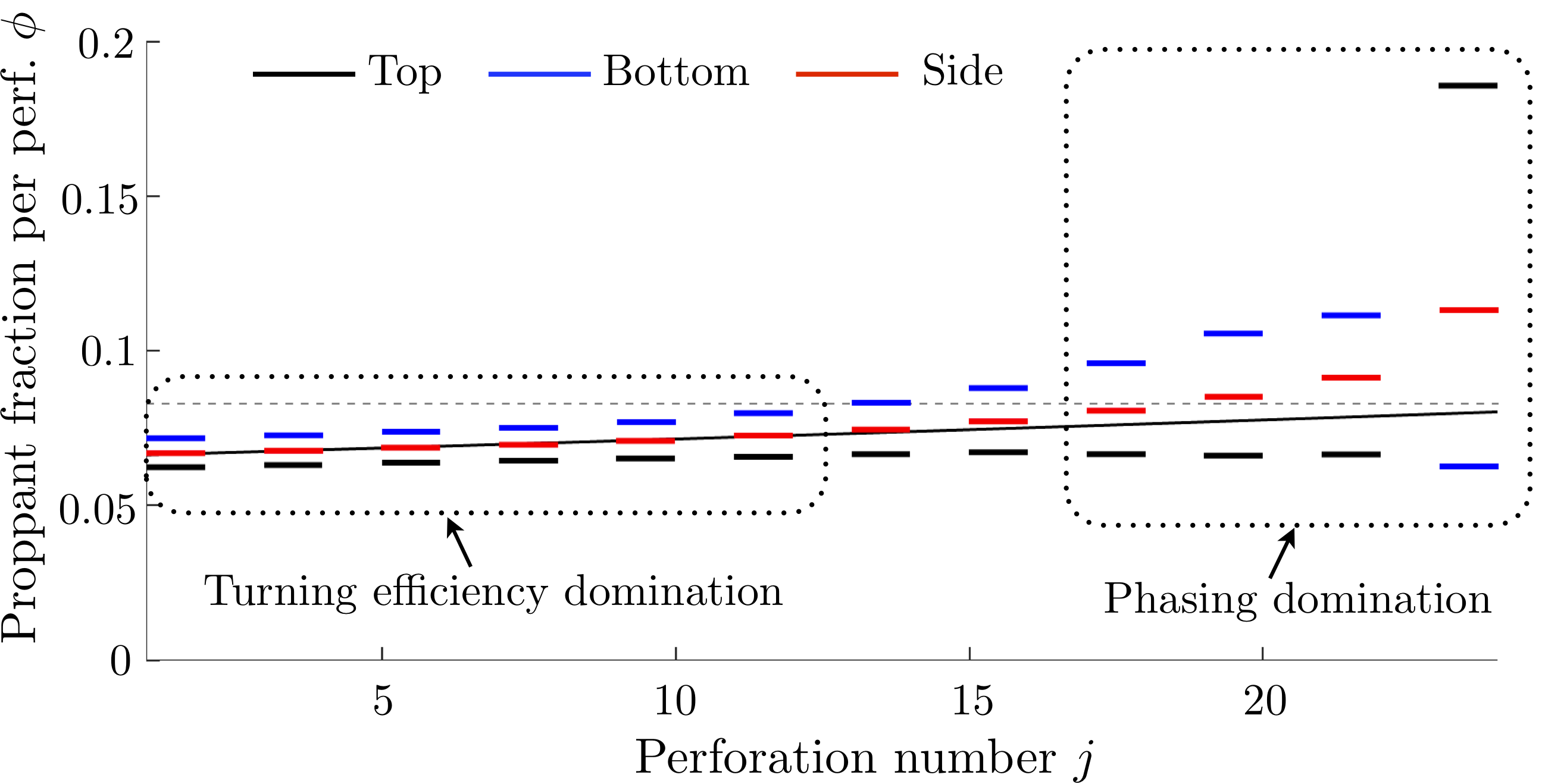}
\caption{Schematics of proppant distribution between clusters for a field case (data is the same as in Fig.~\ref{fig29}$(c)$).}
\label{fig30}
\end{figure}

The developed model suggests that there are two primary phenomena that affect the result. Namely, the non-uniform particle distribution within the wellbore, and non-trivial particle turning efficiency. In general, for field scale data, the particle concentration is relatively uniform within the first part of the stage. Therefore, the result is affected primarily by the turning efficiency there. In the second part of the stage (i.e. closer to the toe), the wellbore velocity drops and the particle distribution within the wellbore's cross-section becomes non-uniform. That is where the perforation orientation starts to play a significant role, but the turning efficiency becomes secondary. Note that based on the results shown in Figs.~\ref{fig25} and~\ref{fig26}, the addition of HVFR tends to increase the turning efficiency in the second part of the stage, which does not significantly affect the result since the overall effect of turning efficiency is relatively small. The bottom line here is that for the field scale parameters, the first part of the stage is primarily affected by the turning efficiency and does not depend on perforation phasing, while the second part of the stage is primarily affected by perforation phasing and does not significantly depend on the turning efficiency. This observation can be used to optimize particle distribution between perforation clusters and affect different parts of the stage independently. As an illustration, Fig.~\ref{fig30} schematically plots the solution for particle distribution for the cases shown in Fig.~\ref{fig29}$(c)$. The data corresponds to a representative field case, i.e the value of $G$ is very small for the first part of the stage and very large for the second part of the stage. Three perforation orientations are compared: top or $\theta=0^\circ$, bottom or $\theta=180^\circ$, and side or $\theta=90^\circ$. For the case of constant turning efficiency and negligible $G$, the slurry and proppant flow rates before and after the $j$th perforation are related as 
\begin{equation}\label{apprsol}
q^s_{j+1} = q^s_j-\dfrac{q^s_0}{N},\qquad q^p_{j+1} = q^p_j-\eta \phi_{w,j}\dfrac{q^s_0}{N},\qquad \phi_{w,j}=\dfrac{q^p_j}{q^s_j}.
\end{equation}
This system of equations can be solved to find $\phi_{w,j+1}=(1+(1\!-\!\eta)/(N\!-\!j))\phi_{w,j}$. Here $N$ is the total number of perforations, while $\phi_{w,j}$ is the particle concentration in the wellbore prior to $j$th perforation. The particle concentration entering the $j$th perforation is simply $\eta\phi_{w,j}$. This approximate solution is shown by the black line in Fig.~\ref{fig30} for $\eta=0.8$ (this number is taken from Fig.~\ref{fig29}$(d)$). The values of $G$ are small in the first part of the wellbore and the result there is dominated by the turning efficiency. The result for the sideways oriented perforation follows the approximate solution, constructed from~(\ref{apprsol}). There is a small effect of $G$ and the associated slight asymmetry of the particle distribution, so that the bottom perforations receive a little bit more proppant than the top. The perforation located on the side effectively averages the asymmetry in particle distribution and hence matches the approximate solution better. In the second part of the wellbore, there is a strong dependence on perforation orientation, as is indicated on the figure. The effect of turning efficiency is smaller in that zone. There is also a transition zone, in which both effects are important and are comparable in magnitude. 

Another important observation is that particle concentration varies significantly versus perforation orientation for the toe clusters. Thus, this can lead to screenout of some of the perforations. It can also cause very different rates of perforation erosion. This highlights the necessity of modeling each perforation independently, as well as to perform laboratory measurements for each perforation independently. To this end, only~\cite{Ahmad2021,Liu2021} considered analyzing the amount of proppant received by each individual perforation. 

The model considers horizontal wellbore, but can be easily extended for inclined wellbores. The simplest way to do that is to replace the gravitational constant $g$ by $g\cos(\psi)$, where $\psi$ is the angle between the wellbore and the horizontal axis. Therefore, the effect of the inclined wellbore can be simply summarized as the reduced value of the dimensionless parameter $G$.

Another important observation is related to the problem of proppant turning. As is evident from Figs.~\ref{fig15}-\ref{fig19}, the values of the perforation flow ratio $q^s_p/q^s_i$ are $O(1)$. At the same time, for the multi-cluster cases, this ratio is equal to $q^s_p/q^s_i = 1/(N-j+1)$, where $j$ the perforation number and $N$ is the total number of perforations. Here it is assumed that each perforation takes the same amount of slurry. Therefore, the first perforation has $1/N$, the middle perforation has roughly $2/N$ and the last perforation has $1/2$. Given that $N\approx 40$ for field applications (and that the turning is more relevant for the first part of the stage), then the most relevant values of $q^s_p/q^s_i$ are under $0.1$. As a result, it is recommended to perform calibration for this range of $q^s_p/q^s_i$. This is not done at this point. The calibration for higher ratios and extrapolation to the lower ratios is probably accurate, but it is still worth checking.

The developed model has several fitting parameters. The choice of these parameters is not perfectly unique. In other words, it is possible to change these parameters in certain range and still have an adequate match. Also, the steady-state profile for particle concentration in the wellbore does not depend on particle size, while the time needed to reach this profile depends on it. This may be somewhat counterintuitive and needs experimental validation. To have a better calibration, the following laboratory experiments can be done. First of all, the problem needs to be decoupled into slurry flow and turning the corner problem. For the slurry flow, the goal is to have measurements similar to~\cite{GilliesPhD}. One drawback of the results in~\cite{GilliesPhD} is that particles with a wide range of sizes were mixed. Therefore, repeating the results for nearly mono-disperse particles can improve the calibration procedure. Also, it is important to experimentally measure the settling dynamics in the wellbore with the goal of calibrating the effective viscosity parameter $p_V$. At this point, there is no such measurement and therefore the calibration of $p_V$ is indirect. In other words, the goal is to measure the transition from nearly uniform particle concentration at the inlet to some steady-state profile further away from the inlet. The effect of fluid rheology was also not considered in~\cite{GilliesPhD}, and needs to be investigated. Regarding the particle turning, it would be better to perform it for vertical wells to eliminate the effect of gravity. In this case, this problem will be decoupled from the flow problem and the issue of the non-uniform particle distribution is not going to be present. As was mentioned above, the focus should also be made on measurements for small ratios $q^s_p/q^s_i$ (the ratios between the perforation and wellbore slurry flow rates).

Finally, CFD simulations provide a very useful tool, which allows to model dynamics of slurry in a perforated wellbore. For this tool to be viable, it needs to be calibrated against laboratory data. In particular, it is important to calibrate it for two problems: slurry flow in the wellbore and the associated particle concentration and velocity profiles, and turning the corner problem. Specifically, the calibration to match particle concentration profile is crucial and it has not been done in the current CFD modeling studies.

\section{Summary}\label{secsumm}

This paper considers the problem of slurry flow in a perforated wellbore. The aim is to develop a model to predict the proppant distribution between perforations for the given problem parameters. This problem has two sub-problems. The first one is related to the fully turbulent flow of slurry in a pipe. The solution of this problem is particle concentration and velocity profiles within the wellbore and their time dependence. The second sub-problem is related to the particle making a sharp turn from the wellbore to perforation. This leads to some particles missing the perforation entry and effectively lower particle volume fraction entering the perforation. Generally speaking, there is also erosion of perforations, which can also affect the result. But, at this stage, perforation erosion is not included.

A mathematical model for the turbulent flow of slurry is developed. It has fitting parameters that are calibrated against available experimental data. The latter experimental results consist of particle concentration and velocity profiles for various particle sizes, velocities, concentrations, and pipe diameters. The model predicts steady-state solution for the particle concentration and velocity profiles, as well as the time dependent transition to it. The primary dimensionless parameter that affects the result is the dimensionless gravity $G$. For small values of $G$, the particles are distributed nearly uniformly. At the same time, for very large values of $G$, the suspension flow resembles the flowing bed state, in which most of the particles flow at the bottom of the horizontal well.

A mathematical model for the proppant turning problem is developed. It also has a fitting parameter and is calibrated against a series of laboratory experiments as well as CFD simulations. The model predicts that there are different limits of the solution, which includes the inertial limit, in which the drag force does not affect the mechanics of turning. Particle size and fluid viscosity do not enter the result in this case. At the same time, another limit is drag dominated. The drag force dictates the turning ability in this case and particle size and fluid rheology may affect the result. Overall, the parameter that quantifies the turning ability is turning efficiency $\eta$. If $\eta=1$, then all particles flowing to the perforation actually enter the perforation. If $\eta=0$, then only fluid enters the perforation.

The two aforementioned models are combined to solve the problem of slurry flow in a perforated wellbore. It is shown that the two primary parameters that affect the behavior are the dimensionless gravity $G$ and turning efficiency $\eta$. Based on the values of these parameters, the particle distribution between the clusters can be nearly uniform, toe biased, heel biased, or be more complicated. Comparison with numerous laboratory and field scale experimental results is performed. In addition, a comparison with available CFD simulations is made. Note that the calibration was done previously for the two sub-problems, while here the comparison is made to evaluate accuracy of the model. Results demonstrate an overall good agreement between the model and the available data, with the exception of some CFD results. One important conclusion that is relevant for optimization of field scale problems is that the turning phenomenon is more relevant within the first or heel part of the stage, but the perforation phasing is almost irrelevant there. At the same time, the non-uniform particle distribution in the wellbore provides the dominant effect in the second or toe part of the wellbore. Therefore, there is strong sensitivity to perforation phasing in this part of the stage, while the effect of turning efficiency is secondary. Also, there is a strong variation of the amount of proppant with respect to perforation orientation in this zone. 

Potential drawbacks of the developed model are also discussed. A series of laboratory experiments is suggested to clarify all the issues. These experiments would fill the gaps in the currently available data and allow to have a more accurate model. With regard to CFD modeling, it is necessary to calibrate the model for the problem of turbulent slurry flow in a wellbore. This is something that the authors of current studies did not do thoroughly. The non-uniform particle distribution in the wellbore strongly affects the result, especially in the second part of the stage. As such, CFD models need to be calibrated against available experimental data to be reliable.

\section*{Acknowledgements}

I would like to sincerely thank Dr. Mark McClure (ResFrac Corporation) for useful recommendations regarding this paper.



\end{document}